\documentclass[pre,twocolumn,preprintnumbers,amsmath,amssymb]{revtex4}

\usepackage{graphicx}
\usepackage{dcolumn}
\usepackage{bm}
\newcommand{\beq}{\begin{equation}}
\newcommand{\eeq}{\end{equation}}
\newcommand{\abs}[1]{|#1|}
\newcommand{\A}{{a^{\dagger}}}

\begin{document}
\title{Self Consistent Proteomic Field Theory of Stochastic Gene Switches}
\author{Aleksandra M. Walczak$^a$}
\email{walczak@physics.ucsd.edu}
\author{Masaki Sasai$^c$}
\email{sasai@info.human.nagoya-u.ac.jp}
\author{Peter G. Wolynes$^{a,b}$}
\email{pwolynes@ucsd.edu}
\affiliation{$^a$ Department of Physics and Center for Theoretical Biological Physics,
$^b$ Department of Chemistry and Biochemistry\\
University of California at San Diego, La Jolla, CA 92093, USA \\
$^c$ Department of Complex Systems Science, Graduate School of Information Science, Nagoya University, Nagoya 464-8601, Japan\\}
\date{\today}
\begin{abstract}
We present a self-consistent field approximation to the problem of the genetic switch composed of two mutually repressing/activating genes. The protein and DNA state dynamics are treated stochastically and on equal footing. In this approach the mean influence of the proteomic cloud created by one gene on the action of another is self-consistently computed. Within this approximation a broad range of stochastic genetic switches may be solved exactly in terms of finding the probability distribution and its moments. A much larger class of problems, such as genetic networks and cascades also remain exactly solvable with this approximation. We discuss in depth certain specific types of basic switches, which are used by biological systems and compare their behavior to the expectation for a deterministic switch.
\end{abstract}
\maketitle
\subsection*{Introduction}
Genetic switch systems are an elementary means of regulatory control present in every living organism. Their complexity and details differ, but the general mechanism of the expression of a given gene being regulated by proteins, is believed to be universal (Ptashne and Gann, 2002). They are building blocks of larger regulatory elements: genetic networks and signaling cascades. The pathways by which these systems operate is passed on from generation to generation. Understanding their stability and characteristics is therefore fundamental. A lot of previous work has considered a deterministic description of genetic switches (Shea and Ackers, 1982), (Hasty et al., 2001). The need for a stochastic treatment of genetic switches due to the single copy of the DNA molecule and multiple protein molecules in the cell, has been largely recognized (Sneppen and Aurell, 2002), (Kepler and Elston, 2001). \\
The most general way of accounting for non deterministic processes is to write down the master equation for a given system. To define the state of the switch one must specify the DNA binding states of particular genes and the number of proteins of each type. The probability distribution even of a single switch consisting of two genes, the product proteins of which act as regulator proteins for the system, may not be determined exactly and approximations must be considered (Bialek, 2001), (Hasty et al., 2000), (Sneppen and Aurell, 2002).\\
Several approaches to account for the probabilistic nature of chemical reactions have been undertaken, ranging from the Langevin description of single genes (Bialek, 2001), and two interacting gene switches (Hasty et al., 2000), to the master equation reduced to a Fokker-Planck equation  considerations (Kepler and Elston, 2001), (Hasty et al., 2001a). A dynamical action formulation has also been used (Sneppen and Aurell, 2002) to determine the lifetimes of states of the switch. A popular alternative to purely analytical methods, which often need to make approximations or are limited to very simple model systems, has been to conduct stochastic simulations of genetic switches. Two types of simulations are mostly used. In the first the randomness of the system is introduced by means of a Monte Carlo algorithm with fixed time step (Paulsson et al., 2000). The second is based on the Gillespie algorithm (Gillespie, 1977) to predict the probability of a given reaction occurring (Arkin et al., 1998). For single gene systems, stochastic simulations have shown that stochasticity in the system is responsible for the bimodal probability distributions (Cook et al., 1998), observed experimentally. These methods prove very useful, as they allow us to test the theoretical predictions on model systems, which might be hard to build experimentally. However this approach often does not enable us to gain intuition or insight into the mechanisms behind the functioning of the system. The aim of the present work is to gain a better and deeper understanding of the device physics of genetic switches. We therefore, contrary to many important previous discussions (McAdams and Arkin, 1997), (Aurell et al., 2002), (Vilar et al., 2003) do not present a specific concrete biological system, but discuss generic behavior and try to understand its sources. Our approximation also allows for an exact solution of a broad class of genetic switch systems without any further assumptions and with little computational effort. Hasty et al (Hasty et al., 2001b) present an overview of the existent theoretical approaches.\\
A popular approximation, assumes the DNA binding state reaches equilibrium much faster than the protein number state. Therefore the adiabatic approximation is often considered (Shea and Ackers, 1982), (Sneppen and Aurell, 2002), (Darling et al., 2000), allowing for a thermodynamic treatment (Shea and Ackers, 1982) of the DNA binding state. The protein number fluctuations are then treated stochastically. Even before the statistical thermodynamics approach of Shea and Ackers (Shea and Ackers, 1982) using partition functions, much previous work assumed the DNA binding and unbinding can simply be accounted by an equilibrium constant, since the relaxation timescales for equilibration of the DNA state are much larger than those of the protein numbers, which require protein synthesis and degradation to change. The partition function approach has also been successful at looking at logic gates build from switches (Buchler et al., 2003). The adiabatic approximation is believed to hold true in many cases, judging by the experimental parameters of biological switches (Darling et al., 2000). But as the experiments of, for example Becskei et al (Becskei et al., 2001) show, not all switches need function in the adiabatic limit and the non-adiabatic limit may result in new phenomena. We therefore consider a wide range of parameter ratios in our discussion. \\
In this paper we explore more fully an approximation, previously used by Sasai and Wolynes (Sasai and Wolynes, 2003) for the variational treatment of the problem, the self-consistent proteomic field (SCPF) approximation. Within this approximation one assumes the probability of finding the switch in a given state is a product of probabilities of states of individual genes. One can then solve the steady state master equation for the probability distribution of many regulatory systems exactly. We discuss the approximation and present a detailed study of different classes of genetic switches, some of which have never previously been considered theoretically. We consider several particular features of such systems, found in known switches, separately to be able to characterize their contributions to the behavior of the whole system. To be specific, starting from a symmetric toggle switch, we go on to compare the effects of multimer binding and of the production of proteins in bursts on the stability of the switch.\\
The stochastic effects prove to be modest for symmetric switches without bursts, especially if the genes have a basal production rate. We find the deterministic and stochastic SCPF solutions to have similar probabilities of particular genes to be on and mean numbers of proteins of a given species in the cell. However in the non-adiabatic limit, when the unbinding rate from the DNA is smaller than the death rate of proteins, the probability distributions have two well defined peaks, unlike in the deterministic approximation or adiabatic limit of the stochastic SCPF solution. \\
We also show the effect of stochasticity on the observables becomes more apparent when proteins are produced in bursts. In these types of switches, the definition of the adiabatic limit, which was clear for the switches in which proteins are produced separately, is no longer simple. Our discussion shows that the properties of genes often analyzed in the deterministic limit, may be strongly influenced by stochasticity in this case. Randomness in a biological reaction system leads to quantitative and in many examples even qualitative changes from predictions of deterministic models. \\
We also discuss the differences in the behavior of an asymmetric and symmetric switch. We point to the mechanisms resulting in different types of bifurcations and show how they are influenced by noise. Within the SCPF approximation switches that are regulated by binding and unbinding of monomers, do not have regions of bistability. This holds true for both symmetric and asymmetric switches. When proteins are produced individually rather than in bursts, fast unbinding from the DNA can effectively minimize the destructive effect of protein number fluctuations on the stability of the DNA binding state. Furthermore a detailed analysis of the probability distributions show they have long tails and are far from Poissonian in both the adiabatic and non-adiabatic limit. We discuss the properties of the system in terms of clouds of proteins buffering the DNA. We show how fast  or slow DNA binding characteristics and protein number fluctuations influence the stability of the buffering clouds leading to specific emergent behavior of observables. Throughout the paper a comparison is made between results of the exact stochastic solution with solutions of deterministic kinetic equations for the system, within the self-consistent proteomic field approximation. \\
We establish a base of potential building blocks of more complicated switches and systems, such as networks and signaling cascades, for which an exact solution within the present approximation can also be obtained. A detailed discussion of these larger systems will be the topic of another paper. We also present limitations of the present style of analysis where exact solutions are not possible.\\
There are two aims of this paper. The first is to discuss the self consistent field approximation and show that it has an exact solution which may be extended to a large class of systems. This approximation lets one deal in a straightforward and computationally inexpensive manner with the effect of random processes on genetic networks. The second is to discuss the many components of biological switches present in nature and in engineered systems, in the necessary stochastic framework. 
\subsection*{The Self-Consistent Proteomic Field Approximation}
The basic mechanism of gene transcription regulation in prokaryotes may be reduced to the binding and unbinding of regulatory proteins, repressors and activators, to the operator site of the DNA. If we use this simplified treatment, which neglects extra levels of regulation, such as the binding of RNA polymarase, effectively each gene can be described as being either in an active (on) state, when the repressor is unbound (activator bound), or in an inactive (off) state, with the repressor bound (activator unbound). The stochastic system of a single gene and its product proteins is described by the joint probability distribution $\vec{P}(n,t)=(P_1(n,t),P_2(n,t))$ of the number of product proteins in the cell $n$, and the DNA binding site state: on (protein not bound)- $1$, or off (protein bound) - $2$. To conserve probability $\sum_n \vec{P}(n,t)=1$.\\
If one considers two interacting genes, the description in terms of a joint probability vector needs to be extended to four states: both genes may be on, or off, or one of the genes may be on, the other off. If the two genes do not interact, as would be the case for two self regulatory proteins, the probability of a finding the two gene system in a given state, defined by both the number of product proteins and the DNA binding site state, would be the the product of the states of particular genes ${P}_{jj'}(n_1,n_2;t)=P_j(n_1;t) P_{j'}(n_2;t)$. This is generally not true for two interacting proteins, as is the case in a genetic switch. However, as a first approximation to the problem, one can ignore correlations between the spaces of the two genes and assume the space of the switch is a sum of spaces of the genes that compose it. Since we are looking for solutions in which the symmetry of the system is broken and different behaviors of the on and off state of a gene are possible, we must allow for different probability distribution functions for the on and off states. This is analogous to the unrestricted Hartree approximation in quantum mechanics, where allowing different spatial functions for spin up and spin down states results in breaking of the symmetry of the bound molecular orbital solution to the dissociated solution of two separate hydrogen atoms with opposite spin states for large internuclear distances. We therefore allow for multiple solutions for a given set of parameters. The total probability of having a given gene state $i$ and $n_i$ proteins of that type is simply given by $P_j(n_i,n_{i'})=P_{j,j'=0}(n_i,n_{i'})+P_{j,j'=1}(n_i,n_{i'})$.\\
The self-consistent approximation is a crude approximation since in the case of the genetic switch, the state of a given gene is determined by the number of protein products of the other gene. However, within this approximation, one can solve the master equation for the probability distribution exactly without any further approximations. This yields a powerful computational tool, which simultaneously gives useful insight.
\begin{figure}
\includegraphics[height=5.cm,width=6cm]{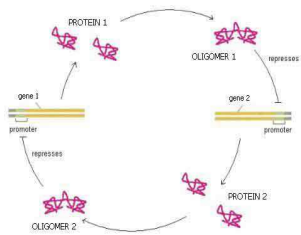}
\caption{A schematic representation of the toggle switch. Gene 1 produces proteins of type 1 which repress gene 2 and gene 2 produces proteins of type 2 which repress gene 1.}
\label{tswitch}
\end{figure}
\subsection*{The Toggle Switch}
For clarity of exposition, we show how the problem may be solved exactly within the self-consistent field approximation on a well defined system of the toggle switch. We then expand the method to apply to other systems. The elementary system we use as an example is composed of two genes, labeled 1 and 2, as presented in Fig. \ref{tswitch}. Gene 1 produces proteins of type 1 which, act as regulatory proteins, say repressors, on gene 2. The product of gene 2, proteins of type 2, in turn repress gene 1. In this simplified model, we assume that protein production occurs instantaneously upon unbinding of the repressor. For now, we assume that repressor proteins bind as dimers, since that is a common scenario in biological systems, but we do not treat dimerization kinetics explicitly. For simplicity the coupling form between the genes responsible for binding will be taken to be of the form $h_i n^p_{3-i}$, where $p$ is the order of the multimerization of the repressor. This form is a small approximation to the more exact $h_i n_{3-i} (n_{3-i}-1)...(n_{3-i}-p+1)$. We have checked that using the simpler monomial  does not influence the results in any regime discussed. We also do not account for the existence of mRNA molecules and the consequent time delays owing to their synthesis as intermediates. The extensions of the model are discussed later. Within the self-consistent field approximation the set of master equations for the corresponding system is of the form:
\begin{eqnarray*}
\frac{\partial P_1(n_i)}{\partial t}&=&g_1(i)[P_1(n_i-1)-P_1(n_i)]+\\
&&+k_i [(n_i+1)P_1(n_i+1)-n_iP_1(n_i)]+\\
&&-h_i n^2_{3-i} P_1(n_i)+f_iP_2(n_i) \\
\frac{\partial P_2(n_i)}{\partial t}&=&g_2(i)[P_2(n_i-1)-P_2(n_i)]+\\
&&+k_i [(n_i+1)P_2(n_i+1)-n_iP_2(n_i)]+\\ 
&&+h_i n^2_{3-i} P_1(n_i)-f_iP_2(n_i)
\end{eqnarray*}
for $n \geq 1$ where the $i=1,2$ refers to the gene label. $P_1(n_1)$ describes the probability of gene 1 being in the on state and there being $n_1$ protein molecules of type 1 in the cell. The first term on the right hand side of the equations describes the production of proteins of type $i$ with a production rate $g_j(i)$, where j=1,2, depending on whether the gene is in the on or off state. The second term accounts for the destruction of proteins with rate $k_i$. The binding of repressor proteins produced by the other gene is proportional to the number of dimer molecules present in the system $n_{3-i}$ with rate $h_i$. We assume unbinding occurs with a constant rate $f_i$. Binding and unbinding contributes to the kinetics of the DNA binding states, as described by the last two terms. This set is supplemented by the $P_j(n_i=0)$ equations to account for boundary conditions.\\
\begin{eqnarray*}
\frac{\partial P_1(n_i=0)}{\partial t}&=&-g_1(i)P_1(n_i=0)+k_i P_1(n_i=1)
\nonumber \\
&&-h_i n^2_{3-i} P_1(n_i=0)+f_iP_2(n_i=0)
\nonumber \\ 
\frac{\partial P_2(n_i=0)}{\partial t}&=&-g_2(i)P_2(n_i=0)+k_iP_2(n_i=1)
\nonumber \\ 
&&+h_i n^2_{3-i} P_1(n_i=0)-f_iP_2(n_i=0)
\end{eqnarray*}
For convenience, let us define $\sum_{n_i}P_j(n_i)=C_j$, the probability of finding the DNA binding site in a given state. One can now sum the $P_j(1)$ equations over the number states of the 2-the proteins with $P_1(2)+P_2(2)$, and likewise the $P_j(1)$ equations. Due to the SCPF approximation, the only term affected is the repressor binding term $h_1(n^2_2)$, and since $\sum_{n_2} P_1(2)+P_2(2)=1$, the summation results in $\sum_{n_2}h_1(n^2_2)(P_1(2)+P_2(2))=h_1(C_1(2)<n^2_{12}>+C_2(2)<n^2_{22}>) = h_1 F(2)$, where $<n^2_{j2}>$ is the second moment of the number distributions of type 2 proteins produced when gene 2 is in the j-th state. The equations of motion of the moments of the probability distribution are of the form:
\begin{eqnarray*}
\frac{\partial C_j(i) <n^k_{ji}>}{\partial t}=g_j(i)[<(n_{ji}+1)^k>-<n^k_{ji}>]C_j(i)&+& \\
+k_i [<n_{ji}(n_{ji}-1)^k>-<n^{k+1}_{ji}>] C_j(i)&+& \\
+ (-1)^jh_i F(3-i) <n^k_{1i}>C_1(i)&+&  \\
+(-1)^{j+1}f_i <n^k_{2i}>C_2(i)&&
\end{eqnarray*}
 The steady state equations for the moments of the distributions that follow are closed form, the $n_i^{th}$ order moment equation of motion depends only on the lower moments of the $i^{th}$ gene and $n_{3-i}^2$. \\
To analyze the behavior of switches we introduce the following scaled parameters: the adiabaticity parameter $\omega_i=f_i/k_i$, which represents the characteristic rate of change of the DNA state compared to the characteristic rate of change in protein number, $X_i^{eq}=f_i/h_i$ measures the tendency for proteins to be unbound from the DNA, $X_i^{ad}=\frac{1}{2}(g_1(i)+g_2(i))/k_i$ the effective production rate and $\delta X_i^{sw}=\frac{1}{2}(g_1(i)-g_2(i))/k_i$ distinguishes between the two DNA states in terms of protein dynamics. Furthermore, in the operator formalism developed for classical diffusion by Doi (Doi, 1976) and Zeldovich' and Ovchinikov (Zeldovich and Ovchinikov, 1978), the number operator may be written in terms of number state creation $\A$ and annihilation $a$ operators, as $n=\A a$. It is then particularly easy to write down the equations for the $a$ moments instead of the $n$ moments. Setting the left hand side to zero one obtains the steady state equations:
\begin{eqnarray*}
0=-\omega_i[\frac{F(3-i)}{X^{eq}_i}C_1(i)-C_2(i)]&&
\nonumber \\ \\
0=k[(X^{ad}_i+(-1)^j \delta X^{sw}_i)<a_{ji}^{k-1}>-<a_{ji}^k>]C_j(i)&+&
\nonumber \\
+(-1)^j \omega_i[\frac{F(3-i)}{X^{eq}_i}<a_{1i}^k> C_1(i)-<a_{2i}^k> C_2(i)]&&
\end{eqnarray*}
Using the probability conservation relation $C_1(i)+C_2(i)=1$, the zeroth order equations become:
\begin{eqnarray}
C_1(i)=\frac{X_i^{eq}}{X_i^{eq}+F(3-i)} &\nonumber   & C_2(i)=\frac{F(3-i)}{X_i^{eq}+F(3-i)}\\
\label{c1c2eqns}
\end{eqnarray}
Dividing the higher order $a_j(i)$ moment equations by $C_j(i)$ and using the relation $\frac{C_1(i)}{C_2(i)}=\frac{F(3-i)}{X^{eq}_i}$ from the zeroth order equations one can calculate 
\begin{eqnarray*}
<a^k_{1i}-a^k_{2i}>=((X^{ad}_i+\delta X^{sw}_i)<a^{k-1}_{1i}>+&& \\
-(X^{ad}_i-\delta X^{sw}_i)<a^{k-1}_{2i}>)\frac{k C_j(i)}{\omega_i+k C_j(i)}&&
\end{eqnarray*}
 which depends only on $a$ moments of lower order than the $k^{th}$ moment. This allows one to obtain the following form for the higher order $a$ moments
\begin{eqnarray*}
<a^k_{1i}>&=&(X_i^{ad}+\delta X_i^{sw})(1-\frac{\omega_i C_2(i)}{\omega_i+kC_1(i)})<a^{k-1}_1>+\\ \\
&&+(X_i^{ad}-\delta X_i^{sw})\frac{\omega_i C_2(i)}{\omega_i+kC_1(i)}<a^{k-1}_2>
\\ \\
<a^k_{2i}>&=&(X_i^{ad}-\delta X_i^{sw})(1-\frac{\omega_i C_1(i)}{\omega_i+kC_1(i)})<a^{k-1}_2>+ \\ \\ 
&&+(X_i^{ad}+\delta X_i^{sw})\frac{\omega_i C_1(i)}{\omega_i+kC_1(i)}<a^{k-1}_1> \\ \\
\end{eqnarray*}
Going back and forth between the two types of moments is straightforward. The n-moment equations have however more complicated forms:
\begin{eqnarray*}
<n^k_{1i}>&=&\frac{1}{k}\Big[\sum^{k-1}_{s=0}\big[\frac{k!}{s!(k-s)!}(X_i^{ad}+\delta X_i^{sw})\\
&&(1-\frac{\omega_i C_2(i)}{\omega_i+C_1(i)k})<n^s_{1i}>+\\ 
&&+(X_i^{ad}-\delta X_i^{sw})\frac{\omega_i C_2(i)}{\omega_i+C_1(i)k}<n^s_{2i}>\big]+ \\ 
&&+\sum^{k-2}_{s=0}\frac{k!}{s!(k-s)!}(-1)^{k-s}\\
&&\big[(1-\frac{\omega_i C_2(i)}{\omega_i+C_1(i)k})<n^{s+1}_{1i}>\\
&&+\frac{\omega_iC_2(i)}{\omega_i+C_1(i)k}<n^{s+1}_{2i}>\big]\Big]
\end{eqnarray*}
\begin{eqnarray*}
<n^k_{2i}>&=&\frac{1}{k}\Big[\sum^{k-1}_{s=0}\big[\frac{k!}{s!(k-s)!}(X_i^{ad}-\delta X_i^{sw})\\
&&(1-\frac{\omega_i C_1(i)}{\omega_i+C_1(i)k})<n^s_{2i}>+ \\ 
&&+(X_i^{ad}+\delta X_i^{sw})\frac{\omega_i C_1(i)}{\omega_i+C_1(i)k}<n^s_{1i}>\big]+ \\ 
&&+\sum^{k-2}_{s=0}\frac{k!}{s!(k-s)!}(-1)^{k-s}\\
&&\big[(1-\frac{\omega_i C_2(i)}{\omega_i+C_1(i)k})<n^{s+1}_{2i}>+\\ 
&&+\frac{\omega_iC_2(i)}{\omega_i+C_1(i)k}<n^{s+1}_{1i}>\big]\Big]
\end{eqnarray*}
The resulting equations for the zeroth moments couple to the higher moments by the interaction function $F(i)$. These lower moments can be solved self-consistently. The resulting solution predetermines all the other moments, which completely describe the probability distribution. Each gene therefore couples to the other gene by the influence of the self-consistently generated proteomic field. One could define the generating function and calculate the probabilies of having a given DNA binding state $j$ for the $i^{th}$ gene when there are $n_i$ proteins of type $i$ in the cell. In practice, it is easier to go back to the steady state master equation and solve directly for the probability distributions than sum an infinite number of moments. Rewriting the steady state master equation one gets:
\begin{eqnarray*}
P_1(n_i)&=&\frac{1}{X_i^{ad}+\delta X_i^{sw}+\omega_i\frac{F(3-i)}{X_i^{eq}}+n}\\
&&[(X_i^{ad}+\delta X_i^{sw})P_1(n_i-1)+\\ 
&&+(n_i+1)P_1(n_i+1)+\omega_iP_2(n_i)]\\ 
P_1(n_i=0)&=&\frac{1}{X_i^{ad}+\delta X_i^{sw}+\omega_i\frac{F(3-i)}{X_i^{eq}}}[P_1(n_i=1)+\\
&&\omega_i P_2(n_i=0)]\\ 
P_2(n_i)&=&\frac{1}{X_i^{ad}-\delta X_i^{sw}+\omega_i+n}\\
&&[(X_i^{ad}-\delta X_i^{sw})P_2(n_i-1)+\\ 
&&+(n_i+1)P_2(n_i+1)+{\omega}_i  \frac{F(3-i)}{X_i^{eq}}P_1(n_i)]\\ 
P_2(n_i=0)&=&\frac{1}{X_i^{ad}-\delta X_i^{sw}+\omega_i}[P_2(n_i=1)+\\
&&\omega_i \frac{F(3-i)}{X_i^{eq}} P_1(n_i=0)]
\end{eqnarray*}
These sets of equations give recursion relations for $P_j(n_i)$ which one can use to express $P_j(n)$ as a function of $P_1(0)$ and $P_2(0))$. The normalization condition $\sum_{n_1=0}(P_1(n_1)+P_2(n_1))=1$ gives $P_j(0)$ in term of constants and the result is the probability function $P_j(n)$ as a series. The SCPF approximation reduces the two gene problem to a one gene problem parametrized by the moments of the second gene, which can be worked out independently, as we have already shown and are represented by $F(2)$, which is a constant in terms of this calculation.\\
To see the effect of the stochastic nature of the system we compare the exact solutions of the self consistent field approximation equations to the results that would follow from deterministic kinetic rate equations for the number of proteins of each type and the fraction of on/off DNA binding states for each gene:
\begin{eqnarray*}
C_1(1)&=&\frac{X_1^{eq}}{X_1^{eq}+n^2(2)}
\nonumber \\ 
C_1(2)&=&\frac{X_2^{eq}}{X_2^{eq}+n^2(1)}
\nonumber \\ 
n(1)&=&X_1^{ad}+\delta X_1^{sw} (C_1(1)-C_2(1))
\nonumber \\ 
n(2)&=&X_2^{ad}+\delta X_2^{sw} (C_1(2)-C_2(2))
\end{eqnarray*}
where $n(i)$ is the number of proteins of type $i$ present in the cell. The exact SCPF equations reduce to the deterministic kinetic equations in the limit of large $\omega$ and $X^{ad}$ for the case discussed above. The $F(3-i)$ term in the stochastic SCPF equations is replaced by the $n^2(3-i)$ term in the deterministic kinetic rate equations. For the toggle switch, where repressors bind as dimers it is easily shown that the interaction functional may be rewritten in the form:
\begin{eqnarray*}
F(i)&=&(X^{ad}_i)^2+X_i^{ad}+(\delta X_i^{sw})^2+\\
&&\delta X_i^{sw} (C_{1}(i)-C_{2}(i))(1+2X_i^{ad})+\\
& &-4 \omega_i (\delta X_i^{sw})^2 \frac{C_1(i)C_2(i)}{\omega_i + C_1(i)}=\\
&&=<n(i)>^2\frac{\omega_i+1}{\omega_i+C_1(i)}+<n(i)>
\end{eqnarray*}
which in the large $\omega$ limit reduces to $F(i)=<n(i)>^2+<n(i)>$. So for large mean numbers of proteins present in the cell, which corresponds to large effective production rates $X^{ad}$, $<n(i)>$ of the order of hundreds is a small correction to $<n(i)>^2$. We therefore reproduce the deterministic kinetics result.\\
As shown by Sasai and Wolynes (Sasai and Wolynes, 2003) the difference in the probability that gene 1 is active and that gene 2 is active, $\Delta C=C_1(1)-C_1(2)$, plays the role of an order parameter. We can now consider a family of switches and discuss their stability, sensitivity of regions of bistability to control parameters and types of bifurcations. 
\subsection*{The Symmetric Toggle Switch}
\begin{figure}
\begin{minipage}[t]{.43\linewidth}
\includegraphics[height=3cm,width=4cm]{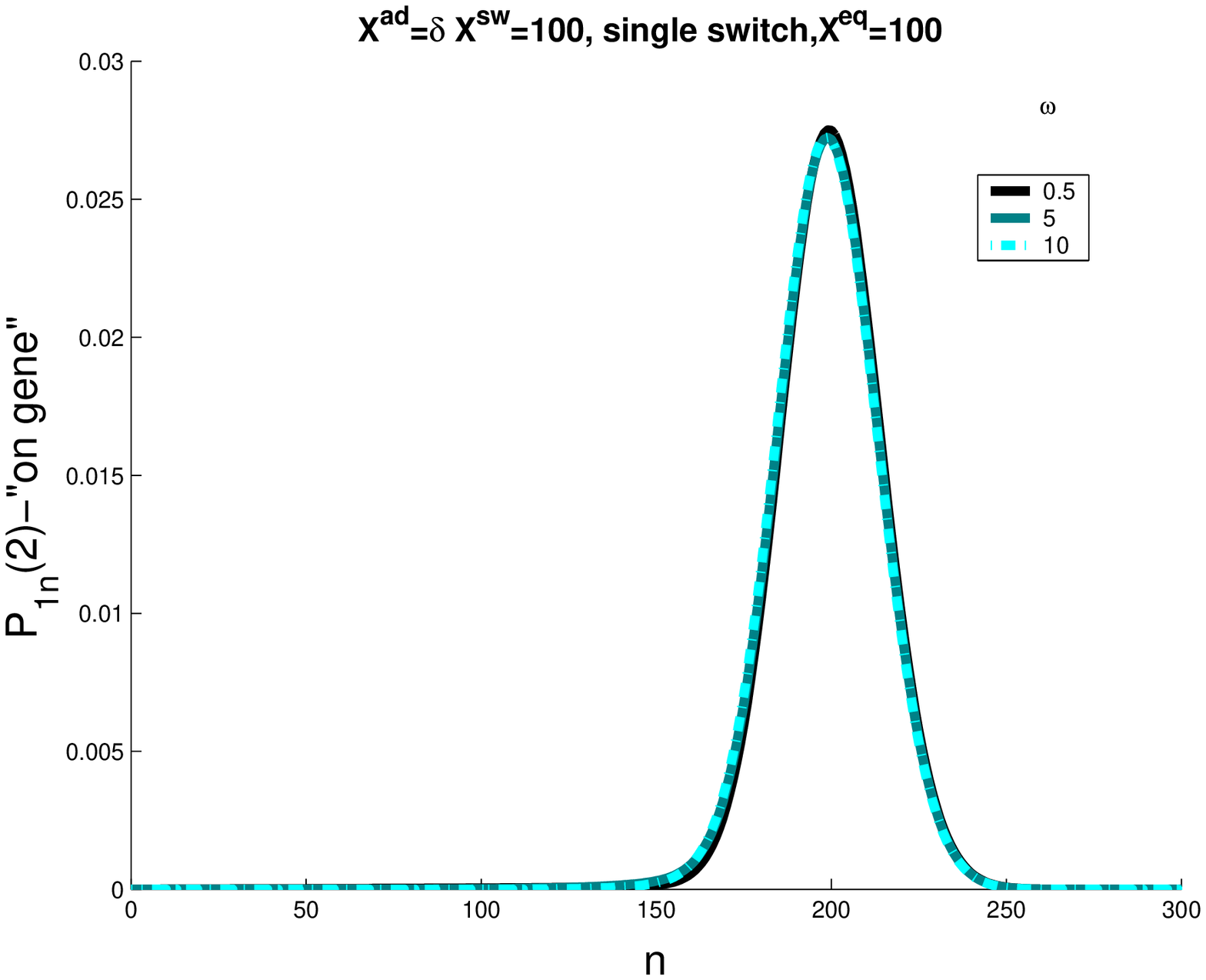}
\end{minipage}\hfill
\begin{minipage}[t]{.5\linewidth}
\includegraphics[height=3cm,width=4cm]{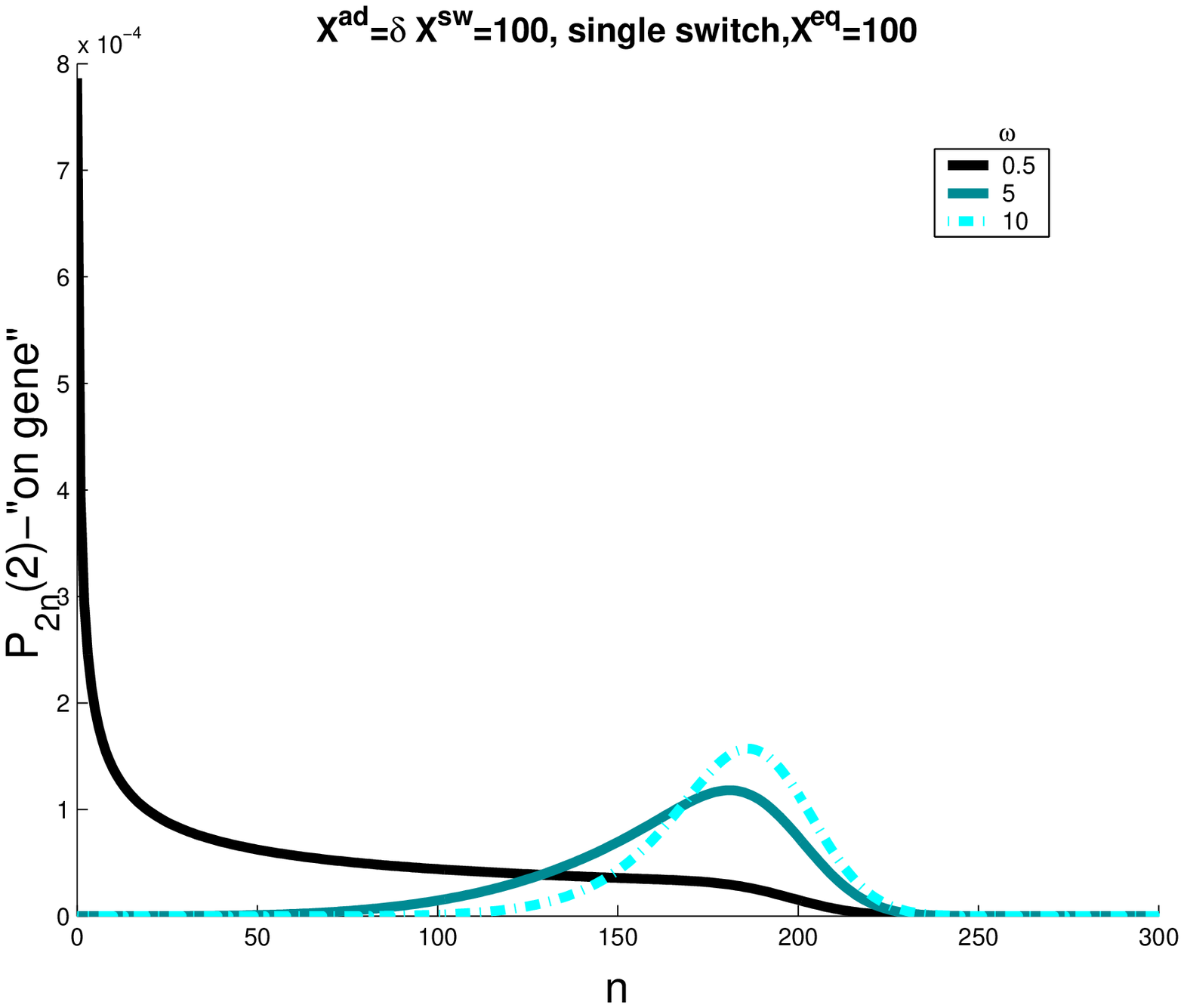}
\end{minipage}
\caption{Probability distributions for the gene to be in the on state (a) and off state (b) for a gene in the active state for different values of the adiabaticity parameter $\omega=0.5,10,100$. $X^{eq}=100$, $X^{ad}=\delta X^{sw}=100$. }
\label{evpoissonom}
\end{figure}

\begin{figure}
\begin{minipage}[t]{.43\linewidth}
\includegraphics[height=3cm,width=4cm]{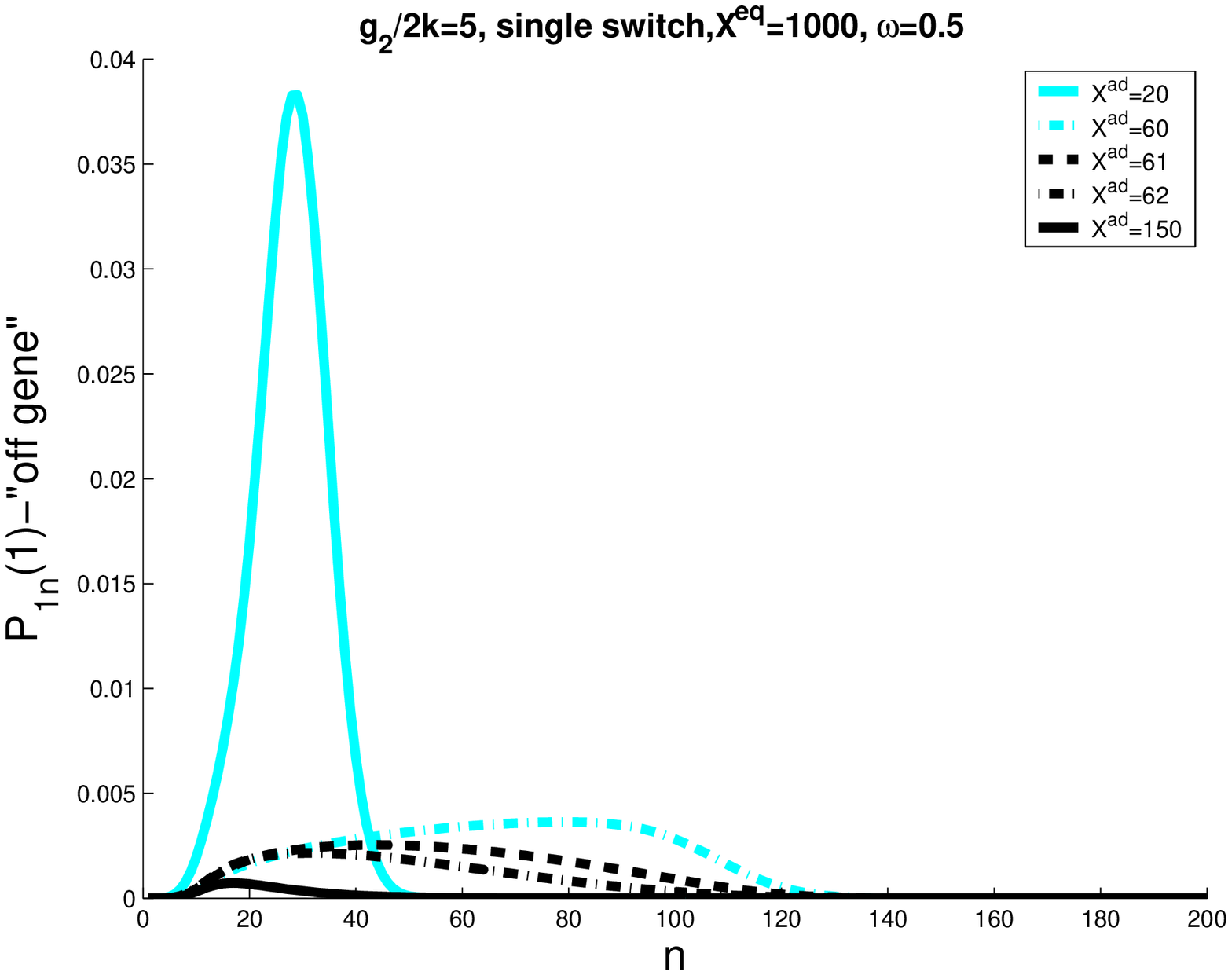}
\end{minipage}\hfill
\begin{minipage}[t]{.5\linewidth}
\includegraphics[height=3cm,width=4cm]{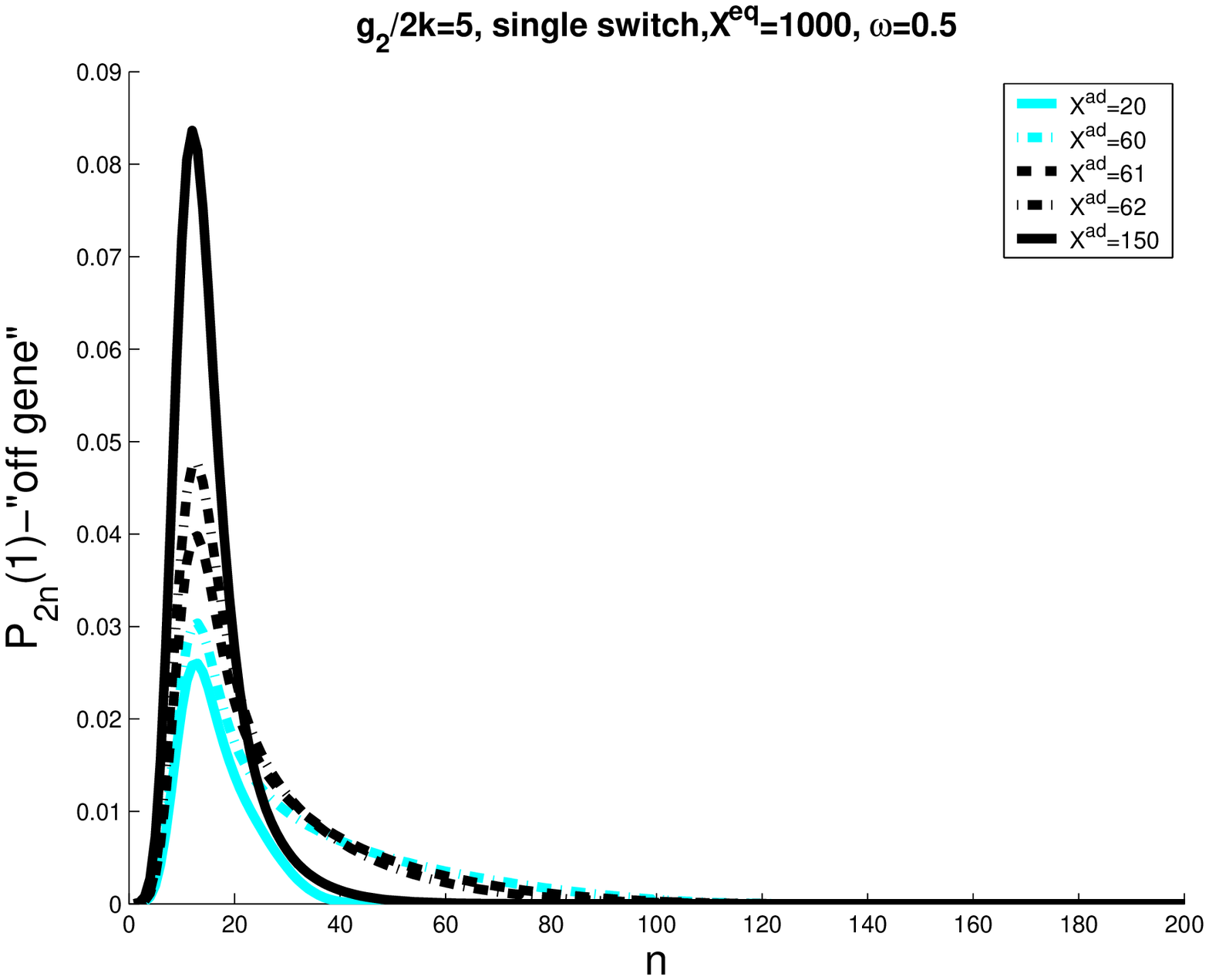}
\end{minipage}
\caption{Evolution of probability distributions for the probability of the gene that will be inactive after the bifurcation to be on (a) and off (b ) as a function of the order parameter $X^{ad}$ for a system with a basal production rate $\frac{g_2}{2k}=5$. The bifurcation occurs at $X^{ad}=61$.}
\label{evpoissonom3}
\end{figure}

For pedagogic purposes, we will start by analyzing the single symmetric toggle switch, such as discussed above in which repressors bind as dimers, with $\omega_1=\omega_2=\omega$, $X_1^{ad}=X_2^{ad}=X^{ad}$, $\delta X_1^{sw}=\delta X_2^{sw}=\delta X^{sw}$ and $X_1^{eq}=X_2^{eq}=X^{eq}$, as it is the most intuitive and shows the most generic behavior. It is an academic example, as even individual genes in switches engineered in the laboratory mostly have different chemical parameters. Yet a lot can be learned from this simple system.\\
\begin{figure}[ptb]
\begin{minipage}[t]{.2\linewidth}
\includegraphics[height=3cm,width=2.7cm]{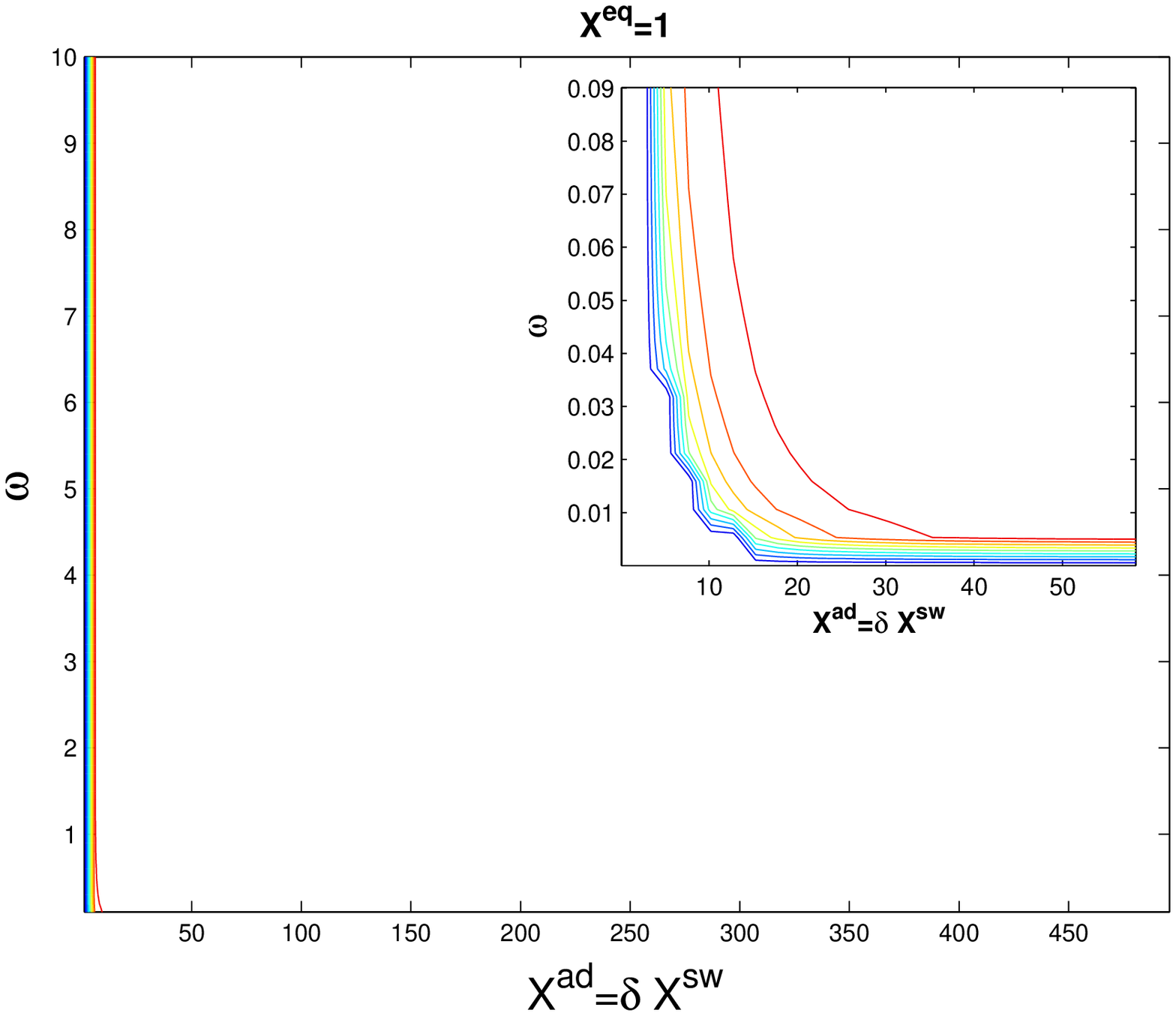}
\end{minipage}\hfill
\begin{minipage}[t]{.2\linewidth}
\includegraphics[height=3cm,width=2.7cm]{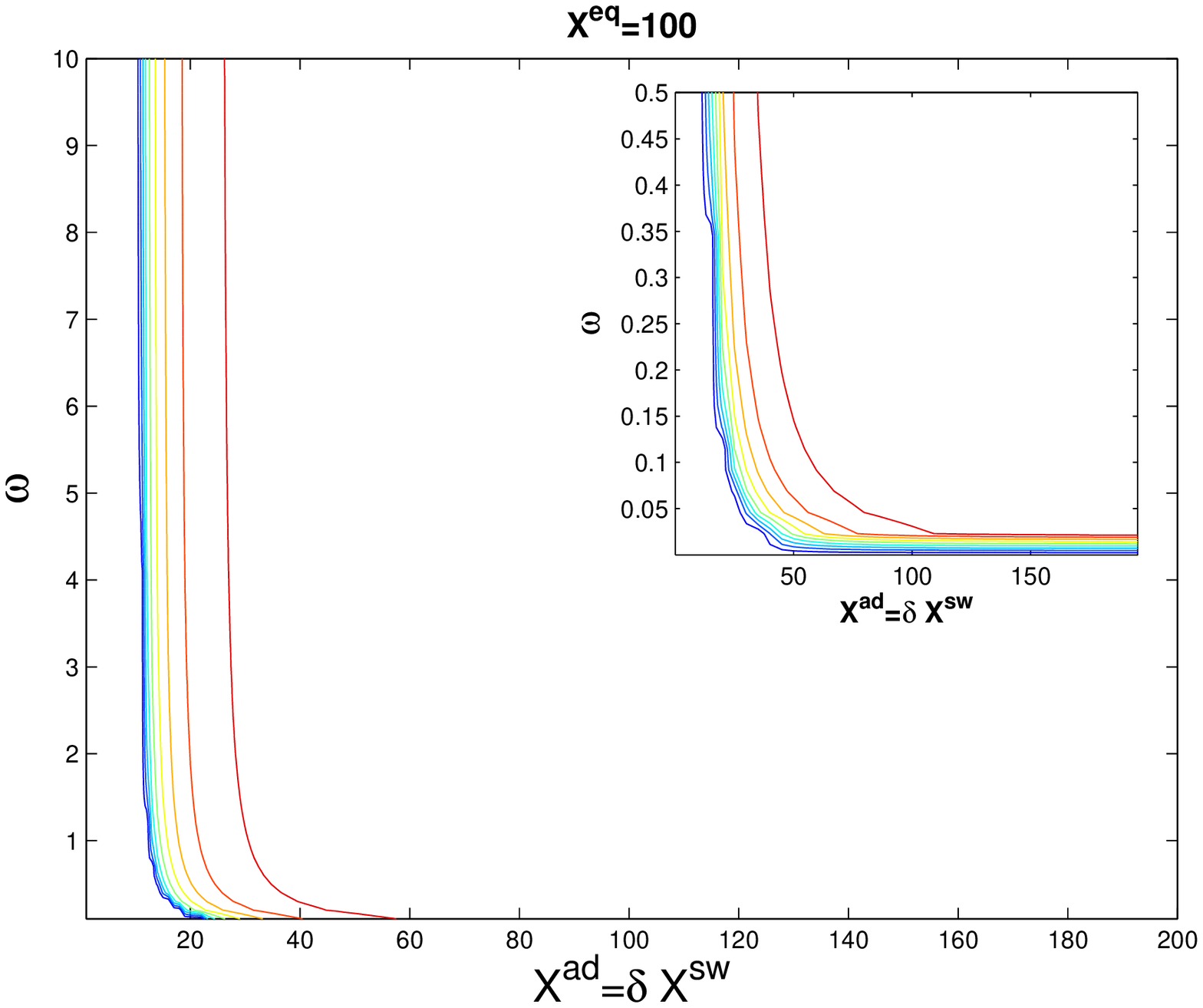}
\end{minipage}\hfill
\begin{minipage}[t]{.35\linewidth}
\includegraphics[height=3cm,width=3.3cm]{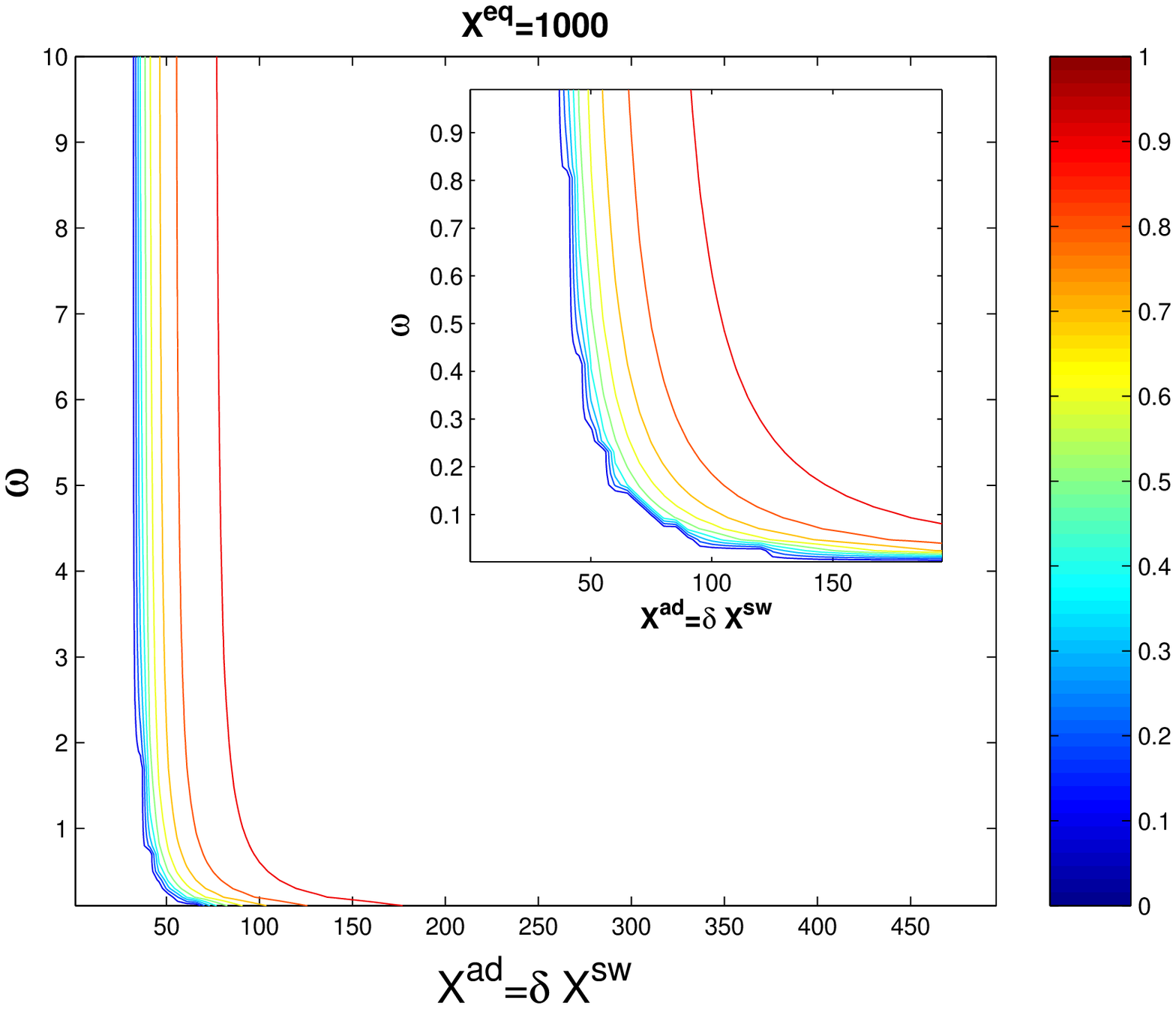}
\end{minipage}
\caption{Phase diagram obtained as an exact solution within the SCPF approximation for the single symmetric switch with $X^{eq}=1$ (a), $100$ (b), $1000$ (c). Contour lines mark values of $\Delta C$.}
\label{phdss}
\end{figure}
\begin{figure}
\begin{minipage}[t]{.225\linewidth}
\includegraphics[height=3cm,width=2.9cm]{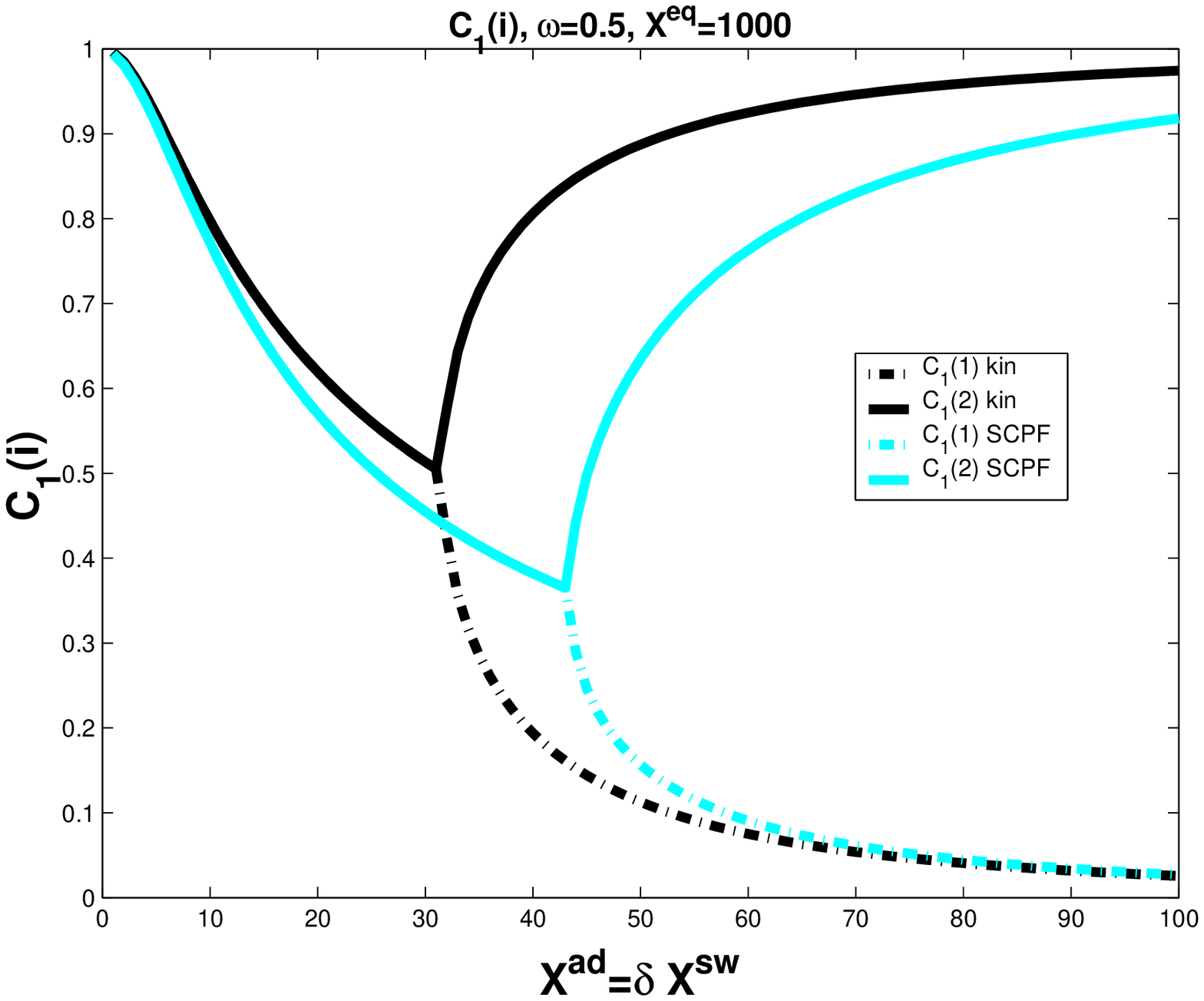}
\end{minipage}\hfill
\begin{minipage}[t]{.2\linewidth}
\includegraphics[height=3cm,width=2.9cm]{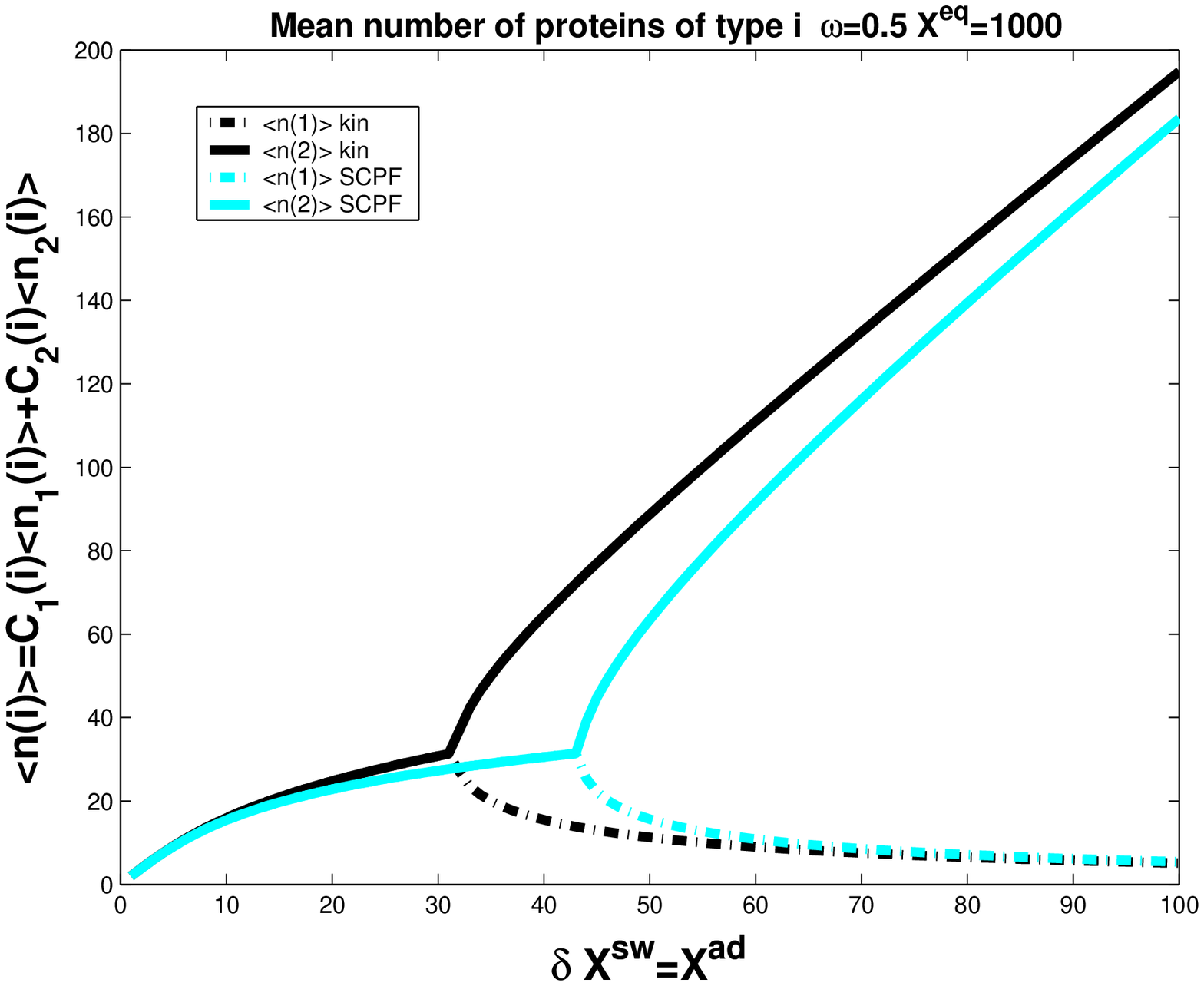}
\end{minipage}\hfill
\begin{minipage}[t]{.35\linewidth}
\includegraphics[height=3cm,width=2.9cm]{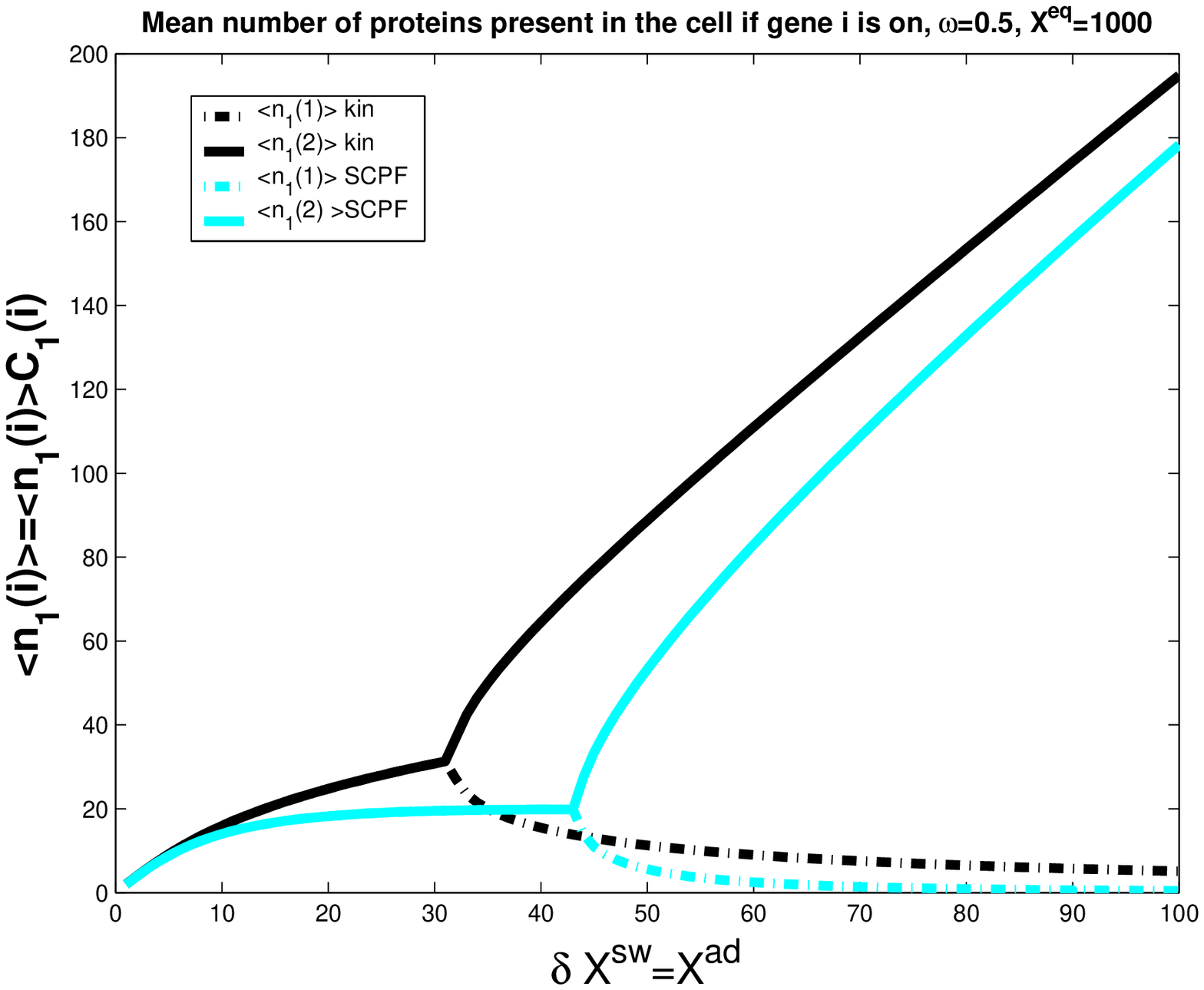}
\end{minipage}
\caption{Probability that genes are in the active state (a), the mean number of proteins of each type present in the cell (b) and the mean number of proteins present in the cell for gene $i$, when it is in the on state (c) as a function of $X^{ad}$. Exact solutions of the SCPF approximation equations compared with deterministic kinetic rate equations solutions, for a single symmetric switch,  $X^{eq}=1000$.}
\label{kcssin}
\end{figure}
\textbf{The general mechanism of the phase transition}.\\
Figure \ref{phdss} shows the phase diagrams for the system, $\abs{\Delta C}$ as a function of reservoir protein number and the adiabaticity parameter for the exact SCPF equations for growing values of the parameter describing the tendency that proteins are unbound from the DNA, $X^{eq}$. The deterministic kinetics and exact SCPF approximations give qualitatively similar results. The analogous deterministic kinetic phase diagrams agree with the SCPF solutions in the large $\omega$ and $X^{ad}$ limit, hence they become more similar with growing $X^{eq}$, as the bifurcation occurs at larger effective production rates for larger $X^{eq}$. For large fluctuations and a small unbinding rate, neither gene 1 nor gene 2 is favoured and the probability of a given gene to be on is determined solely by the effective production rate of the other gene and decreases in a quadratic manner as the number of repressor proteins grow (Fig. \ref{kcssin}). Since the switch is symmetric, the system has one stable state, $\Delta C=0$, where the probabilities of the genes to be on are equal. As the relative protein number fluctuations get smaller and the DNA unbinding rate grows, a proteomic cloud buffers the repressed gene, keeping it repressed. The symmetry of the system is broken and the solution bifurcates into two separate basins of attraction. For the stochastic SCPF equations the bifurcation takes place for larger effective production rates (larger $X^{ad}$), than for the deterministic equations, even in the large $\omega$ limit, which depicts their sensitivity to fluctuations. The critical number of reservoir proteins necessary for the bifurcation of the solution to take place is the same in both approximations and is determined by $<n>_c=(X^{eq})^{\frac{1}{2}}$ (Fig. \ref{kcssin}). In the discussed example $<n>_c=32={1000}^{\frac{1}{2}}$, for $X^{eq}=1000$.
For the deterministic kinetic switch the bifurcation takes place when $C_1(i)=\frac{1}{1+\frac{<n(3-i)>^2}{X^{eq}}}=\frac{1}{2}$, due to the simple form of the interaction function equal to $<n(3-i)>^2=(2 X^{ad} C_1(3-i))^2$. So $C_1(i)=\frac{1}{2}$ is equivalent to the $\frac{<n(3-i)>^2}{X^{eq}}=1$. In a noisy system larger effective production rates are needed to achieve the critical value of proteins. The interaction function in this case may be written as $F(i)=<n(i)>^2 \frac{\omega+1}{\omega+C_1(i)}+<n(i)>$, and $\frac{\omega+1}{\omega+C_1(i)} \geq 1$, always. So at $<n>_c$, $\frac{F(3-i)}{X^{eq}}>1$ and the probability of the genes to be on is smaller than $\frac{1}{2}$, therefore $C^{biff,SCPF}_1(i)<C^{biff,kin}_1(i)$. The mechanism of the bifurcation requires the two genes to be more likely to be unbound than bound for the phase transition to take place. The curvature of the nullclines presented in Fig. \ref{phdss} can be simply worked out to be of the form $\omega=\frac{\zeta_1}{\xi_1 X^{ad 2}+\xi_2 X^{ad}+\zeta_2}-\xi_2$, with $\zeta_i, \xi_i$ constants determined by the specific value of $C_1(1)$, $C_1(2)$.\\
\begin{figure}
\begin{minipage}[t]{.43\linewidth}
\includegraphics[height=3cm,width=4cm]{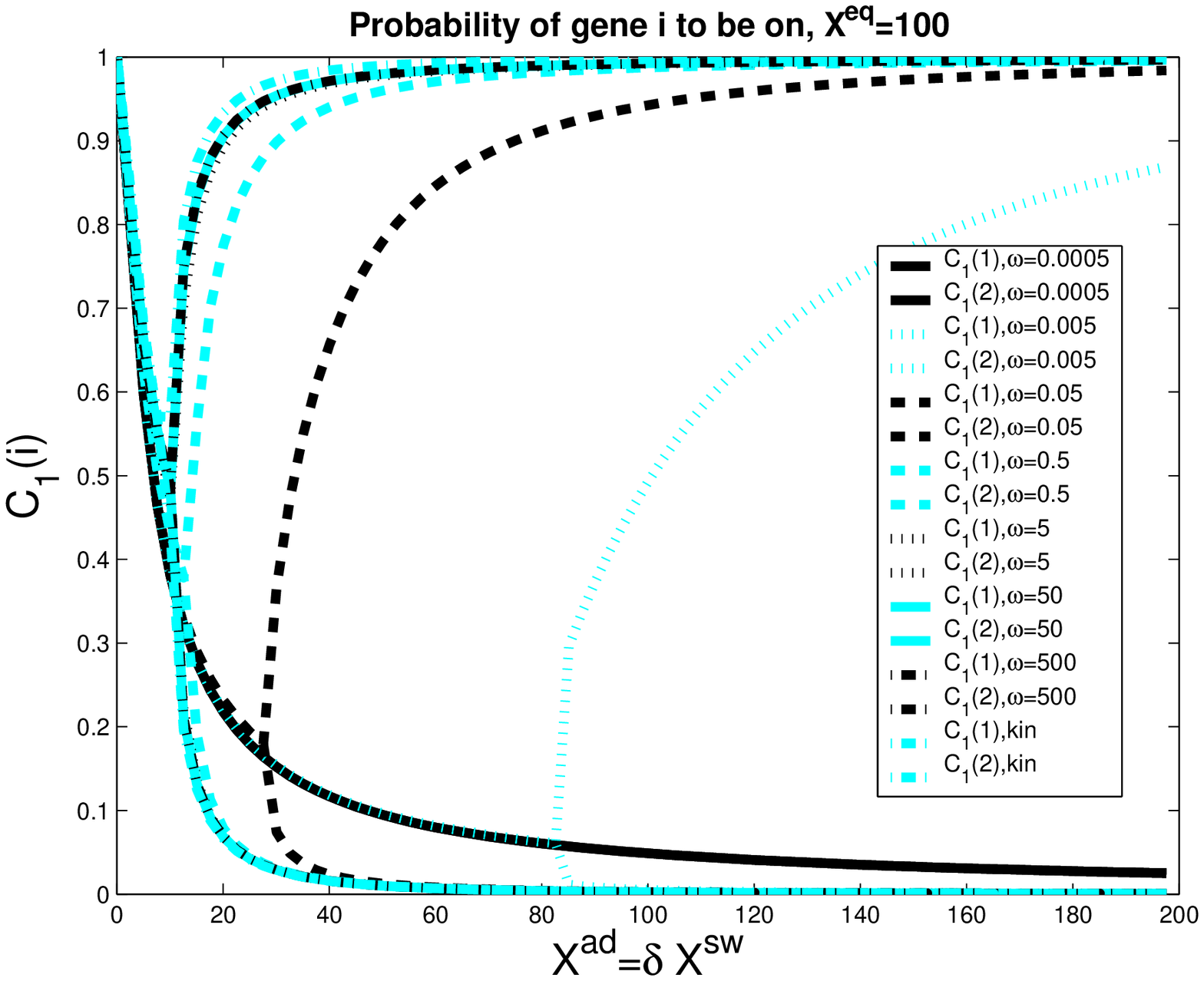}
\end{minipage}\hfill
\begin{minipage}[t]{.5\linewidth}
\includegraphics[height=3cm,width=4cm]{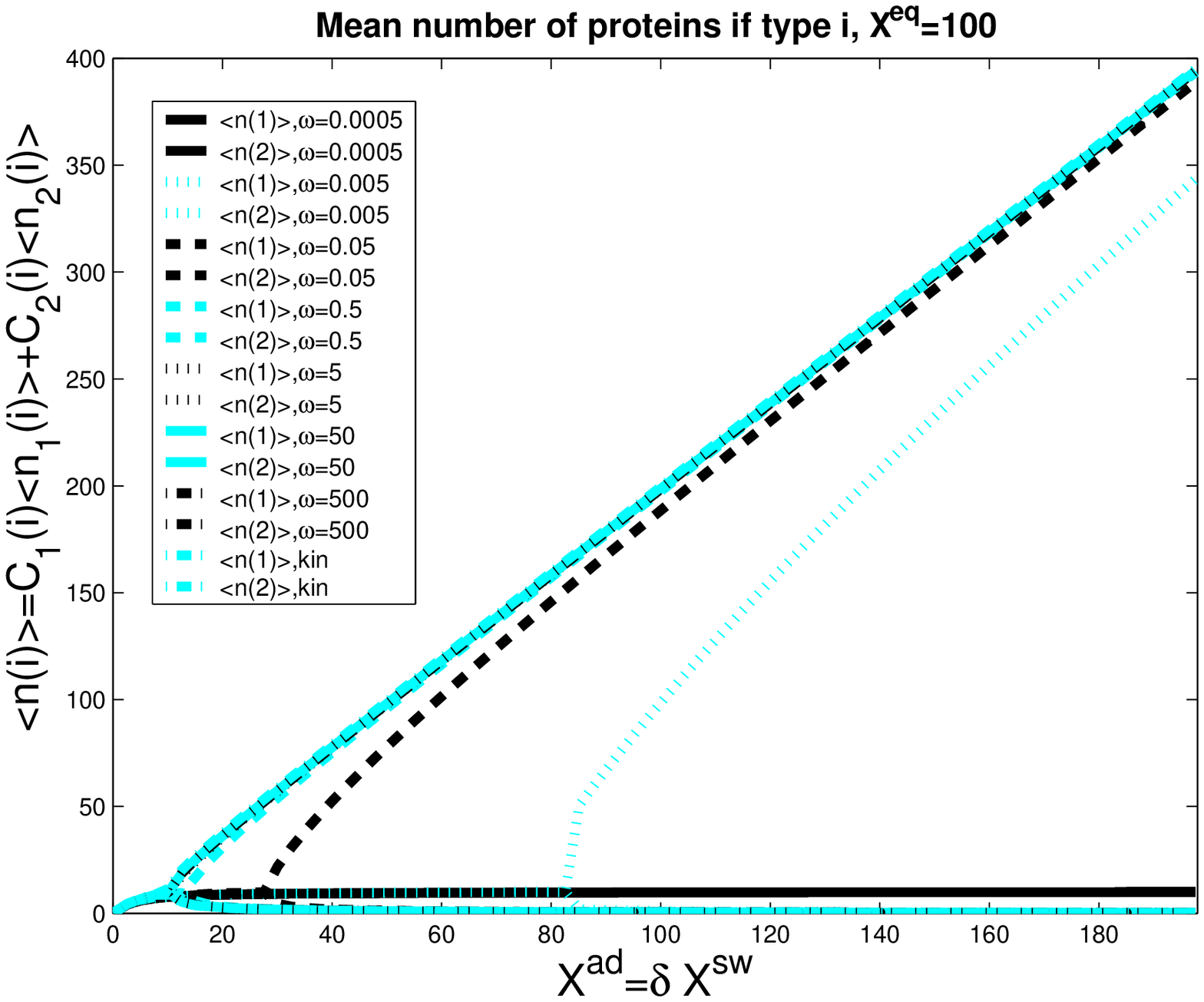}
\end{minipage}
\caption{Probability for each gene to be on (a) and mean number of proteins present in the cell (b) as a function of $X^{ad}$, for different values of $\omega=0.0005,0.005,0.05,0.5,5,50,500$, $X^{eq}=100$ for a symmetric switch.}
\label{sevpr}
\end{figure}
\textbf{Adiabaticity parameter dependence}\\
As the adiabaticity parameter decreases the area of phase space which corresponds to multiple solutions decreases (Figs \ref{phdss}, \ref{sevpr}). For very small values of the adiabaticity parameter, there exists only one solution which corresponds to a state in which the two genes are off. The value of $\omega$ below which only one solution exists decreases with the tendency for proteins to be bound, but exists for all values of $X^{eq}$. Therefore if the two genes have very high repressor binding affinities, the critical number of proteins necessary for the phase transition to take place cannot be formed, even for very high production rates. This region of parameter space where one solution is possible corresponds to a situation in which a buffering proteomic cloud may not form, due to a very fast destruction rate of proteins or a very small unbinding rate from the DNA. The critical number of proteins necessary for the bifurcation to occur grows with the tendency for proteins to be unbound from the DNA ($X^{eq}$), as the cloud buffering the genes needs to be bigger and exhibit smaller relative protein number fluctuations, which effectively decrease with the growth of the adiabaticity parameter. This is further discussed in terms of the probability distributions. Therefore a monostable solution exists at all values of the effective growth rate, $X^{ad}$, for larger values of $\omega$ at large $X^{eq}$ than at smaller $X^{eq}$ values. The bifurcation point is a result of competition between the number of reservoir repressor proteins and the tendency for proteins to be unbound from the DNA. This is clear from the dependence of the number of proteins present in the cell at the bifurcation point on the relative values of $X^{ad}$ and $X^{eq}$, but not the adiabaticity parameter $\omega$, as can explicitly be seen from Fig. \ref{sevpr}. \\
\textbf{Mean protein numbers}\\
The total number of proteins present in the cell, produced both in the on and off state, (Fig. \ref{sevpr}), asymptotically away from the bifurcation points is the same for the deterministic and stochastic approximations, and it is given by $<n(i)>=2 X^{ad}$, when $C_1(1)\approx 1$ the probability of the gene to be on is close to unity. The number of proteins of a given type present in the cell, when the gene that produces them is in the on state is always considerably smaller in the noisy system than the deterministic case (Fig. \ref{kcssin}(c)). Since the production rate in the off state was assumed zero, in the deterministic case no proteins of a given type are present in the cell if the gene is in the off state, unlike in the noisy system. Therefore the number of proteins in the deterministic system is nonzero only if the gene is on. But interaction of the DNA binding state with the proteins buffering it, results in a residual number of proteins present in the off state, for all values of $\omega$. The region of bistability of the switch in parameter space grows as the binding rate increases with respect to the unbinding rate, stabilizing the DNA binding states. As the susceptibility of the system to fluctuations increases, the deterministic equations prove to be a poor approximation to describe the state of the system.\\
\textbf{Gene-buffering proteomic cloud interactions}\\
The stochastic nature of the system manifests itself also at the DNA level (Fig. \ref{phdss}). As the tendency for proteins to be unbound from the DNA grows, the area of parameter space, where multiple solutions are possible decreases, since a larger number of proteins is needed to reach a state in which two genes are more likely to be repressed (protein bound state), than at small $X^{eq}$. For small unbinding rates or large binding rates, regardless of the ratio of the rate of unbinding of repressors from the DNA to protein degradation, bistability requires smaller numbers of proteins, which correspond to larger relative fluctuations, than for large $X^{eq}$. Therefore a larger unbinding rate relative to the binding rate makes the system more susceptible to protein number noise. Competition between $X^{eq}$ and $<n(i)>$ results in $X^{eq}$, for a given nullcline, being a parabolic function of $X^{ad}$, for the dimer binding case, with coefficients determined by $\omega$ and $C_1(i)$. This is easily generalized to higher order functions for higher order ($p$) oligomers, and results in $p$-order dependence. The switching region, by which we mean the region of parameter space between the bifurcation point and $\Delta C>0.9$ decreases as the binding and unbinding rates become comparable ($X^{eq}$ decreases). 
As discussed above, the probability of the genes to be on at the bifurcation point tends to $\frac{1}{2}$ as the adiabaticicty parameter grows (Fig. \ref{sevpr}), therefore the probability to be on has to increase by a smaller $\Delta C$ to reach $C_1(i)=1$. Therefore the switching region decreases also as the unbinding rate from the DNA grows, since smaller effective production rates are needed to reach $\Delta C=1$, than for small $\omega$.  Small values of $\omega$ correspond to large fluctuations in the DNA binding state, as well as the protein number state and result in destabilizing the gene-buffering protein cloud interactions. Hence very large effective production rates are needed for $\Delta C>0.9$. Therefore the DNA unbinding rate must become considerably faster compared to proteins degradation rate for the switch to have two stable solutions in a large region of parameter space.\\
\begin{figure}
\begin{minipage}[t]{.43\linewidth}
\includegraphics[height=3cm,width=4cm]{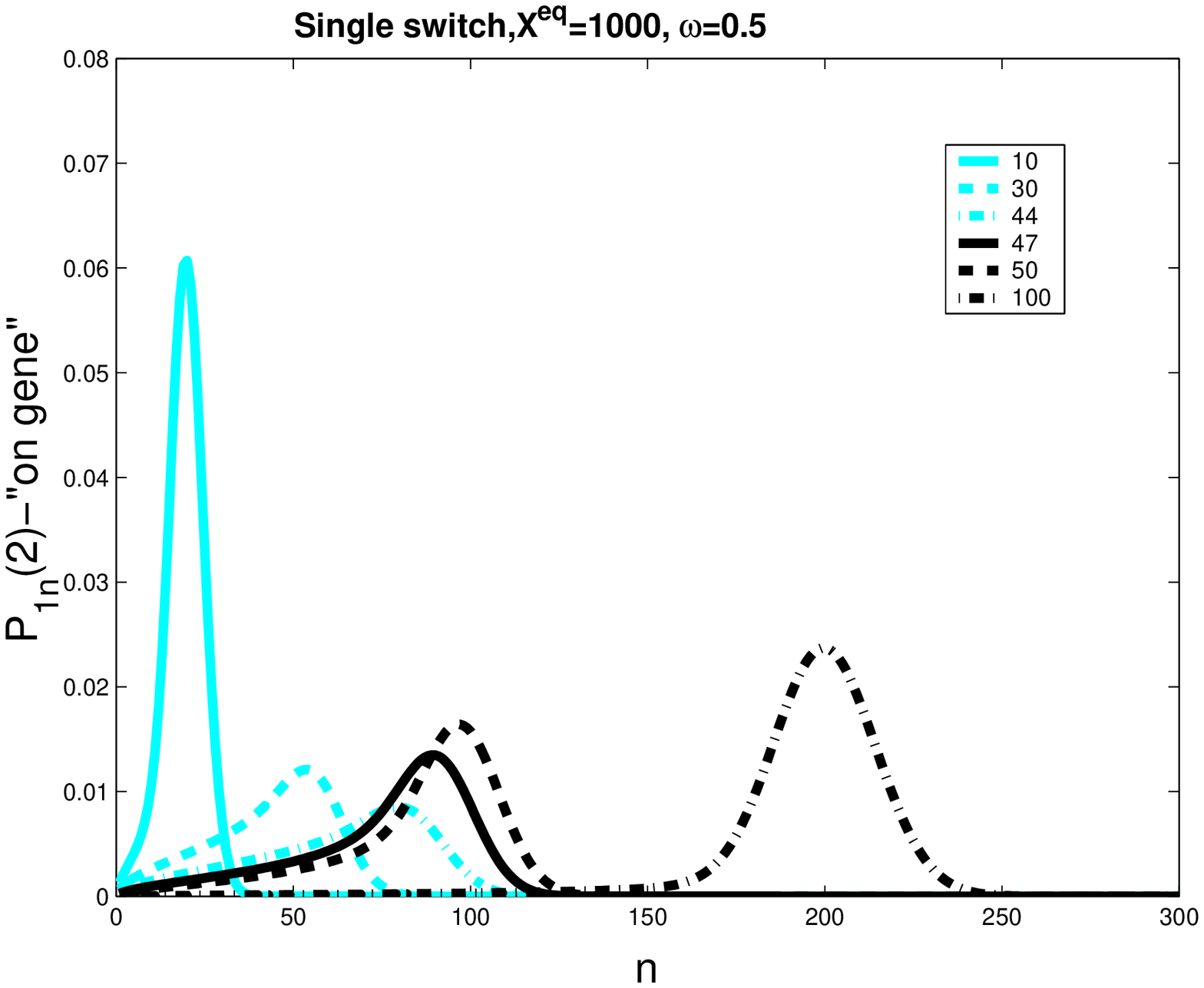}
\end{minipage}\hfill
\begin{minipage}[t]{.5\linewidth}
\includegraphics[height=3cm,width=4cm]{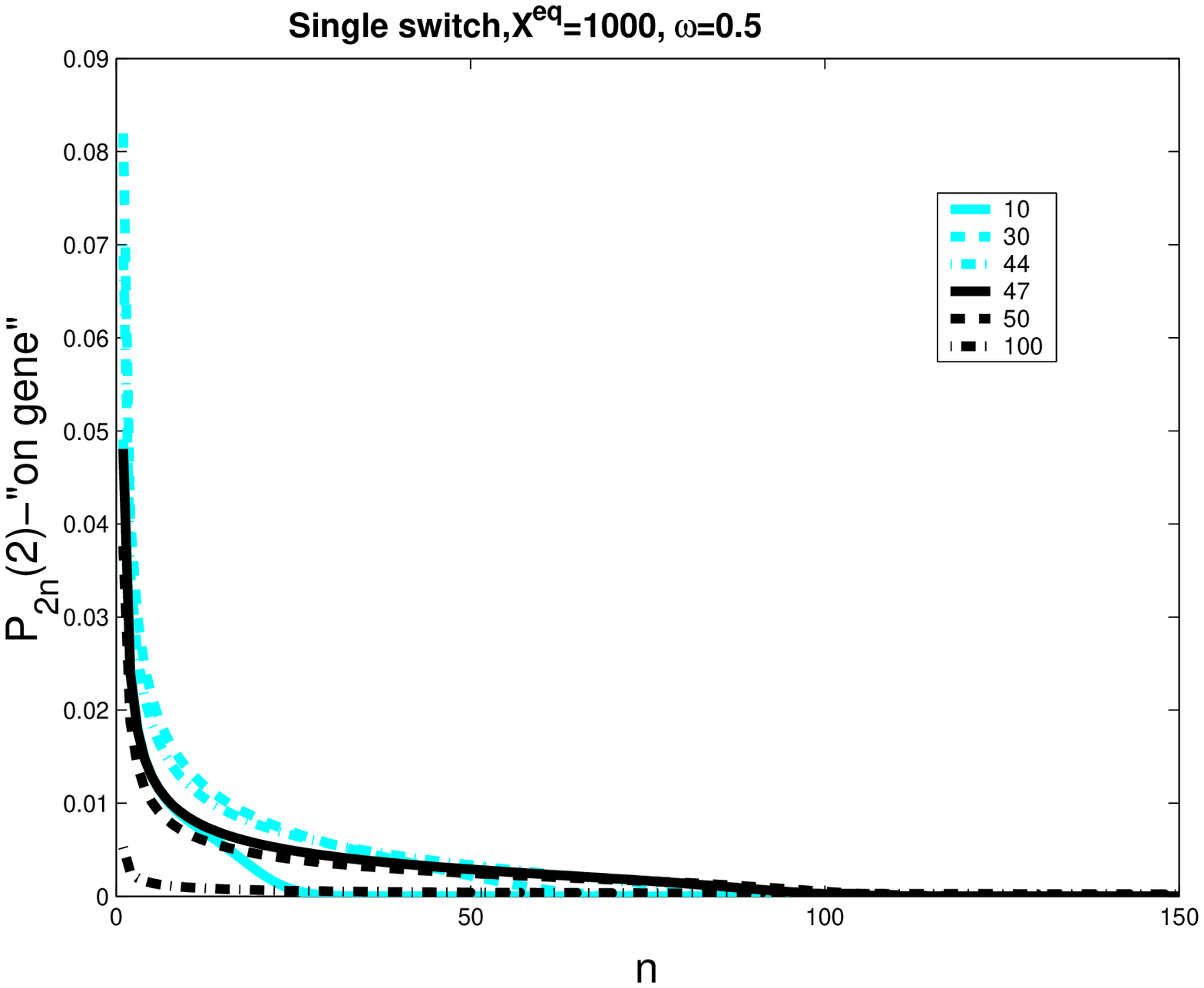}
\end{minipage}
\caption{Evolution of probability distributions for the probability of the gene that will be active after the bifurcation to be on (a) and off (b ) as a function of the order parameter $X^{ad}$. The bifurcation occurs at $X^{ad}=44$.}
\label{sevpr1}
\end{figure}
\begin{figure}
\begin{minipage}[t]{.43\linewidth}
\includegraphics[height=3cm,width=4cm]{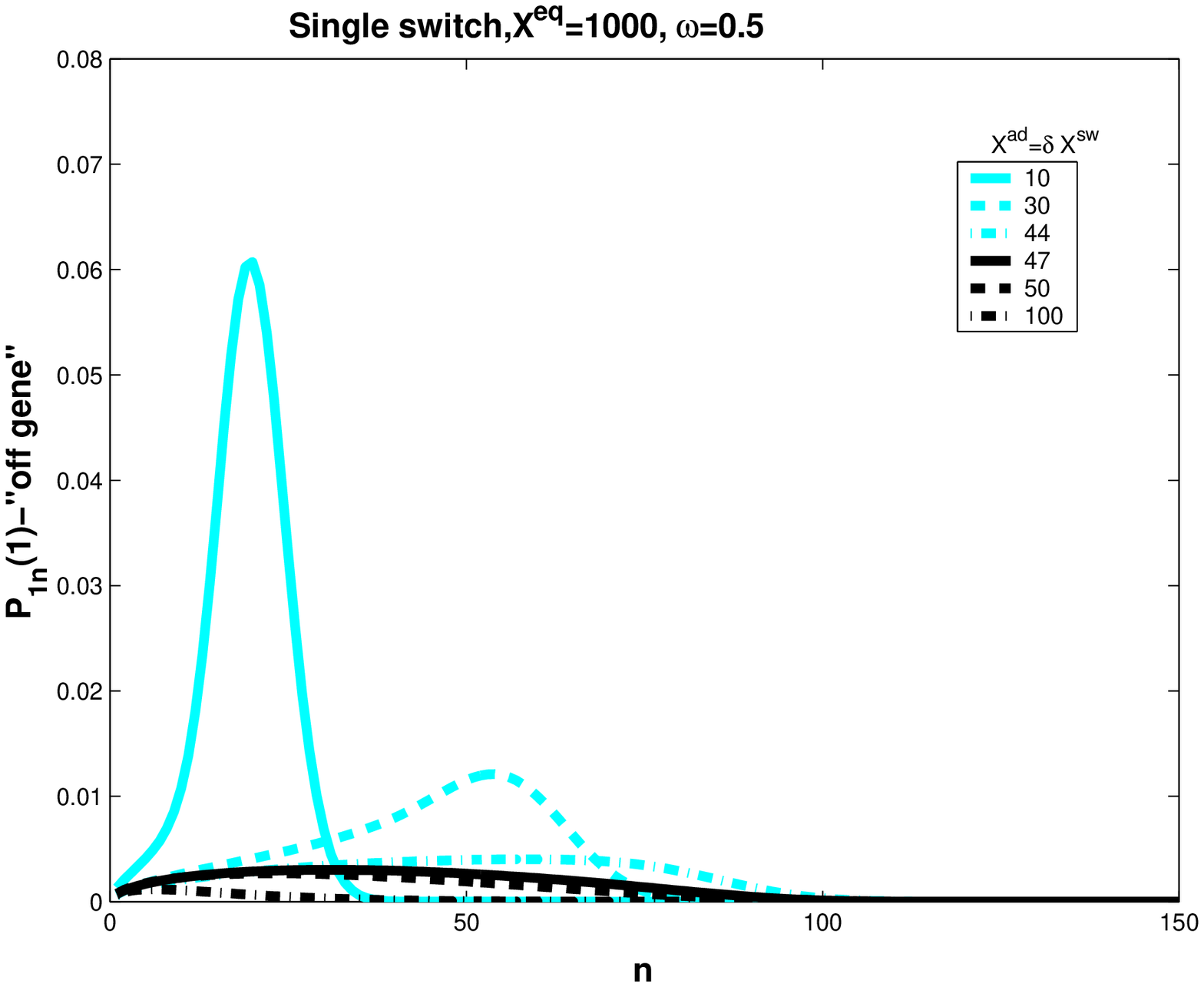}
\end{minipage}\hfill
\begin{minipage}[t]{.5\linewidth}
\includegraphics[height=3cm,width=4cm]{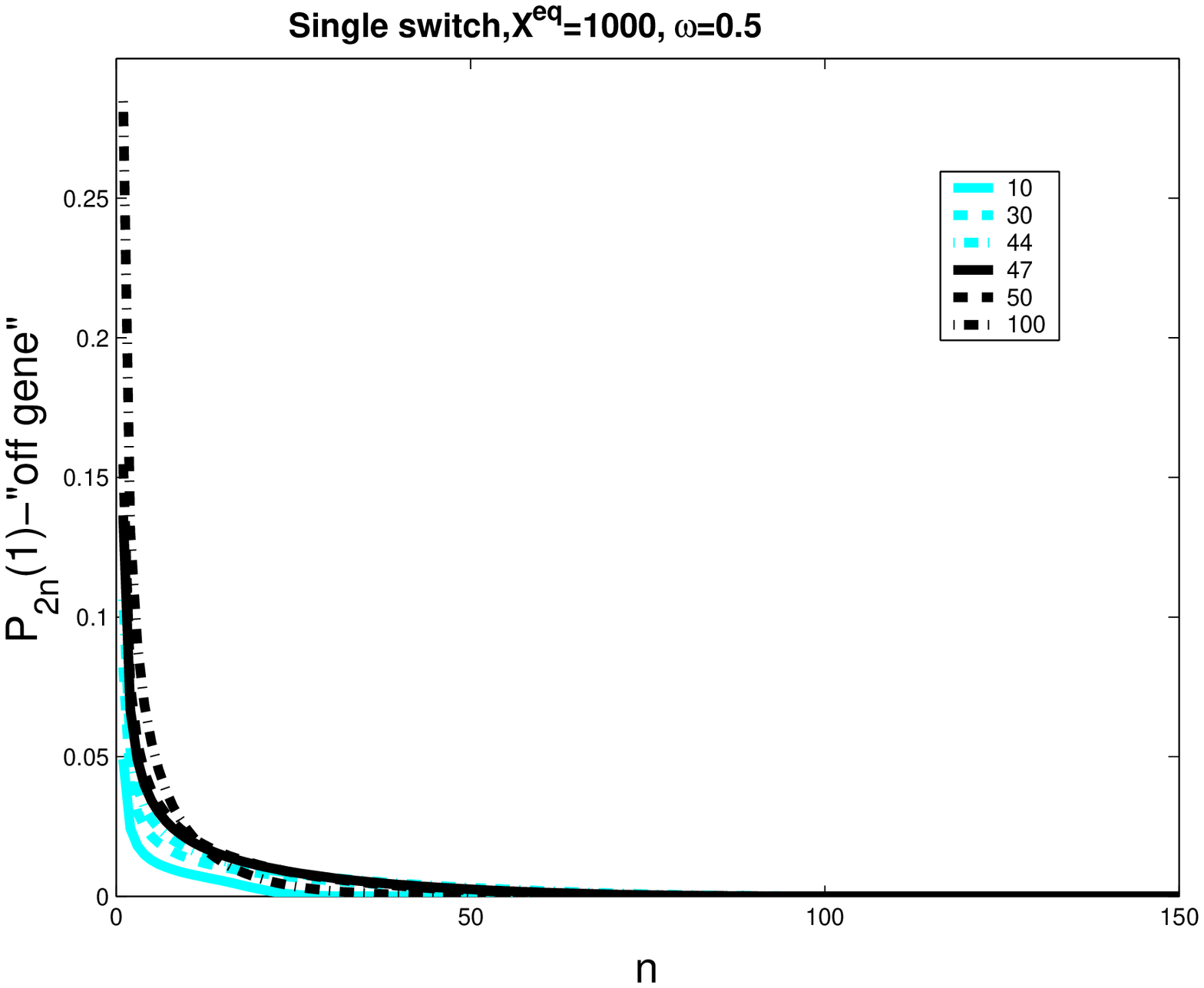}
\end{minipage}
\caption{Evolution of probability distributions for the probability of the gene that will be inactive after the bifurcation to be on (a) and off (b ) as a function of the order parameter $X^{ad}$. The bifurcation occurs at $X^{ad}=44$.}
\label{sevp2r}
\end{figure}
\textbf{The probability distributions}\\
A better understanding of the bifurcation can be gained from examining the probability distributions. Figures \ref{sevpr1} and \ref{sevp2r} show the evolution of the probability distributions of gene 1 and gene 2, respectively, to be on and off as functions of $X^{ad}$. The peak of the distribution decreases and the width spreads out as the control parameter grows, until it reaches the bifurcation point at $X^{ad}=44$. Then the value of the probability function corresponding to the most probable number of proteins grows again. The spread of the functions grows as the effective production rate in the on state increases, however narrows with the increase of the adiabaticity parameter, as would be expected, since the DNA state fluctuations become smaller with $\omega$. The average number of proteins in the cell in the on state ($\Delta C>0.9$) does not show a dependence on $\omega$. Yet as the unbinding rate from the DNA becomes very fast compared to the protein number fluctuations, the system switches often between the two states, hence a large number of proteins is present even in the off state. This results in a two peak - bimodal probability distribution (Fig. \ref{sevpr1}, \ref{sevp2r}). If the DNA unbinding rate is small, the protein number characteristics follow the DNA state having time to reach a steady state within each well, before the DNA binding site switches into the other state, so the number of proteins in the off state falls to zero (Fig. \ref{evpoissonom}). If $\omega$ is large, random fluctuations in the DNA state do not change the effective state of the system, since a residual high mean protein number is present even in the off state. In such a case lower effective production rates than for small $\omega$ result in higher protein yields and what follows smaller switching regions. \\
\begin{figure}
\begin{minipage}[t]{.43\linewidth}
\includegraphics[height=3cm,width=4cm]{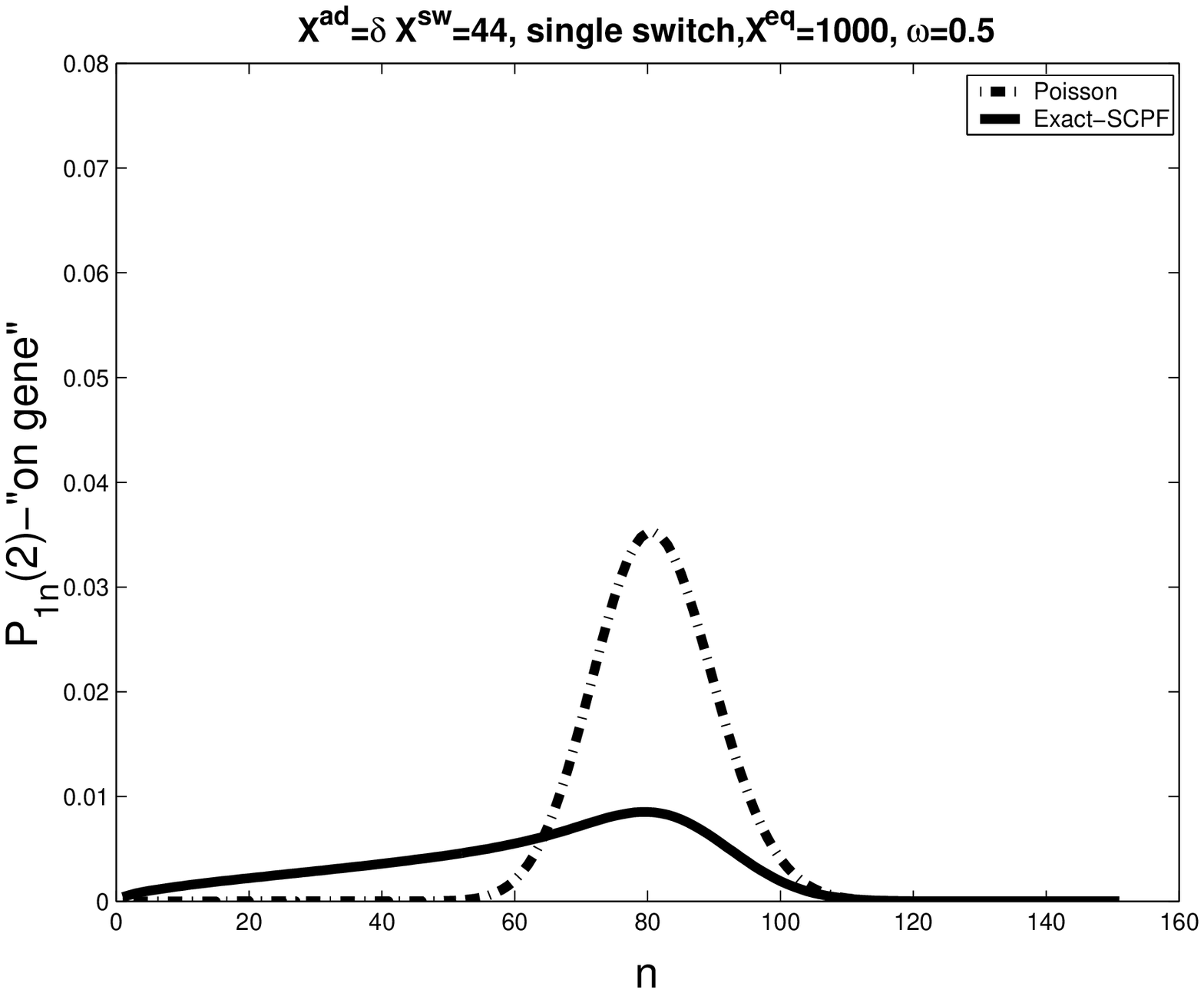}
\end{minipage}\hfill
\begin{minipage}[t]{.5\linewidth}
\includegraphics[height=3cm,width=4cm]{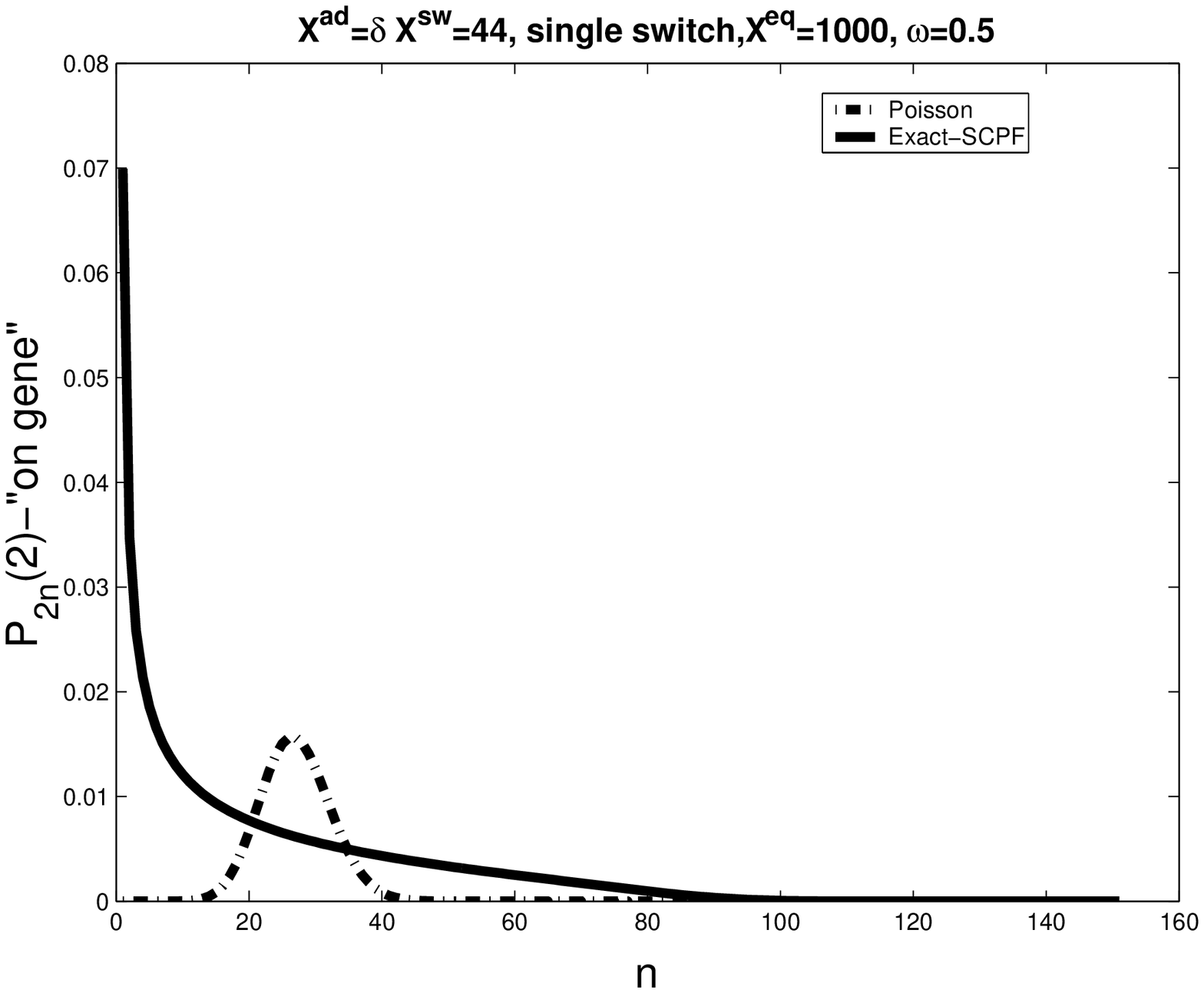}
\end{minipage}
\caption{Comparison of probability distributions obtained by exactly solving the steady state equations in the SCPF approximations with analogous Poissonian distributions.}
\label{evpoisson}
\end{figure}
For small $\omega$ one might expect Poisson distributions of proteins in each of the DNA states, since the unbinding rate from the DNA is smaller than the protein degradation rate, so the proteins may reach a steady state without the DNA state changing. Hence, effectively proteins would feel only one well and be subject to a birth death process. However this is not true. The difference between the exact solution and a solution obtained within a Poissonian approximation to the state of the system is surprisingly large, owing to the skewed tails of these distributions. Figure \ref{evpoisson} compares these probability distributions with distributions for the same system if one assumes a Poissonian probability function. The distributions obtained as an exact solution within the SCPF approximation are clearly not symmetric, but exhibit long tails towards zero. Therefore, although the most probable values of the two types of distributions are similar, noise has a destructive impact on the system, resulting in a larger probability of having a smaller number of proteins in the cell than expected based on a Poissonian distribution, whose higher moments are equal to the mean. Therefore a larger production rate is needed for one of the states to be favoured as a result of noise than predicted from a symmetric probability distribution. The most probable number of proteins in the on state, if the unbinding from the DNA is slow, is zero, unlike predicted by Poissonian distributions. The influence of noise on protein number fluctuations brings the protein number means down, as can also be seen from Fig. \ref{kcssin} (c). Overall the spread of the probability distributions is large, and their characteristics for small values of the control parameters are different from those predicted by Poissonian distributions, let alone by deterministic kinetic equations, therefore the effects of stochasticity may not be neglected.\\
\begin{figure}
\begin{minipage}[t]{.43\linewidth}
\includegraphics[height=3cm,width=4cm]{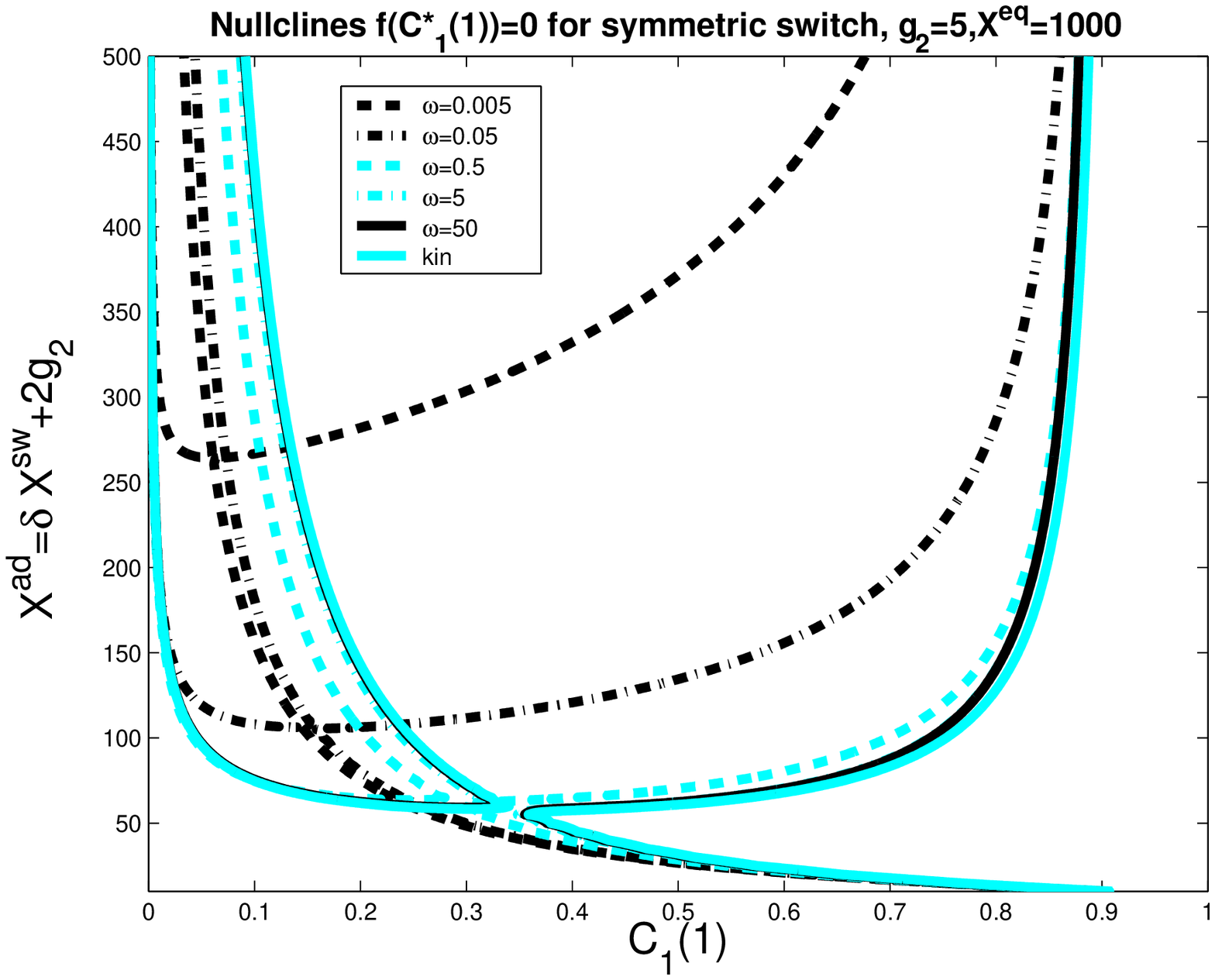}
\end{minipage}\hfill
\begin{minipage}[t]{.5\linewidth}
\includegraphics[height=3cm,width=4cm]{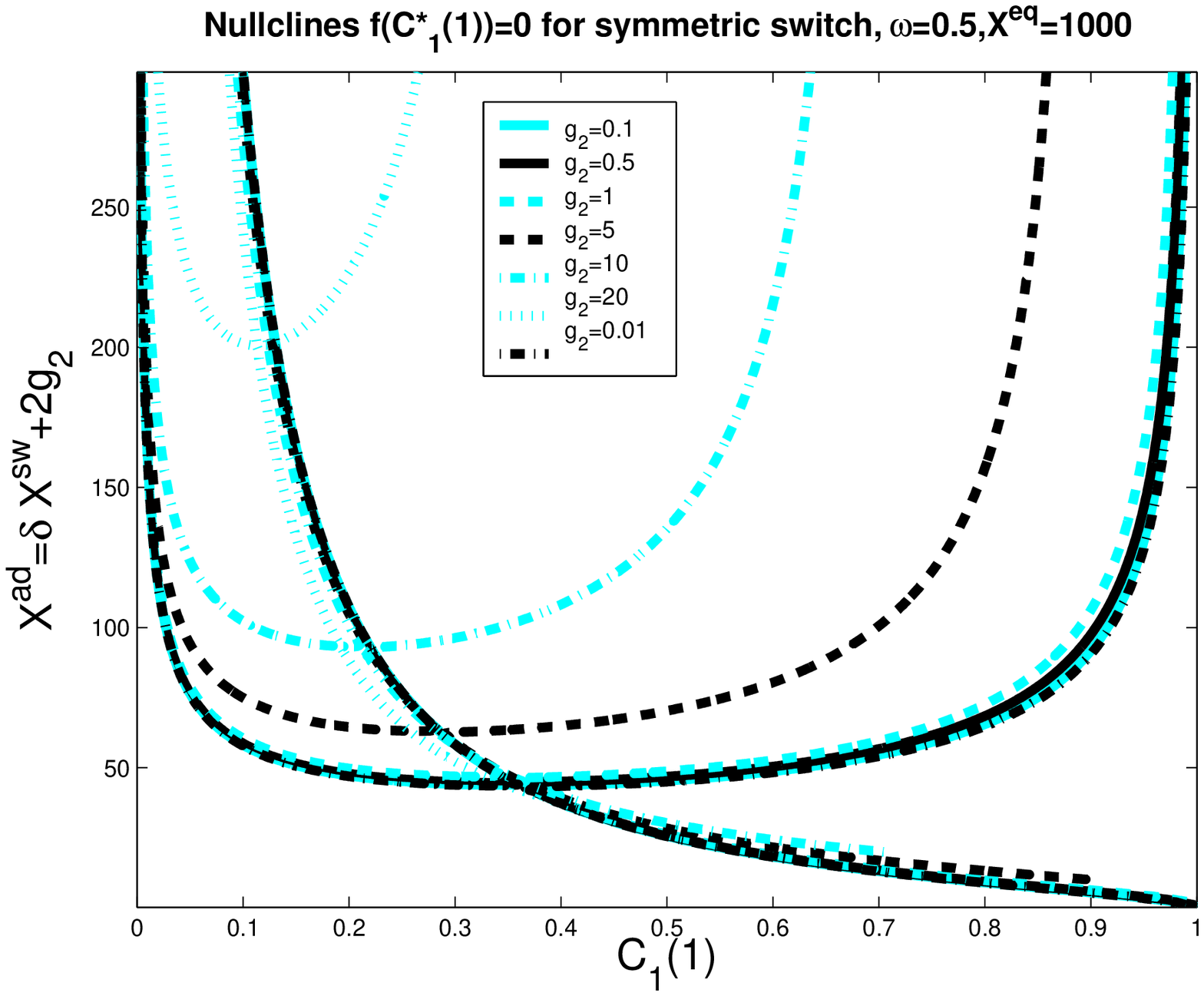}
\end{minipage}
\caption{Nullclines for a symmetric switch when proteins bind as dimers for a symmetric switch when the effective base production rate $g_2/2k \neq 0$. Dependence on the adiabaticity parameter $\omega=0.005,0.05,0.5,5,50$ and the deterministic equations solution (a), $g_2=5$, $X^{eq}=1000$. Dependence on $g_2=0.1,0.5,1.0,5,10,10$, $\omega=0.5$, $X^{eq}=1000$ (b).}
\label{evpoissonom2}
\end{figure}
\begin{figure}
\begin{minipage}[t]{.43\linewidth}
\includegraphics[height=3cm,width=4cm]{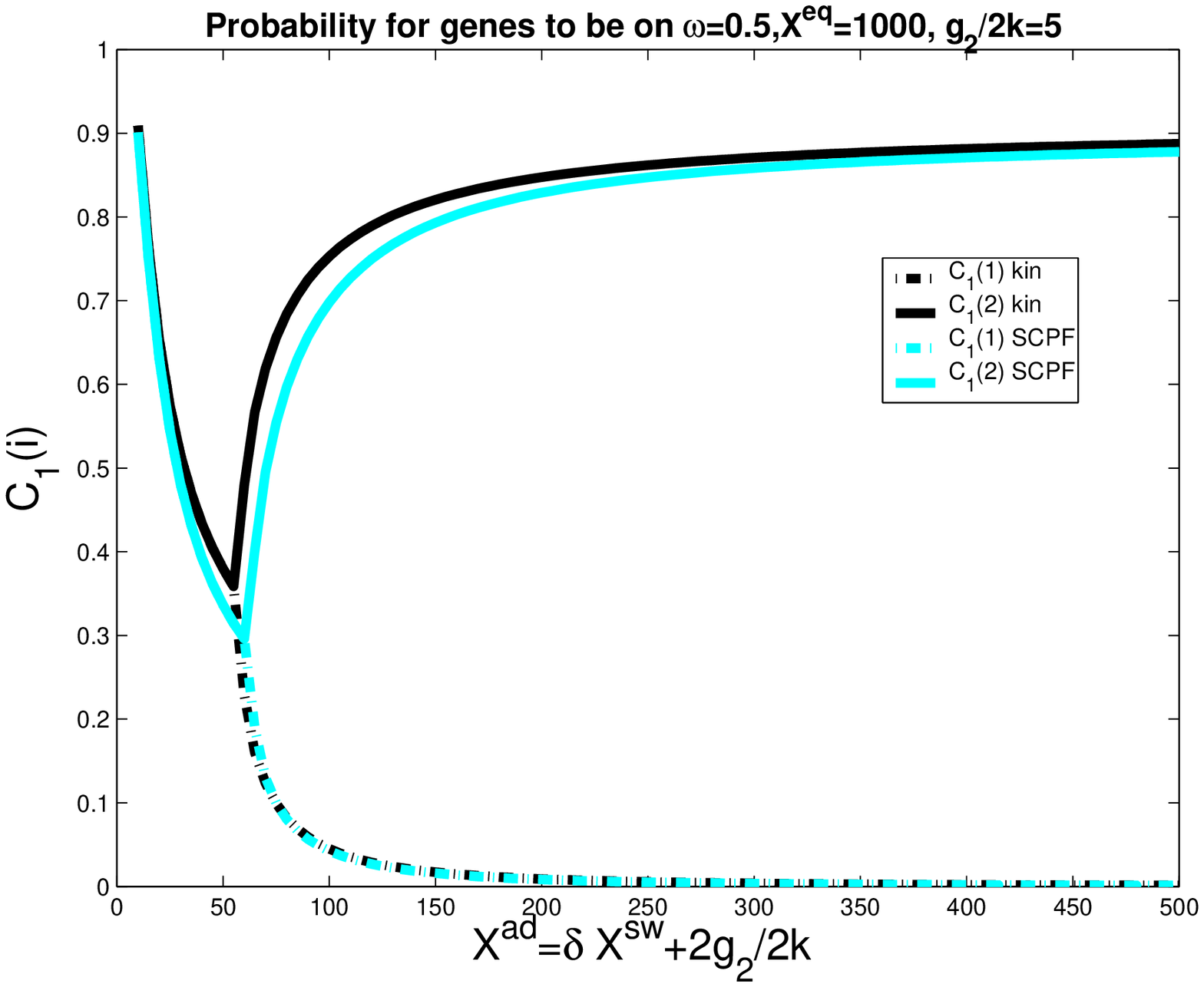}
\end{minipage}\hfill
\begin{minipage}[t]{.5\linewidth}
\includegraphics[height=3cm,width=4cm]{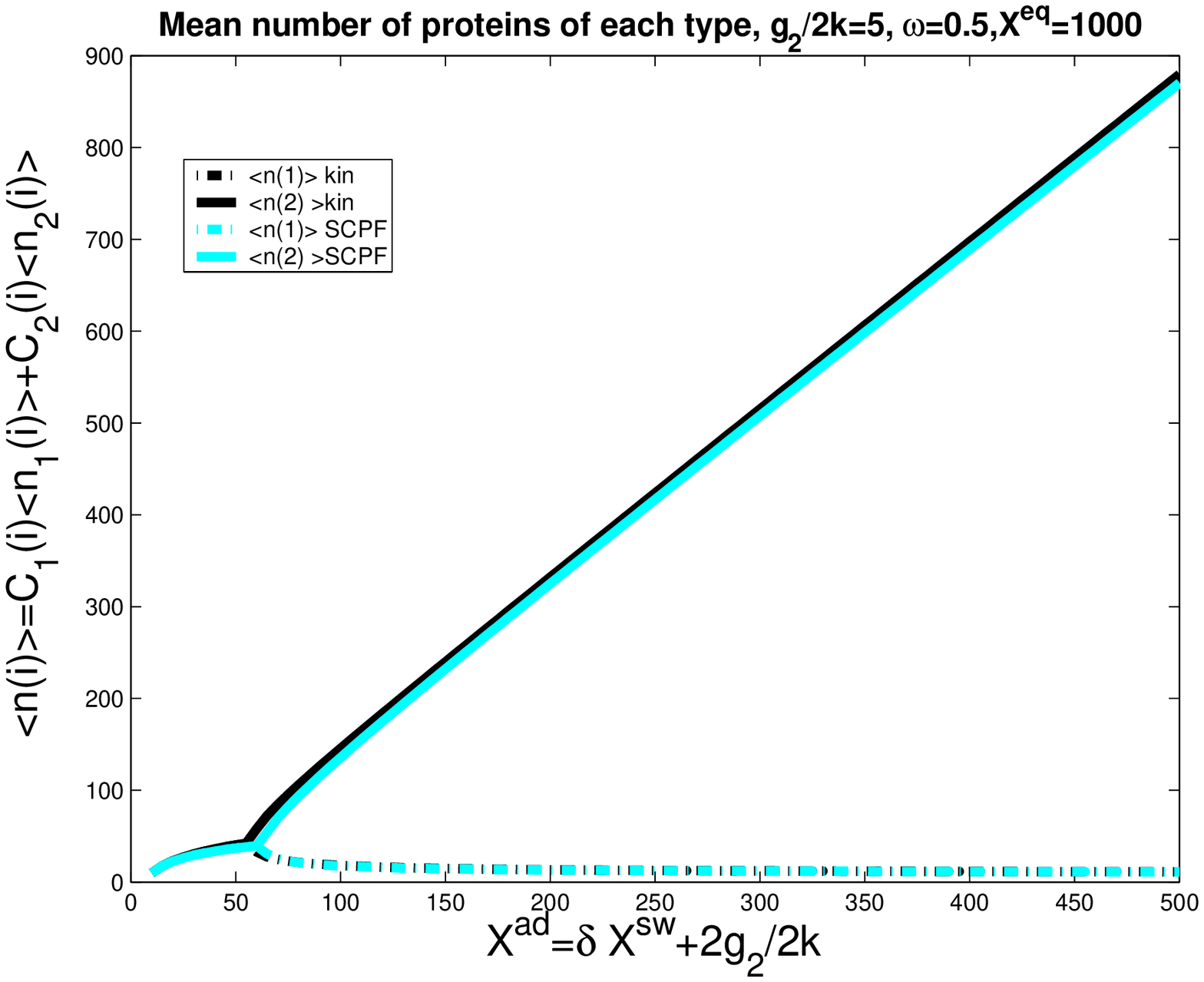}
\end{minipage}
\caption{Probability of genes to be on (a) and mean number of proteins of a given type present in the cell (b) for a symmetric switch with an effective base production rate $\frac{g_2}{2k}=5$, $\omega=0.5$, $X^{eq}=1000$. }
\label{evpoissonom1}
\end{figure}
\begin{figure}
\begin{minipage}[t]{.43\linewidth}
\includegraphics[height=3cm,width=4cm]{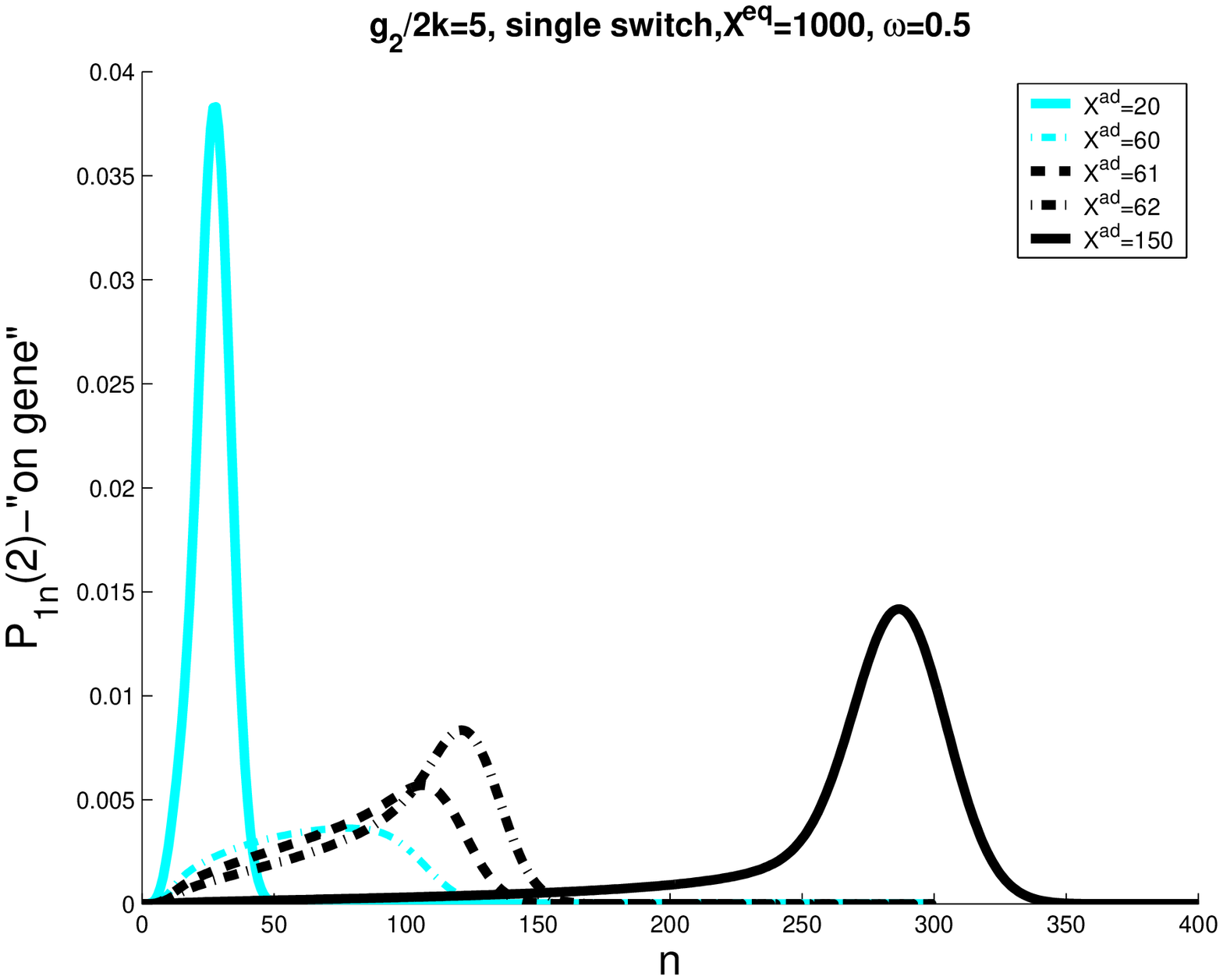}
\end{minipage}\hfill
\begin{minipage}[t]{.5\linewidth}
\includegraphics[height=3cm,width=4cm]{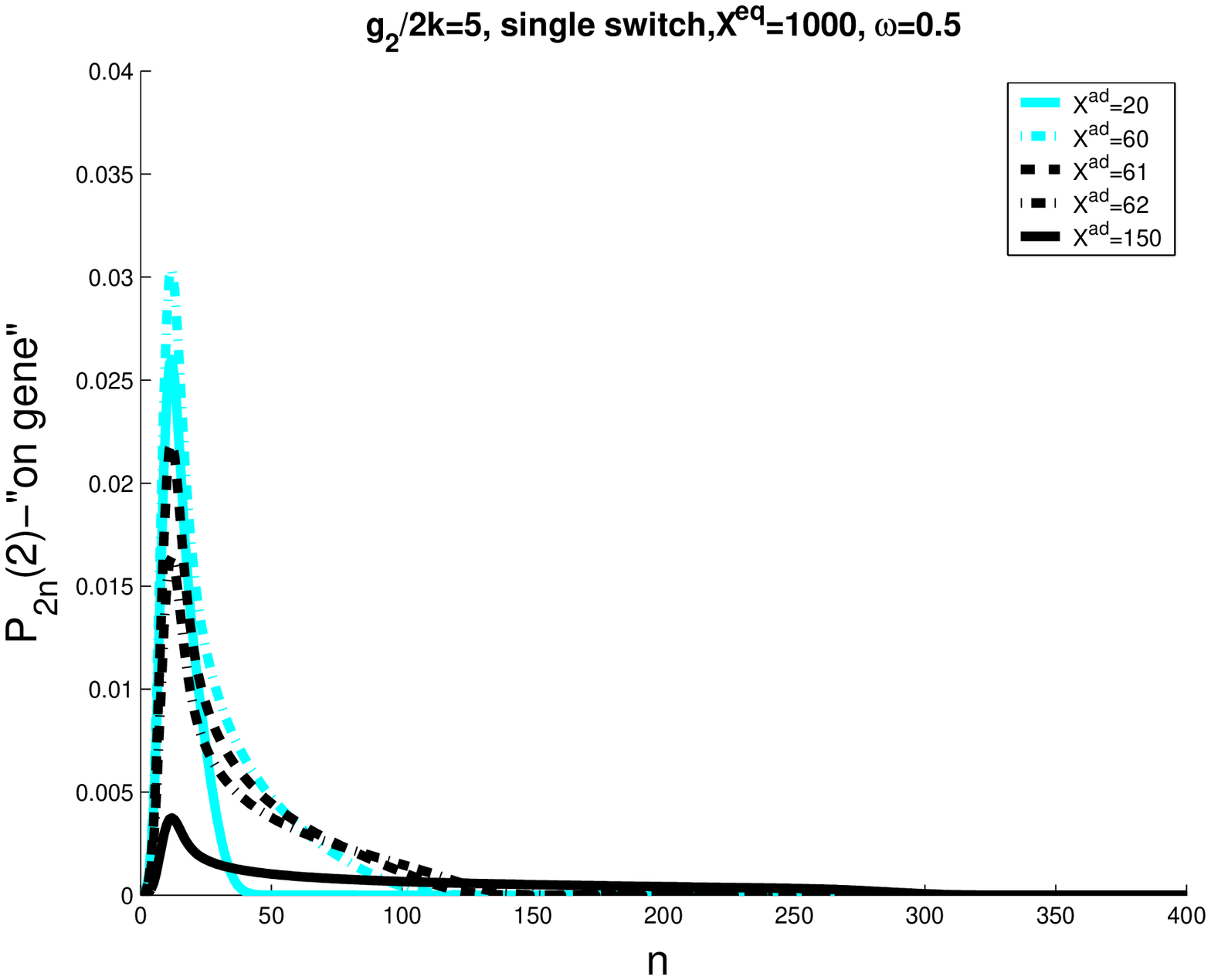}
\end{minipage}
\caption{Evolution of probability distributions for the probability of the gene that will be active after the bifurcation to be on (a) and off (b ) as a function of the order parameter $X^{ad}$ for a system with a basal production rate $\frac{g_2}{2k}=5$. The bifurcation occurs at $X^{ad}=61$.}
\label{pdnbl32}
\end{figure}
\textbf{The nonzero basal effective production rate case.}\\
The above analysis concerns a switch with a zero basal production rate, so proteins were not produced in the off state. In a number of biological systems (Ptashne and Gann, 2002) a non-zero basal production rate exists and we now turn to consider the effect of this on a symmetric switch. Figure \ref{evpoissonom2} (b) shows the dependence of the bifurcation curves for different values of the effective basal production rate $\frac{g_2}{2k}$. Values smaller than one, when the death rate is larger than the production rate, show that for the symmetric switch assuming the effective production rate to be zero in the off state is a reasonable approximation. If the on state has a positive input to the number of reservoir proteins present due to $\frac{2 g_2}{2k}>1$, the probability of the active gene to be on, even for very large on state effective production levels $X^{ad}$ is smaller than one. Hence the off state contributes considerably to the steady state number of proteins. The solution which corresponds to the more active of the two states may effectively be an off state, since it has $C_1(i)<\frac{1}{2}$, although the effective production rate in the on state in the bifurcated region of parameter space is much larger than in the off state (for example cyan line at $\frac{g_2}{2k}=20$ in Fig. \ref{evpoissonom2}). As the effective basal production rate increases, a larger production rate in the on state than for small $\frac{g_2}{2k}>1$ is required to reach the critical number of proteins for the bifurcation to take place, which is given by $<n(i)>=2 X^{ad} C_1(i)-\frac{g_2}{k}(2 C_1(i)-1)$. For this reason even for the deterministic approximation at the bifurcation point, the two genes must be more probable to be off, as can also be seen for the exact SCPF solutions from the probability distributions (Figs \ref{pdnbl32}, \ref{evpoissonom3}). Figure \ref{evpoissonom2} (a) shows the dependence of the bifurcation curves on the adiabaticity parameter, which tend to the deterministic case for large $\omega$.  A closer analysis of the $\frac{2 g_2}{2k}>1$ case, since the $\frac{2 g_2}{2k}<1$ is analogous to the zero basal production rate case which was already discussed, show that mean properties of the system are in even better agreement with the deterministic solution than the $g_2=0$ case (Fig. \ref{evpoissonom1}). The system has a non-zero probability of being in the off state, with the probability distribution of the off gene having a long tail towards higher protein numbers Fig. \ref{pdnbl32}. In the off state the effective production rate $\frac{g_2}{2k}$ is small and the noise input is small, relative to the large protein numbers present in the system. The small effect of stochasticity results in the observed similar mean characteristics. Yet the form of the probability distribution for the gene to be active before the transition is especially broad, with far smaller probability than that of the inactive state (Figs \ref{pdnbl32}, \ref{evpoissonom3}). These clearly show that the two genes are more probable to be in the off state before the bifurcation point. Therefore although the average observables are similar for the deterministic and SCPF stochastic solutions, the predicted distributions are unusual. \\
\textbf{Summary}\\
The symmetric switch is based on a competition between the accessibility of the repressor site and the number of repressor proteins present in the cell. The bifurcation is solely a result of the nonlinearity of the system and introducing noise simply affects the region in parameter space where given states occur. The protein number fluctuations have a destructive role in determining the stability of the bifurcated solution, however fast DNA unbinding rates can compensate for the destabilizing effect of protein number fluctuations. In this region the stochastic solution predicts similar means to the deterministic case, but the form of the probability distributions which depends on a large number of higher moments is non-trivial. It is a result of the interplay of the DNA binding and protein degradation kinetics.\\
\subsection*{The Asymmetric Toggle Switch}
\begin{figure}
\begin{minipage}[t]{.43\linewidth}
\includegraphics[height=3cm,width=4cm]{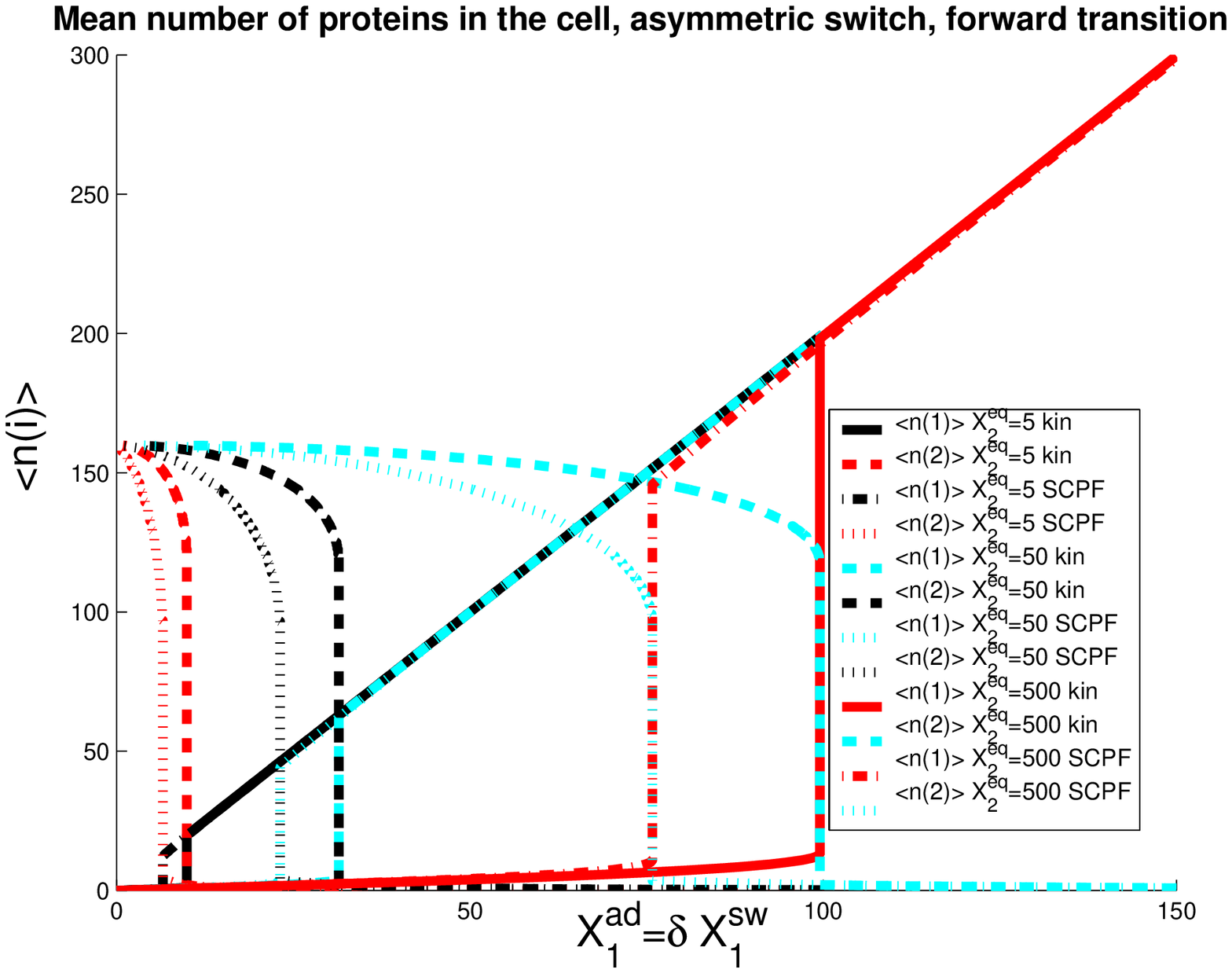}
\end{minipage}\hfill
\begin{minipage}[t]{.5\linewidth}
\includegraphics[height=3cm,width=4cm]{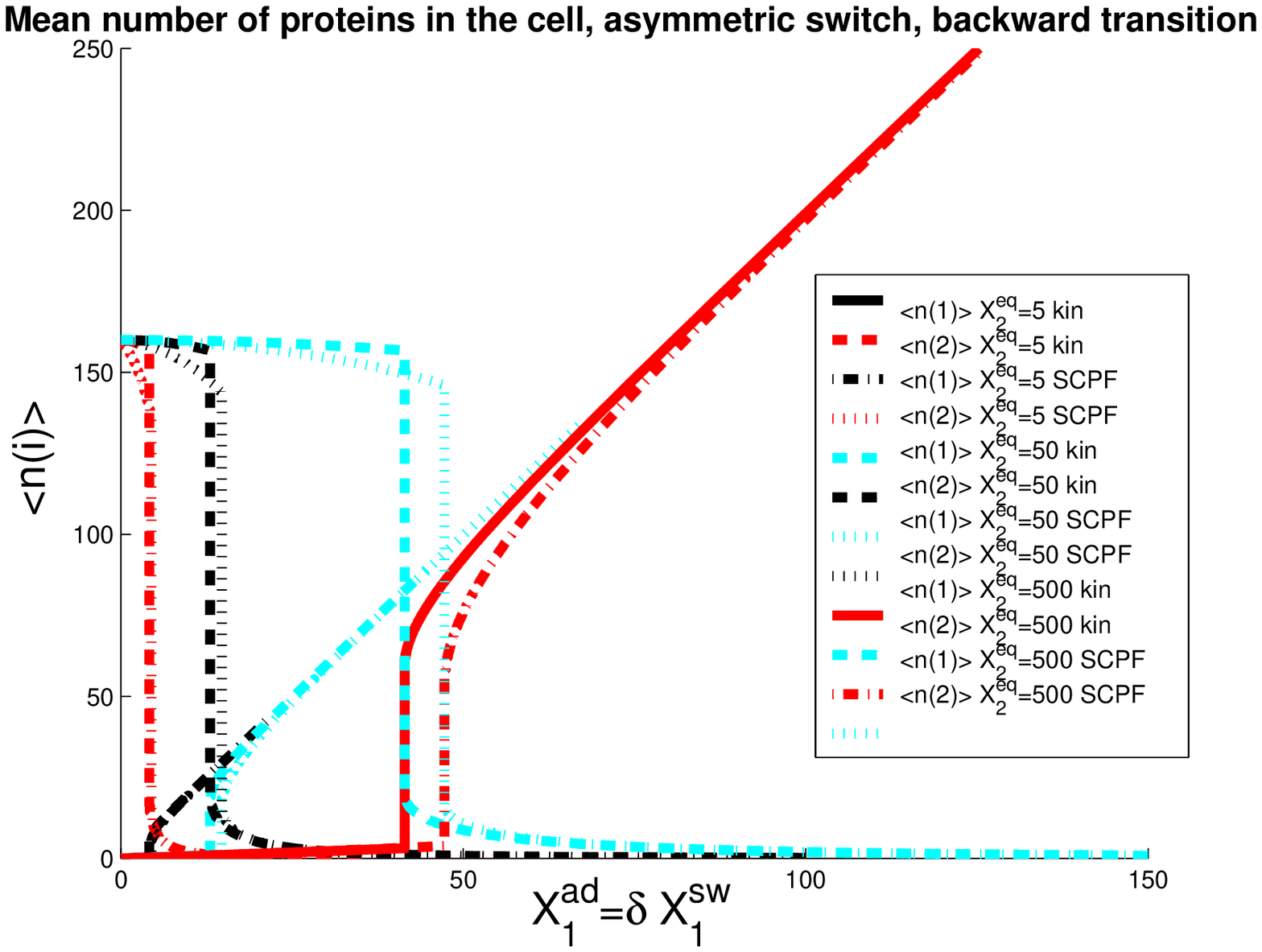}
\end{minipage}
\caption{Mean number of proteins of each type present in the cell, according to exact solutions of the SCPF approximation for an asymmetric switch, with $X^{eq}_1 = 1000;\omega_1=\omega_2=0.5; X_2^{ad}= \delta X_2^{sw}=80$; and $X^{eq}_2 =5,50,500$ during the forward (a) and backward (b) transition in the asymmetric switch.} 
\label{xeqprotein}
\end{figure}

\begin{figure}
\begin{minipage}[t]{.43\linewidth}
\includegraphics[height=3cm,width=4cm]{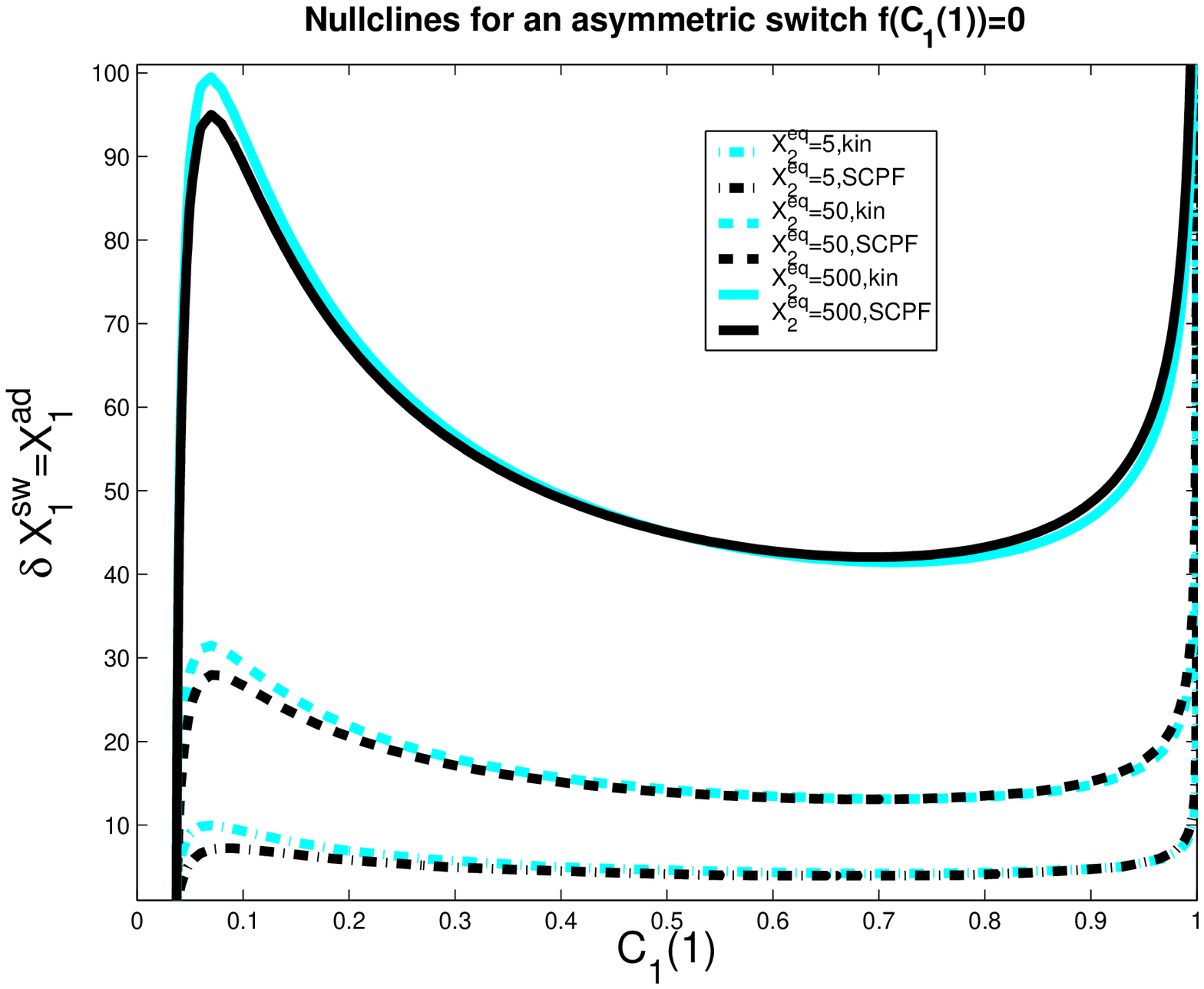}
\end{minipage}\hfill
\begin{minipage}[t]{.5\linewidth}
\includegraphics[height=3cm,width=4cm]{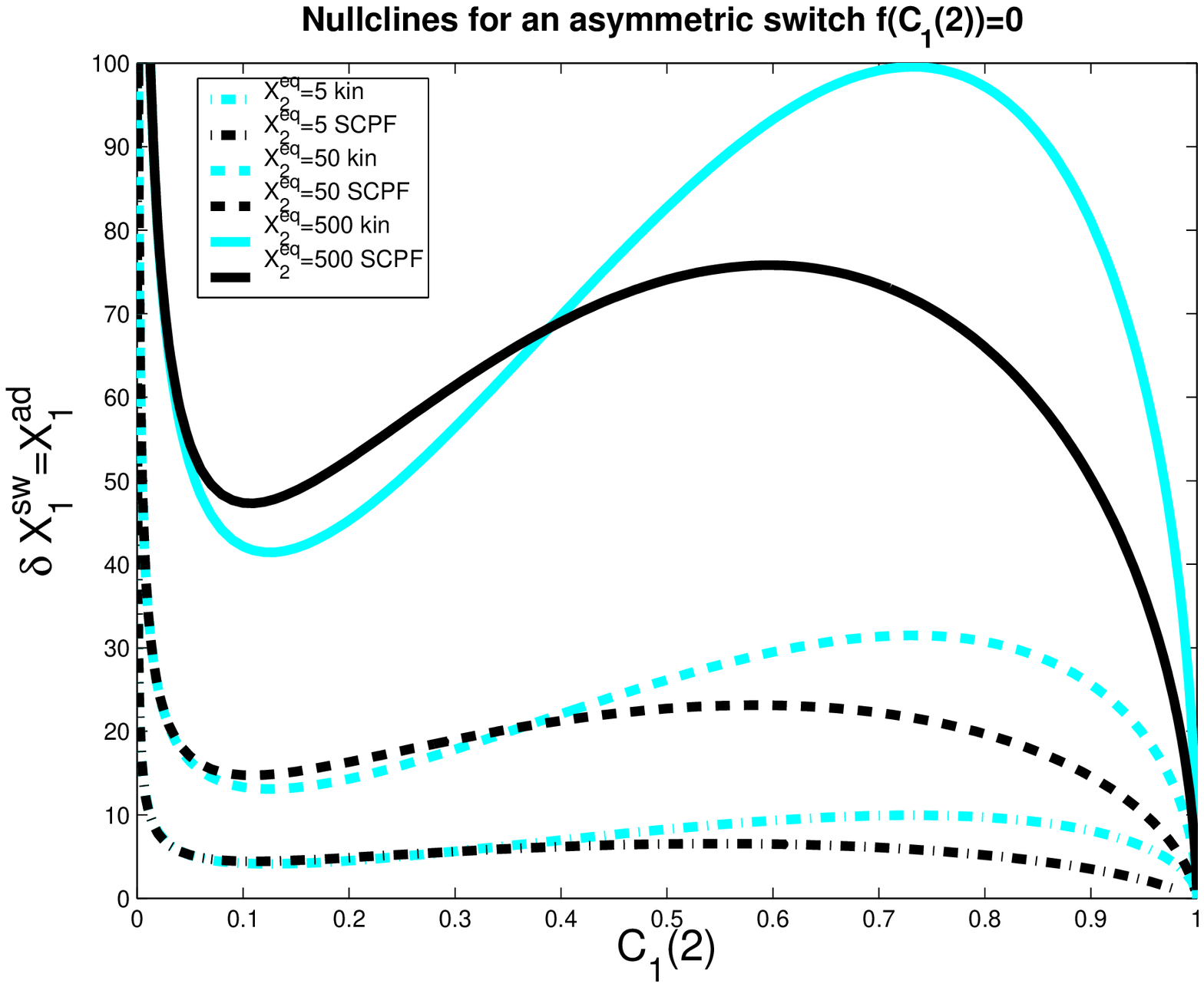}
\end{minipage}\hfill
\caption{Bifurcarion diagrams as a function of $X_1^{ad}=\delta X_1^{sw}$ for $X_2^{eq}=5,50,500$ for exact solution of the SCPF and deterministic kinetic equations for an asymmetric switch. Nullclines for $C_1(1)$ with $\omega=10$ (a) and $C_1(2)$ with $\omega=0.5$ (b) with $X^{eq}_1 = 1000$;$\omega_1=\omega_2=0.5$; $X_2^{ad}= \delta X_2^{sw}=80$. }
\label{asymbif}
\end{figure}

\begin{figure}
\begin{minipage}[t]{.25\linewidth}
\includegraphics[height=3cm,width=2.9cm]{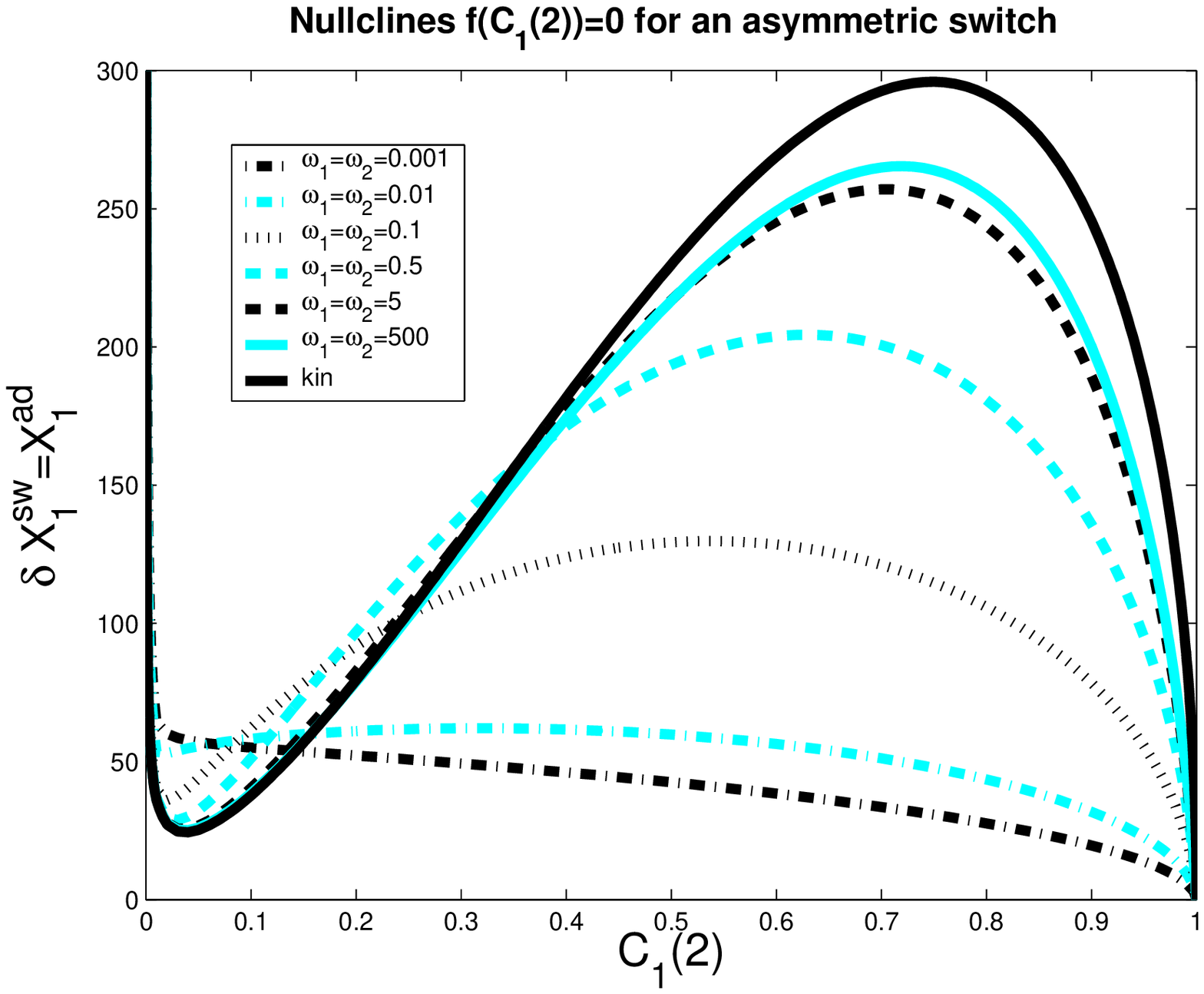}
\end{minipage}\hfill
\begin{minipage}[t]{.25\linewidth}
\includegraphics[height=3cm,width=2.9cm]{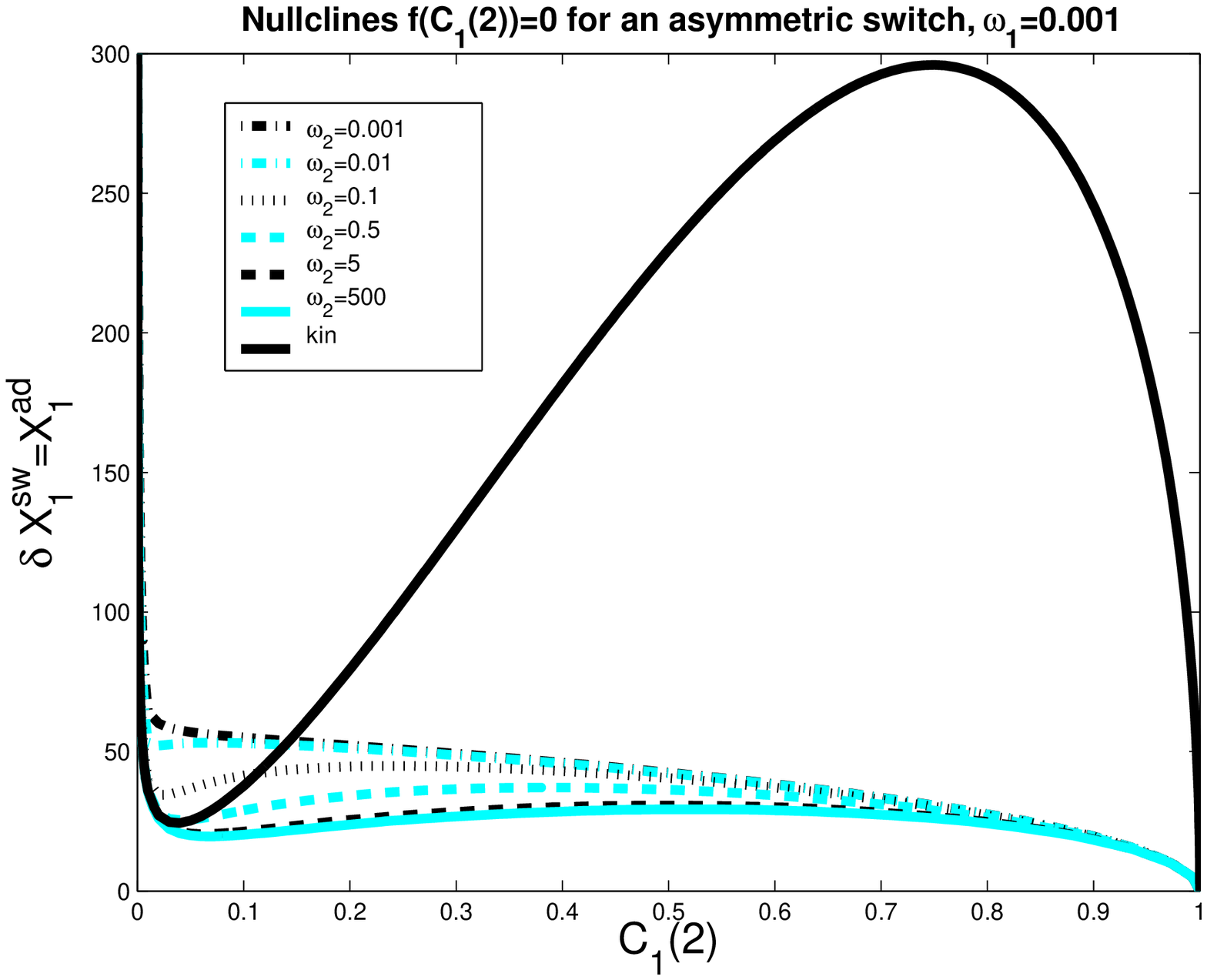}
\end{minipage}\hfill
\begin{minipage}[t]{.35\linewidth}
\includegraphics[height=3cm,width=2.9cm]{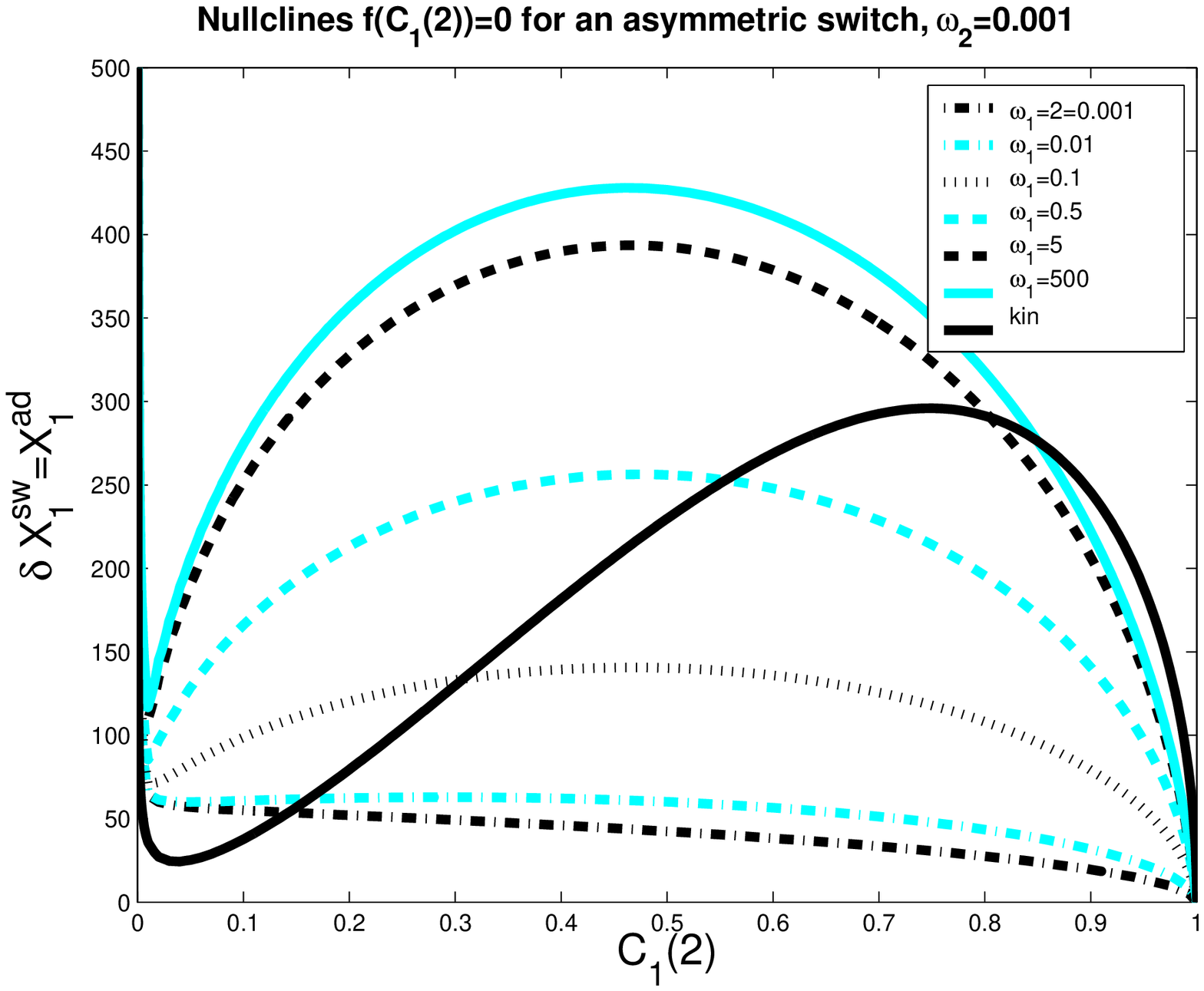}
\end{minipage}
\begin{minipage}[t]{.25\linewidth}
\includegraphics[height=3cm,width=2.9cm]{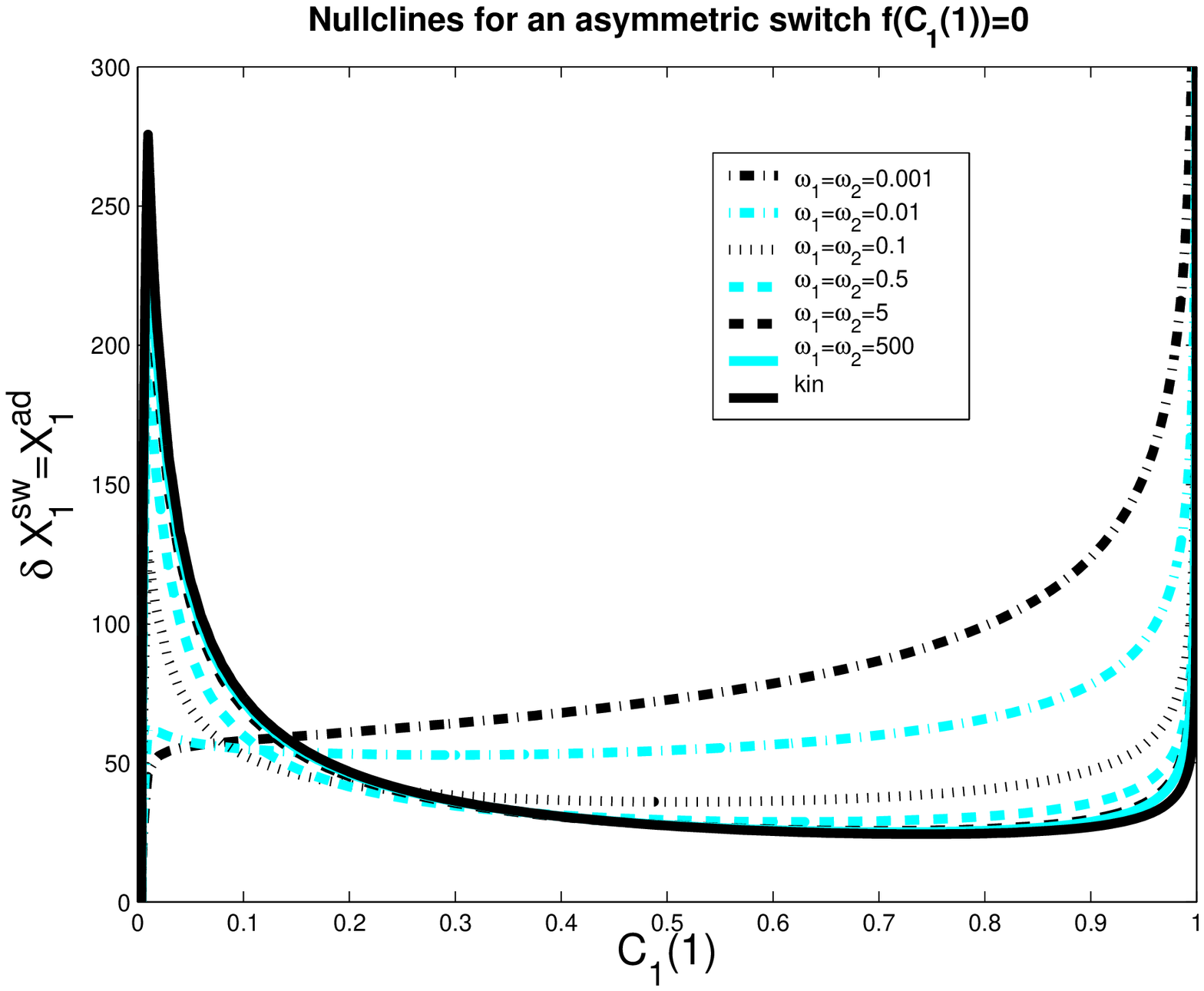}
\end{minipage}\hfill
\begin{minipage}[t]{.25\linewidth}
\includegraphics[height=3cm,width=2.9cm]{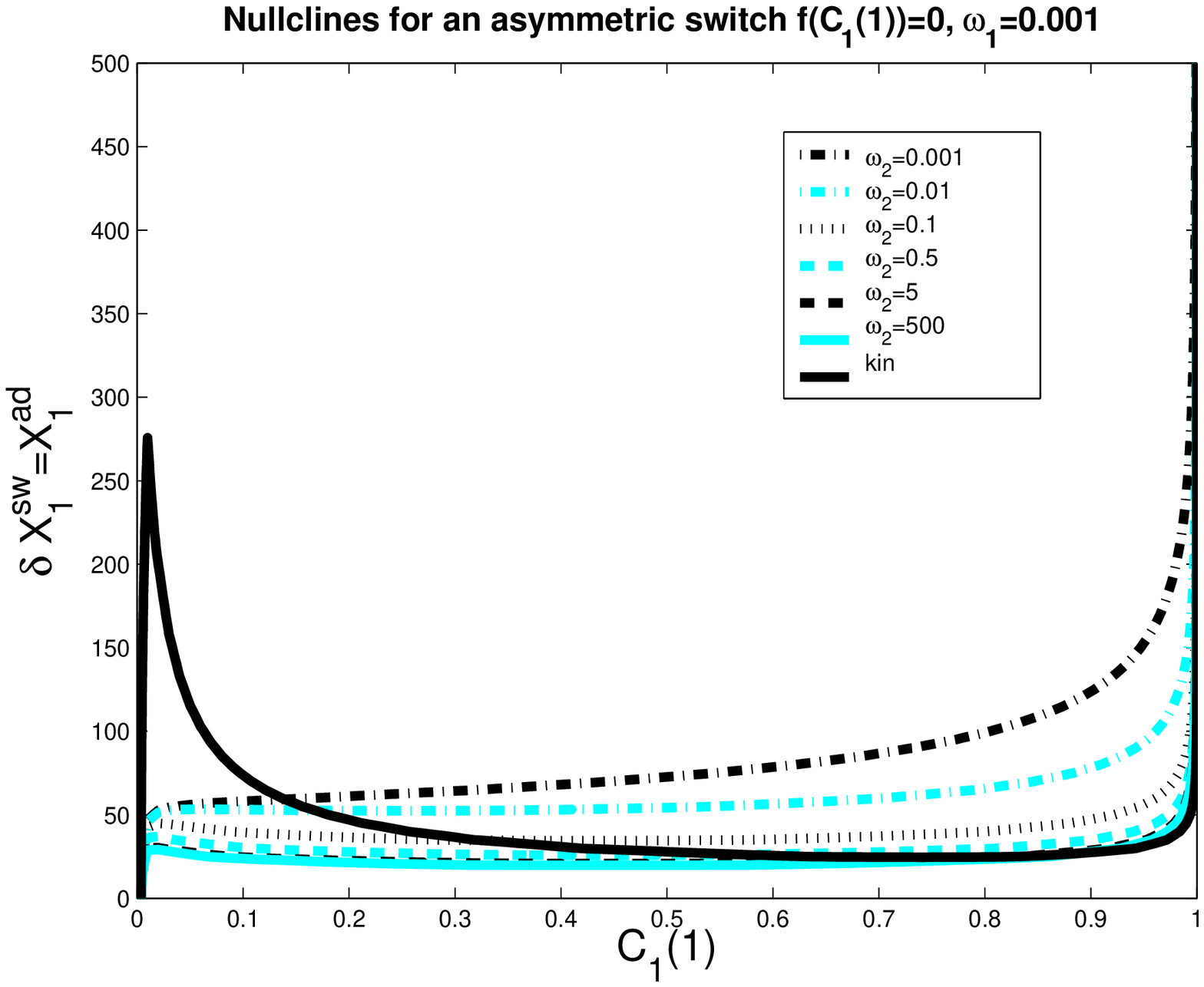}
\end{minipage}\hfill
\begin{minipage}[t]{.35\linewidth}
\includegraphics[height=3cm,width=2.9cm]{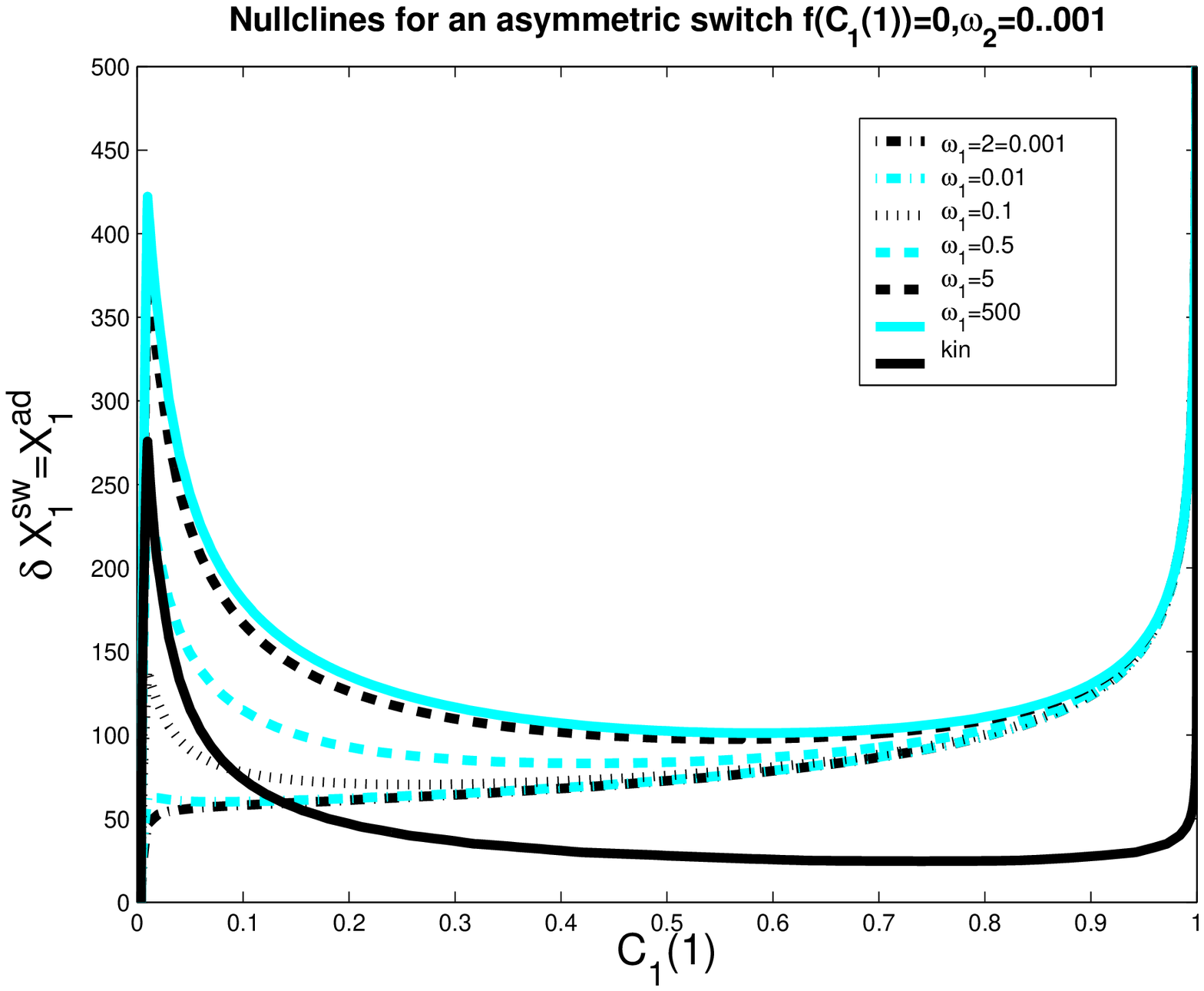}
\end{minipage}
\caption{Bifurcation diagrams for an asymmetric switch, presenting $X_1^{ad}=\delta X_1^{sw}$ as a function of $C_1(2)$ (top), and $C_1(1)$ (bottom) for different values of the adiabaticity parameter: $\omega_1$=$\omega_2$ (a, e), $\omega_2$, with $\omega_1=0.001=const$ (b, d), $\omega_1$, with $\omega_2=0.001=const$ (c, f). $X_1^{eq}=100$,$X_2^{eq}=50$,$X_2^{ad}=\delta X_2^{sw}=80$.}
\label{omdepasym}
\end{figure}

Most switches found in nature are not symmetric. For asymmetric switches, when proteins bind as dimers, the two genes interact, resulting in probabilities to be on, different from those imposed purely by the equilibrium between binding and unbinding. The steady state solution is a compromise between the tendency that repressors are unbound from the initially off gene ($X_1^{eq}$ for the forward transistion, $X_2^{eq}$ for the backward in the following discussion) and the effective production rate of the initially on gene ($X_2^{ad}$ - forward, $X_1^{ad}$ backward transition) (at least for the deterministic case). This results in the characteristic S-curve bifurcation diagram, as presented in, for example Fig. \ref{asymbif}, with possible forward and backward transtions, and what follows hysteresis. We refer to the transition which occurs with increasing $X^{ad}_1$ as the forward transition and that with decreasing  $X^{ad}_1$  as the backwards transition. Since $X_i^{ad}$ is a well defined function of the probabilities that the genes are on, the simplicity of the deterministic equations allows for a completely analytic discussion of the asymmetric switch. The more complicated form of the exact SCPF equations makes this approach impossible. However the deterministic rate solution offers valuable insight into the basic mechanism behind the transition.\\
\begin{figure}
\begin{minipage}[t]{.25\linewidth}
\includegraphics[height=3cm,width=2.9cm]{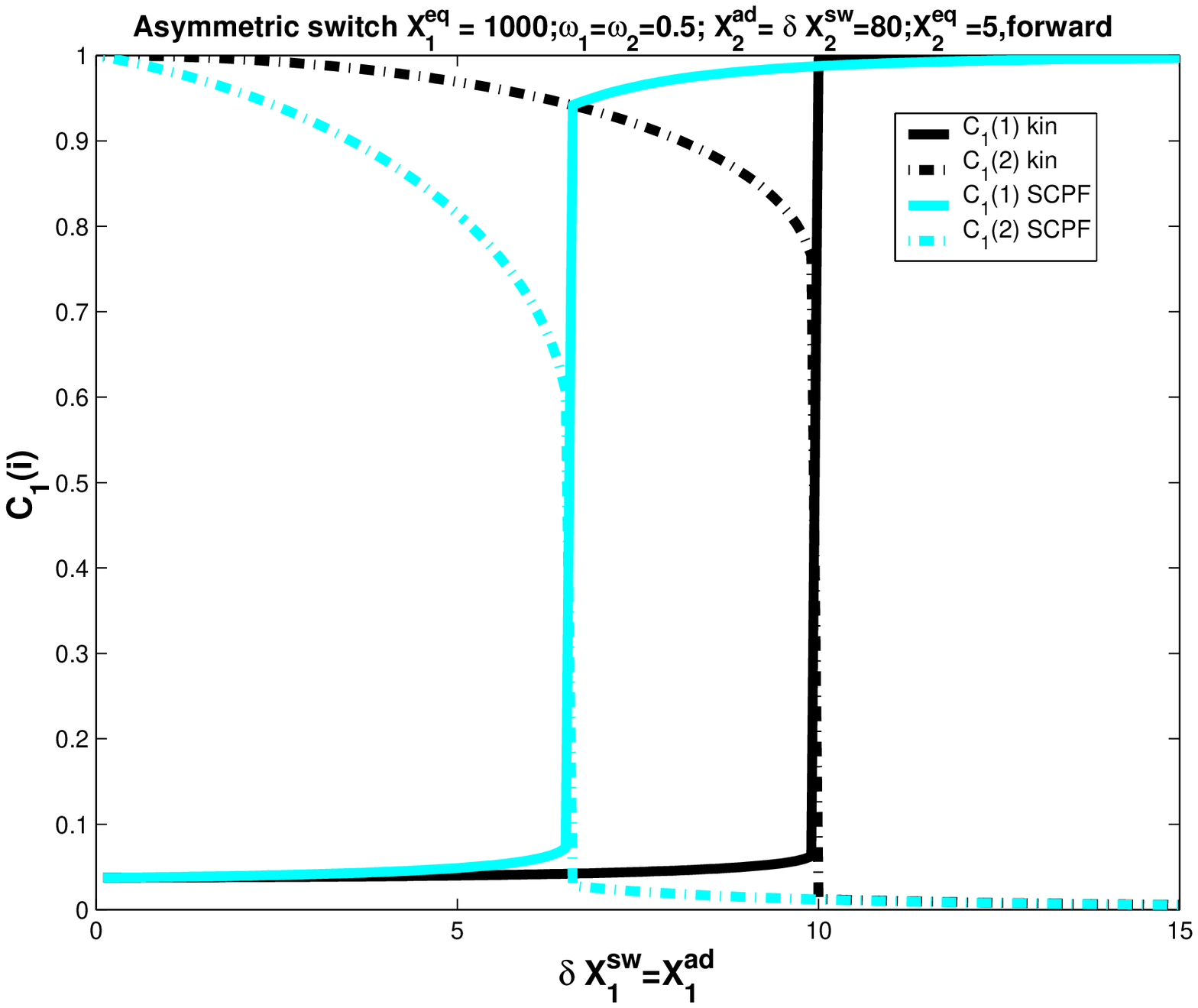}
\end{minipage}\hfill
\begin{minipage}[t]{.25\linewidth}
\includegraphics[height=3cm,width=2.9cm]{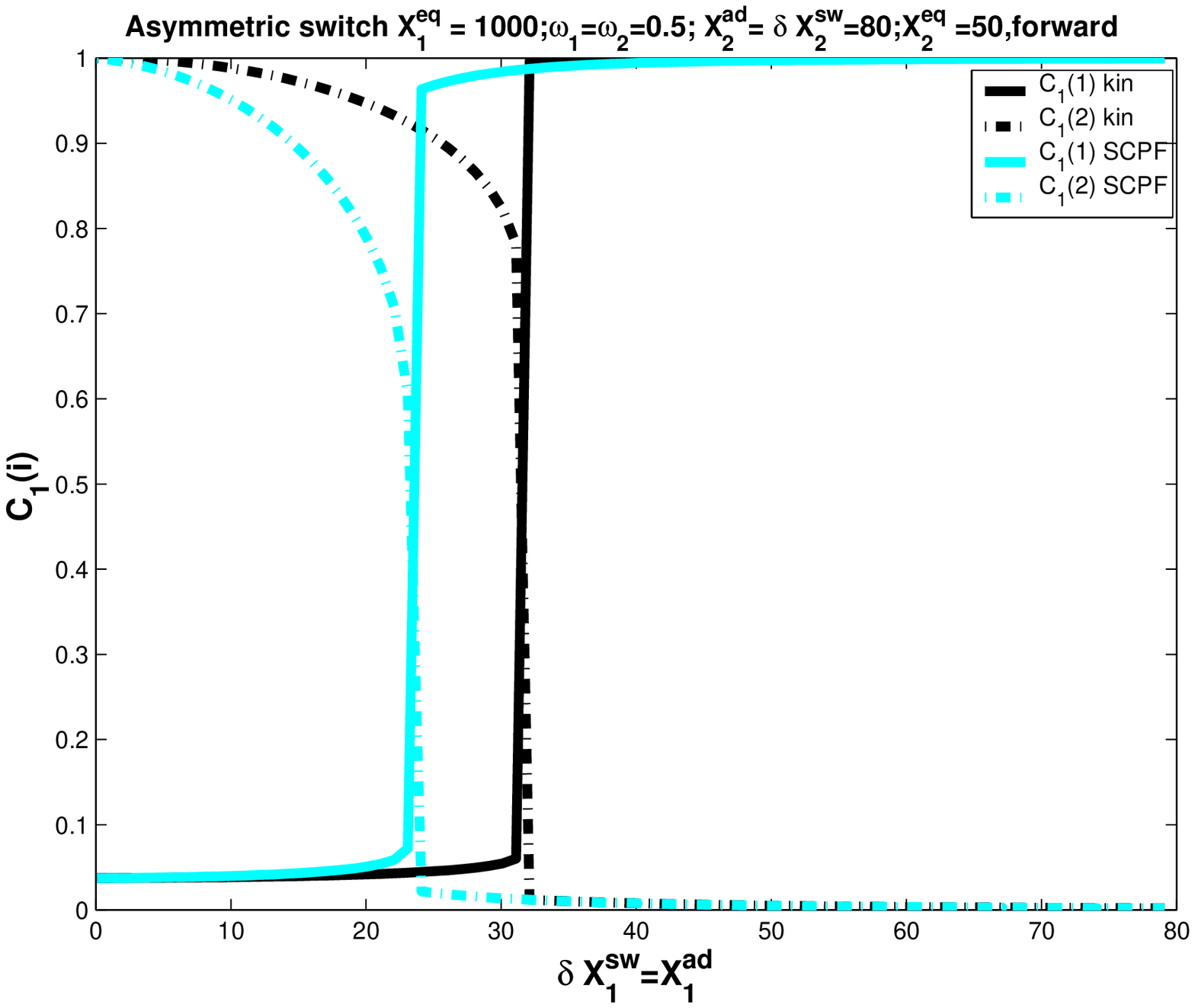}
\end{minipage}\hfill
\begin{minipage}[t]{.33\linewidth}
\includegraphics[height=3cm,width=2.9cm]{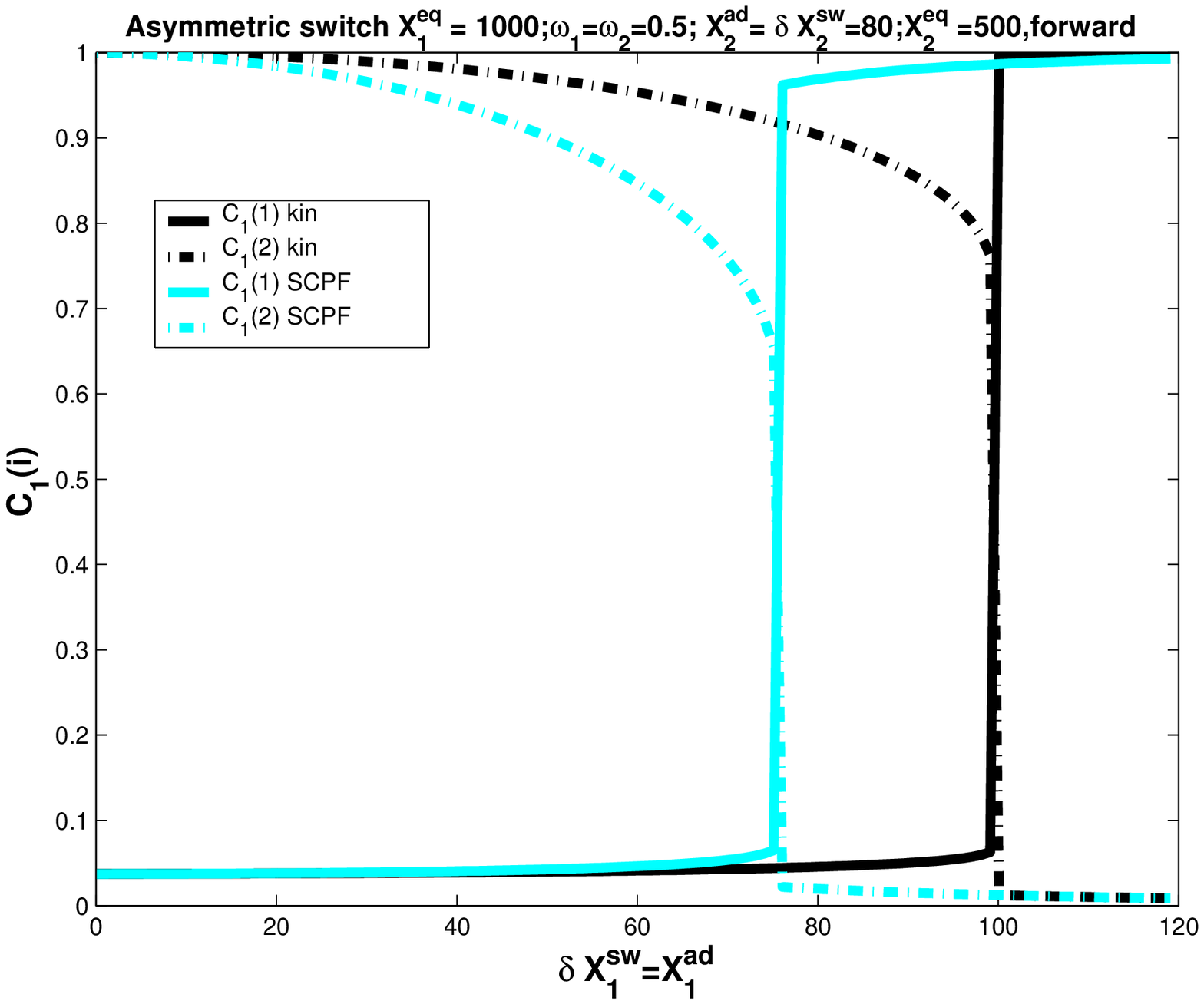}
\end{minipage}
\caption{Dependendce of probability of genes in an asymmetric switch to be on as a function of increasing parameters of one gene $X_1^{ad}=\delta X_1^{sw}$ (forward transition) for different values of $X_2^{eq}$: 5 (a), 50 (b), 500 (c), keeping all other parameters fixed, $X_1^{eq}=1000$, $\omega_1=\omega_2=0.5$, $X_2^{ad}=\delta X_2^{sw}=80$, for deterministic and exact SCPF equations. }
\label{xeqasyma}
\end{figure}
\begin{figure}
\begin{minipage}[t]{.25\linewidth}
\includegraphics[height=3cm,width=2.9cm]{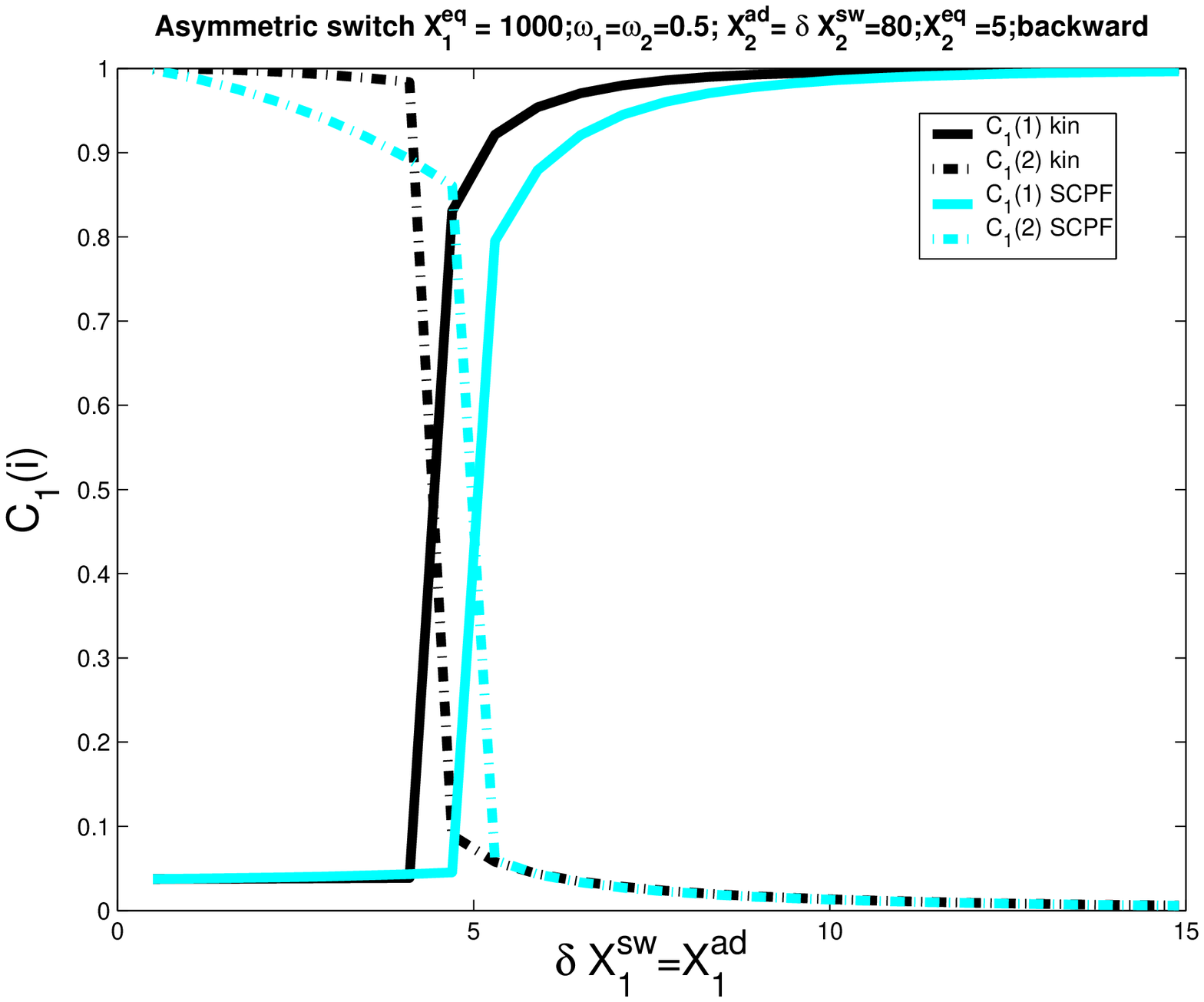}
\end{minipage}\hfill
\begin{minipage}[t]{.25\linewidth}
\includegraphics[height=3cm,width=2.9cm]{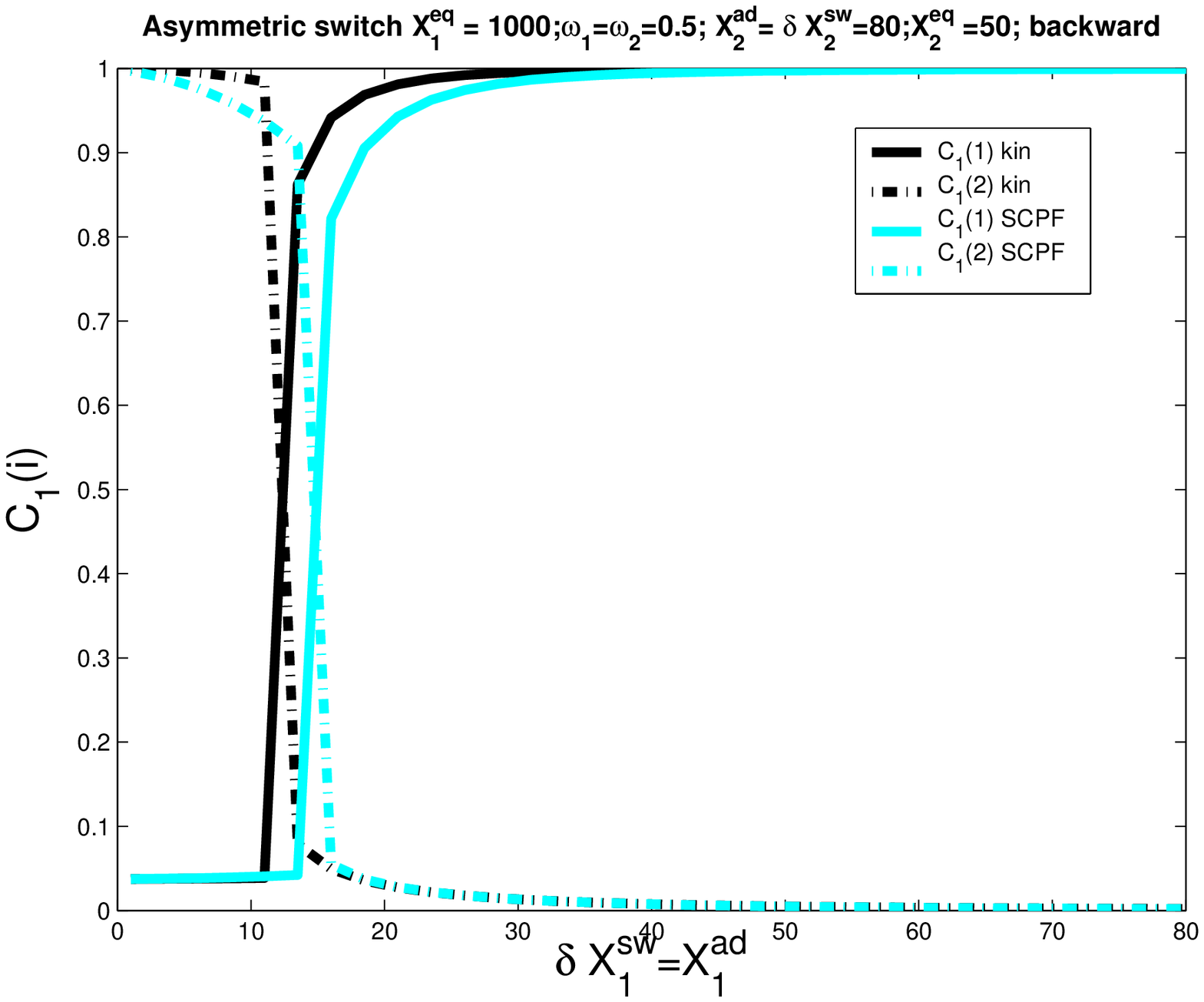}
\end{minipage}\hfill
\begin{minipage}[t]{.33\linewidth}
\includegraphics[height=3cm,width=2.9cm]{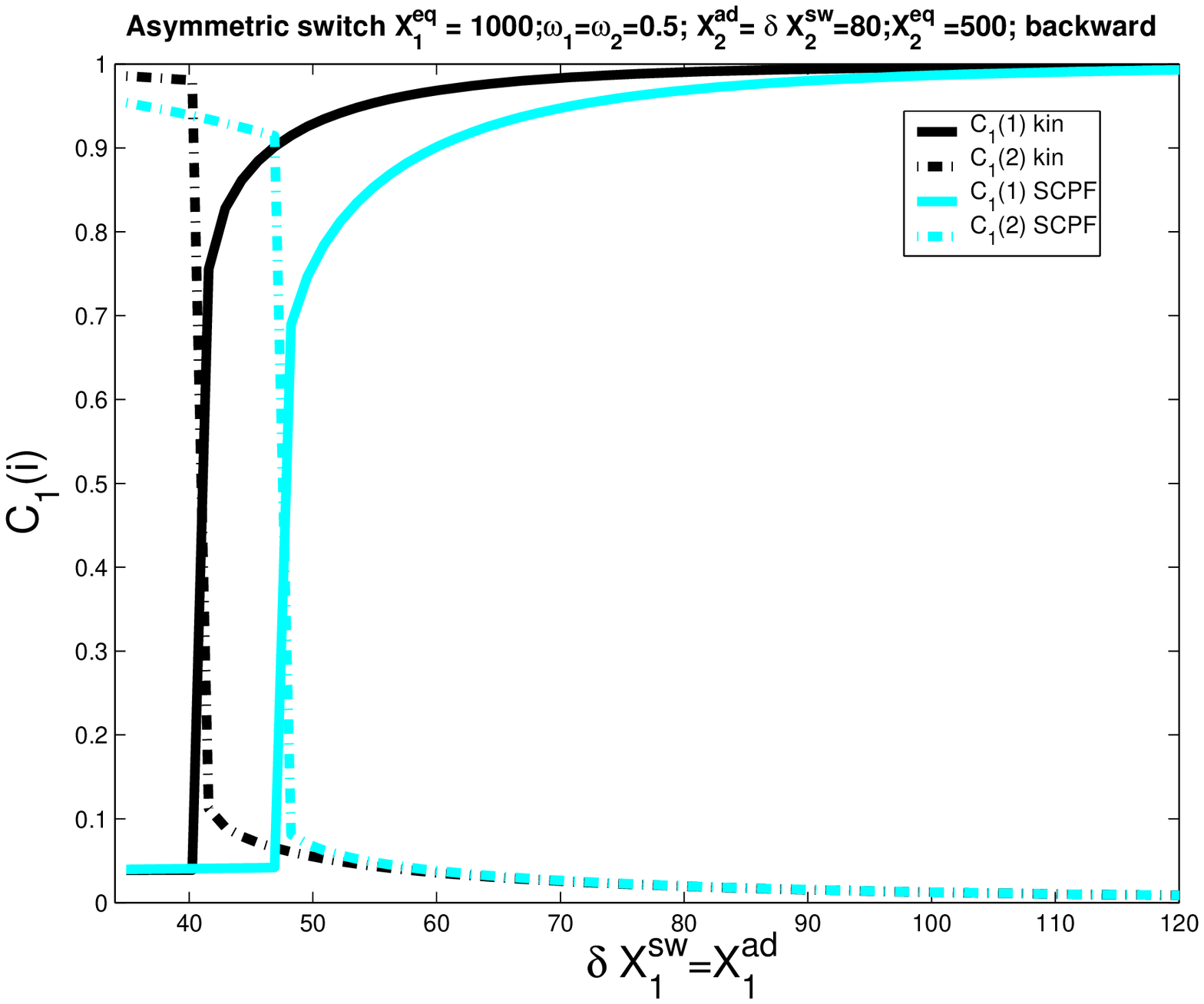}
\end{minipage}
\caption{Dependendce of probability of genes in an asymmetric switch to be on as a function of decreasing parameters of one gene $X_1^{ad}=\delta X_1^{sw}$ (backward transition), for different values of $X_2^{eq}$: 5 (a), 50 (b), 500 (c), keeping all other parameters fixed, $X_1^{eq}=1000$, $\omega_1=\omega_2=0.5$, $X_2^{ad}=\delta X_2^{sw}=80$, for deterministic kinetic rate and SCPF equations.}
\label{xeqasyma1}
\end{figure}
\textbf{The general mechanism}\\
By combining the steady state equations of motion for the probabilities of the two genes to be on Eq. \ref{c1c2eqns} and noting that with a zero basal production rate $<n(i)>=2X^{ad}_i C_1(i)$, one can derive the following form of the deterministic bifurcation curves:\\
\beq
X^{ad}_1(C_1(2))=\frac{{X^{eq}_2}^{\frac{1}{2}}} {2} (1+\frac{( 2 X^{ad}_2 C_1(2))^2}{X^{eq}_1})(\frac{1}{C_1(2)}-1)^{\frac{1}{2}}
\label{bikin1}
\eeq
as a function of $C_1(2)$ and:\\
\beq
X^{ad}_1(C_1(1))=\frac{{X^{eq}_2}^{\frac{1}{2}}} {2 C_1(1)} (\frac{2 X^{ad}_2}{((\frac{1}{C_1(1)}-1)X^{eq}_1)^{\frac{1}{2}}}-1)^{\frac{1}{2}}
\label{bikin2}
\eeq
as a function of $C_1(1)$. The transistion points are determined as the extrema of these functions, which are functions solely of the scaled parameter $X^{ad 2}_2/X^{eq}_1$ and are plotted on the bifurcation graphs. It is worth noticing that the bifurcation points $C_1(i)$ do not depend on the value of $X^{eq}_2$, the parameter describing the gene binding kinetics of the gene that is on initially. This is not true for the exact SCPF solution, which cannot be solved analitically, but the bifurcation curve has the more complex form:
\begin{eqnarray*}
X^{ad}_1(C_1(2))&=&\frac{1}{2}((((\frac{1}{C_1(1)}-1)X^{eq}_2)^{\frac{1}{2}}\frac{\omega_1+C_1(1)}{1+\omega_1}+1)^{\frac{1}{2}}+\\
&&-\frac{\omega_1+C_1(1)}{1+\omega_1})\frac{1}{2 C_1(1)}
\end{eqnarray*}
where $C_1(1)$ is a function of $\omega_2, X^{eq}_1, C_1(2)$ and $X^{ad_2}$. The bifurcation point is therefore determined by the protein ($X^{ad}_i$) and DNA ($X^{eq}_i$) characteristics and their mutual interactions ($\omega_i$) of the two genes. The deterministic approximation therefore greatly simplifies the mathematical mechanism of the transition. This may lead to large errors when studying more complicated biologically relevant systems, where one considers asymmetric switches with non-zero basal production rates and proteins are produced in bursts. The case of the non-zero basal production rate within the deterministic approximation also cannot be solved analytically.\\
\begin{figure}
\begin{minipage}[t]{.43\linewidth}
\includegraphics[height=3cm,width=4cm]{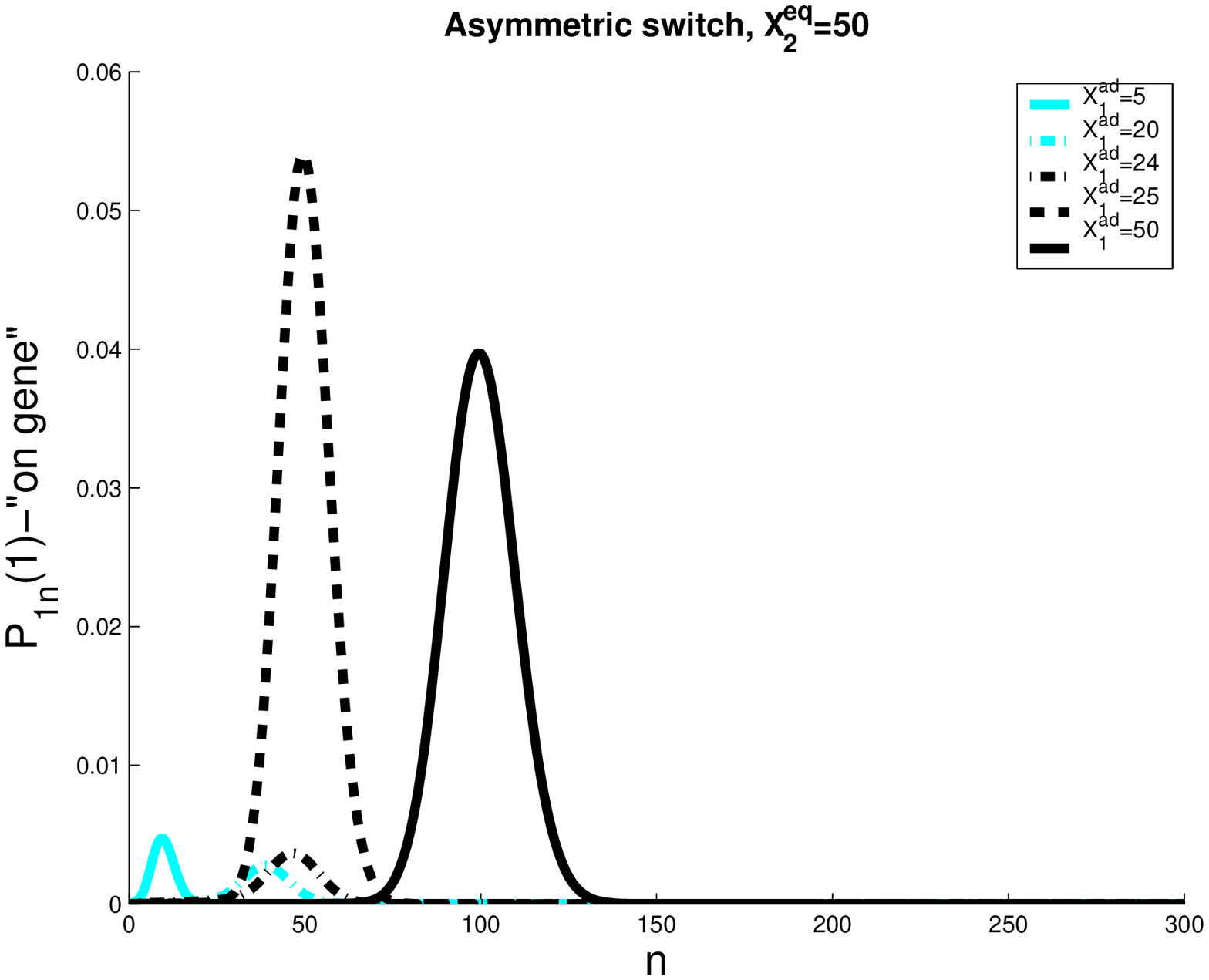}
\end{minipage}\hfill
\begin{minipage}[t]{.5\linewidth}
\includegraphics[height=3cm,width=4cm]{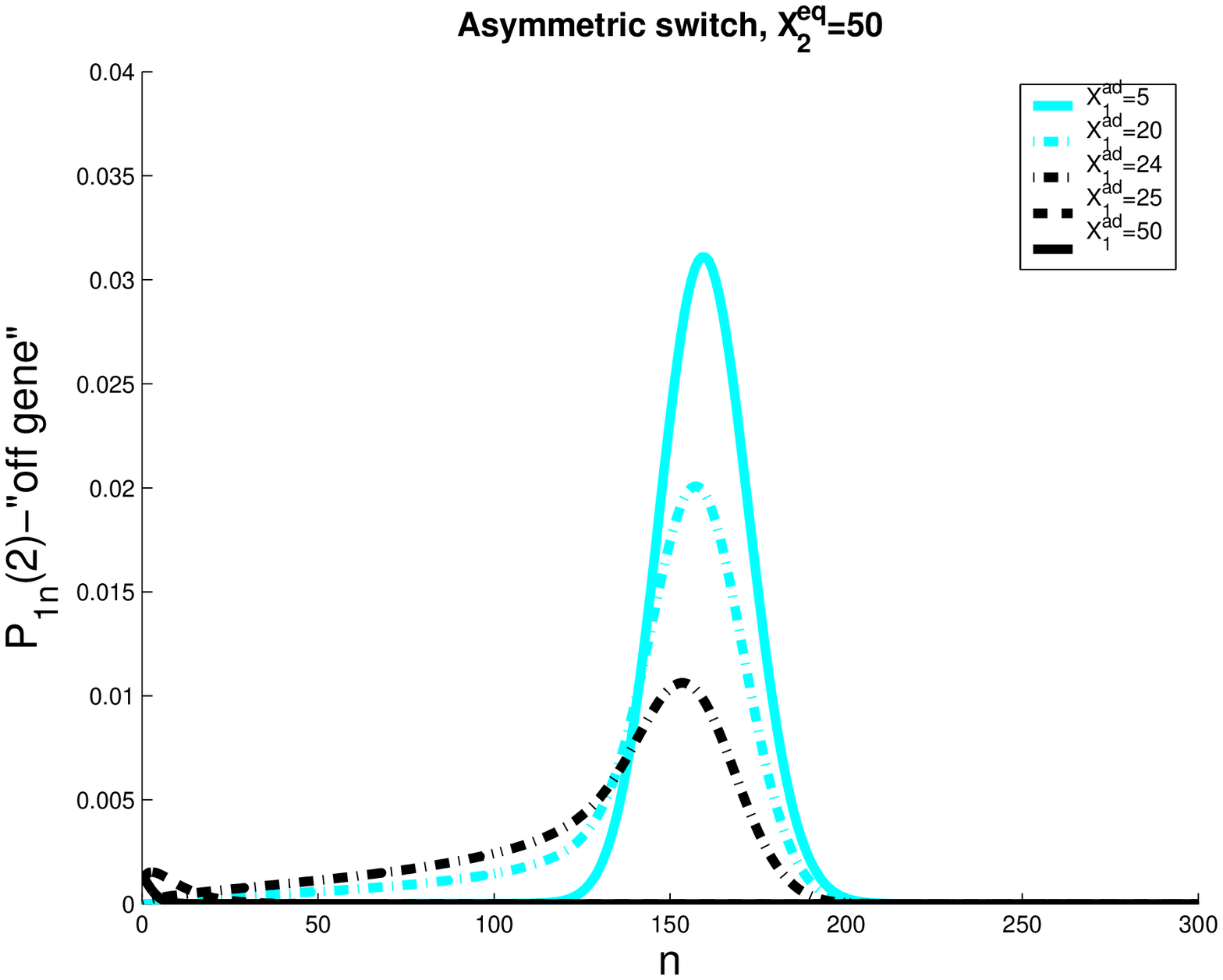}
\end{minipage}
\caption{Evolution of the probability distribution for a forward transition as a function of $X_1^{ad}=\delta X_1^{sw}$ for $X_2^{eq}=50$ with $X^{eq}_1 = 1000$; $\omega_1=\omega_2=0.5$; $X_2^{ad}= \delta X_2^{sw}=80$ for an asymmetric switch.}
\label{asympr}
\end{figure}
\begin{figure}
\begin{minipage}[t]{.43\linewidth}
\includegraphics[height=3cm,width=4cm]{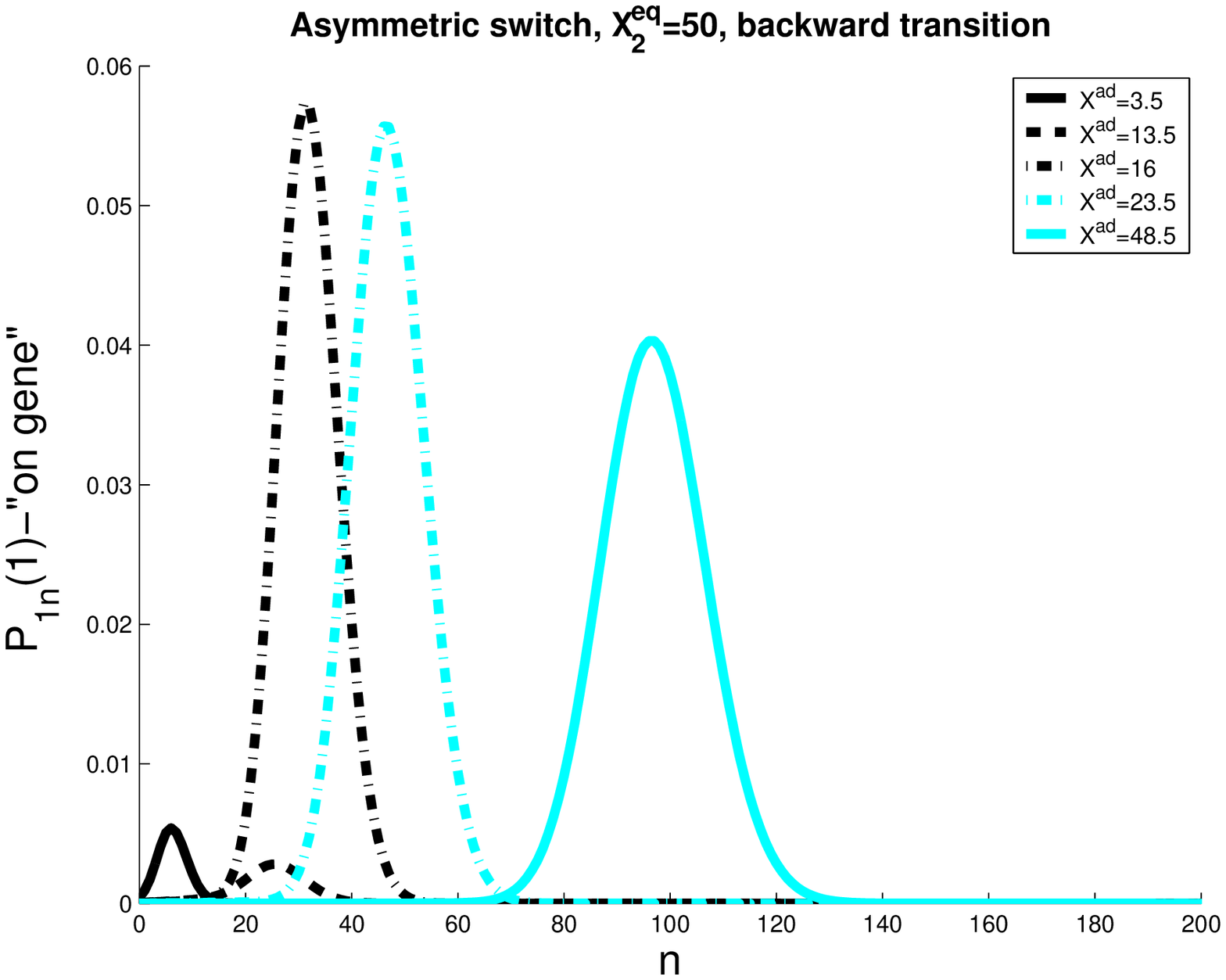}
\end{minipage}\hfill
\begin{minipage}[t]{.5\linewidth}
\includegraphics[height=3cm,width=4cm]{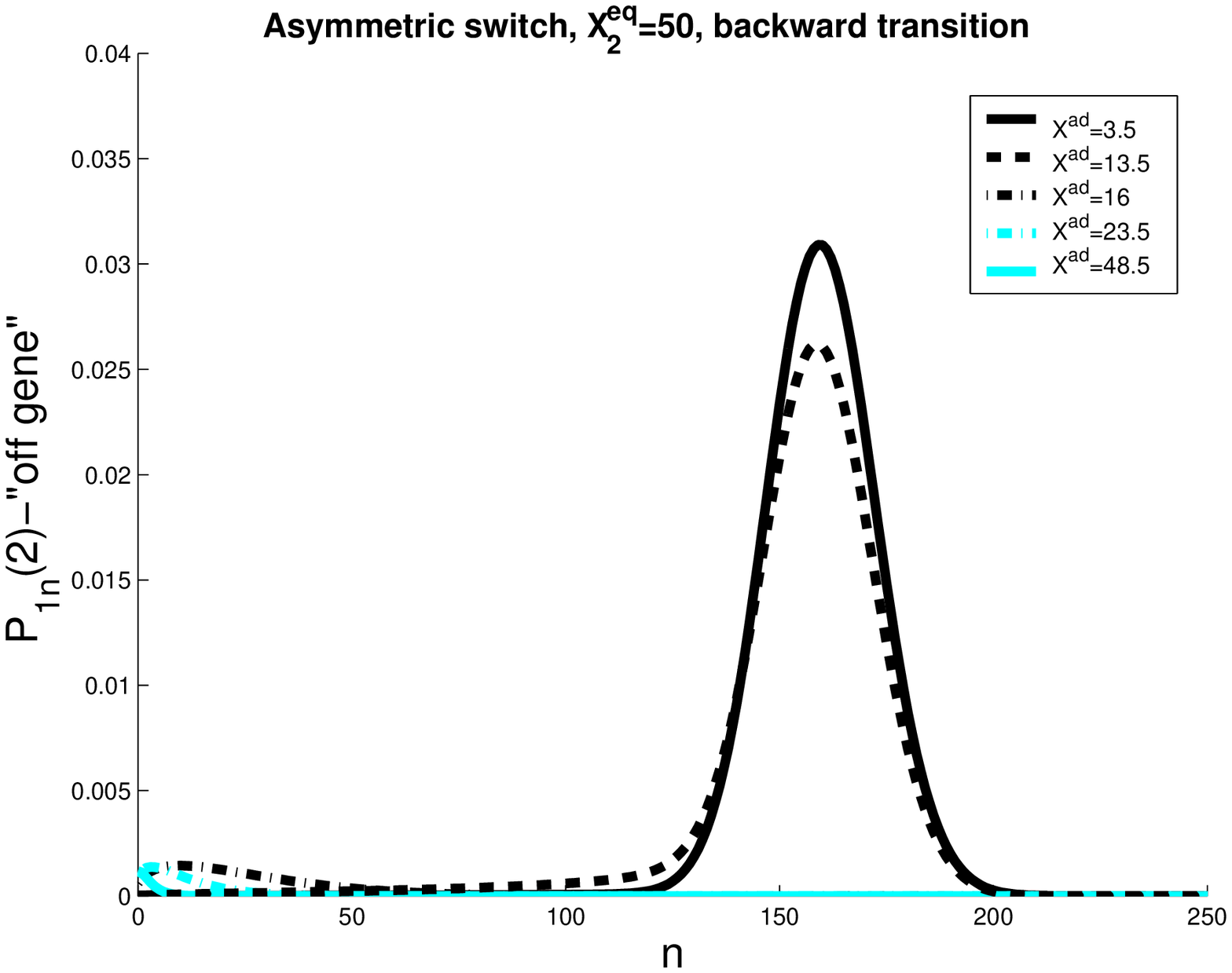}
\end{minipage}
\caption{Evolution of the probability distribution for a backward transition as a function of $X_1^{ad}=\delta X_1^{sw}$ for $X_2^{eq}=50$ with $X^{eq}_1 = 1000$; $\omega_1=\omega_2=0.5$; $X_2^{ad}= \delta X_2^{sw}=80$ for an asymmetric switch.}
\label{asymprback}
\end{figure}
The general picture behind the transition is seen from the deterministic approach. The larger the tendency for proteins to be unbound from the DNA, the larger the effective production rate $X_1^{ad}$ must be for the transition from one gene to be active to the other to be active to take place, since repressor proteins are less likely to bind to the on gene ($i$) at large $X^{eq}_i$ than at small  $X^{eq}_i$. However, if one considers a noisy system, it is effectively harder for proteins to stay bound to the initially off gene due to the destabilizing effect of DNA binding noise (Figs \ref{xeqasyma}, \ref{xeqasyma1}). For the stochastic system, apart from very low values of the adiabaticity parameter ($\omega<0.1$) (Fig. \ref{omdepasym}), there is a threshold number of reservoir proteins which will cause a rapid transition. If we start with a small effective production rate for one type of proteins and increase this rate, keeping the production rate of the other gene fixed at an initially higher value, the proteins produced by the gene with the initially smaller production rate, repress it gradually and ineffectively, until they reduce the probability of the gene to be on to one half, for the exact SCPF solution. The number of proteins present in the on state decreases much more rapidly with the change of $X^{ad}_1$, whether it be increase for the forward transition or decrease for the backwards in the examples presented, than the number of proteins in the off state grows (Fig. \ref{xeqprotein}). Hence the probability to be on of the initially active gene shows a larger sensitivity to the change of $X^{ad}_1$ than the off state probability. This leads to a rapid transition of the previously active gene to an inactive state (Figs \ref{asympr}, \ref{asymprback}). Such behavior is described by Ptashne (Ptashne, 1992), (Ptashne and Gann, 2002) in the $\lambda$ phage switch, who points out its role as a ``buffer against ordinary fluctuations in repressor concentration''. The observed system switches when the ``repression probability'' drops to $50\%$, as in the solutions of this model. Our analysis seconds Ptashne's hypothesis, since the deterministic system lacks this behavior, the transition is rapid and for certain values of parameters takes place when the probability of the initially on gene drops to $~80\%$ (Fig. \ref{asymbif}). The buffering capabilities of the stochastic system are clearly seen in the long tails towards $n=0$ of the probability distibutions of the gene that is switching from the on to the off state (Fig. \ref{asympr}).\\ 
\textbf{The effect of noise on the bifurcation mechanism}\\
The mean number of proteins at the transition point differs for the deterministic and exact SCPF solution (Fig. \ref{xeqprotein}). More repressors are needed to induce the transition in the deterministic approximation than in the stochastic system, since due to the form of the interaction function for the exact case, $F(i)=<n(i)>^2\frac{\omega+1}{\omega+C_1(i)}+<n(i)> > <n{i}>^2$. A smaller number of proteins is therefore needed for the inactive gene to become competitive with the active gene. The mechanism of the transition is different from the symmetric gene case, where a critical number of proteins needs to be reached. The asymmetric switch is based on the competition between the probability that proteins of one kind will repress the opposing genes and the analogous probability for the other kind of proteins. The repression capability is governed by $\frac{X^{ad 2}_{3-i}}{X^{eq}_i}$, which might be looked upon as the product of the probability of having a certain number of repressor proteins ($3-i$) in the cell and the tendency for them to be bound to the opposing gene ($i$). In fact, the transition point in the deterministic case is purely a function of such ratios, $\frac{X^{ad 2}_{3-i}}{X^{eq}_i}=f(\frac{X^{ad 2}_{i}}{X^{eq}_{3-i}})$. In both the stochastic and deterministic cases, the transition points are set by the interaction function which regulates the on and off state probabilities of a given gene $\frac{F(3-i)}{X^{eq}_i}=\frac{C_2(i)}{C_1(i)}$. Inclusion of noise in the system effectively increases the nonlinearity of the system, which results in the already discussed buffering capabilities of the system. Stochasticity alters the very simple competitive mechanism seen in the deterministic kinetics to allow for more levels of control of the stability of the state of the system against random fluctuations.\\
Further comparison of solutions of the deterministic and stochastic equations leads to the same conclusions as for a symmetric switch. As the tendency for proteins to be unbound from the DNA grows, the  difference in the critical number of reservoir proteins necessary for the transition to take place increases for both approximations. The critical number of proteins produced by a given gene necessary for the transition to take place for both genes is, in most cases (see $\omega$ dependence discussion), smaller for the exact solutions of the SCPF equations and the difference between the stochastic and deterministic result grows with both $X^{eq}_i$ and decreases with $\omega_i$ (Fig. \ref{asymbif}). It has a value of $15$ for $X^{eq}_2=500$, $\omega_1=\omega_2=0.5$ and $2$ for $X^{eq}_2=500$, $\omega_1=\omega_2=10$. \\
Consider the forward transition. The initially inactive gene is buffered by a cloud of repressor proteins. As one increases the effective production rate of the proteins produced by the inactive gene ($X^{ad}_1$), the number of proteins which are able to repress gene 2 grows slowly and linearly $<n(i)>=2 X^{ad}_1 C_1(1)$, where $C_1(1) \sim const$ and forms a buffering proteomic cloud around it. In the results presented in the figures of this paper the tendency that proteins are unbound from gene 2, ($X^{eq}_2$), is smaller than $X^{eq}_1$, so gene 1 is able to produce enough repressors to form a stable buffering cloud around gene 2 and turn it into the inactive state at quite modest values of $X^{ad}_1$. If $X^{eq}_1<X^{eq}_2$, gene 1 produces proteins less effectively, as the probability of it being repressed is larger than in the previous case, and larger values of $X^{ad}_1 $ are needed to produce enough repressors to achieve a high effective probability of binding, $\frac{X^{ad 2}_{1}}{X^{eq}_2}$. An example of how $X^{ad, crit}_1$ grows as $X^{eq}_1 \rightarrow X^{eq}_2$, is seen by comparing the $X^{ad}_1 \sim 33$ for $X^{eq}_1=1000$, $X^{eq}_2=50$ in Fig. \ref{asymbif} and $X^{ad}_1 \sim 300$ for $X^{eq}_1=100$, $X^{eq}_2=50$ (Fig. \ref{omdepasym}). \\
\textbf{Adiabaticity parameter dependence}\\
The interaction of the buffering proteomic cloud with the DNA can be altered when the rate of the DNA unbinding rate compared to the protein degradation rate is changed. For small $\omega_i$ values the unbinding rate of repressors to the DNA is slower than the destruction of the produced proteins. Apart from very small $\omega$ values, as long as there is a critical number of repressor proteins in the buffering cloud, the off gene is repressed and it responds by turning on, only once the initially on gene is nearly totally repressed. Large adiabaticity parameters result in the efficient formation of the buffering proteomic cloud. For the initially off gene, a small DNA unbinding rate of the off gene, decreases the effectiveness of the buffering proteomic cloud around it, as the protein number state can reach a steady state before the DNA state does. The hindered DNA reaction to the protein number state effectively increases the tendency of repressor proteins to be unbound from the DNA, for a given $X^{ad}_1$. This in turn decreases the probability of the initially on gene to be on, leading to rapid, switching behavior as an be seen for gene 2 in the forward, or gene 1 in the backward transition for $\omega>0.1$ in Fig. \ref{omdepasym} (a). The initially on gene reacts to the interaction function of the initially off gene, for which $F(i) \rightarrow <n(i)>^2\frac{1}{C_1(i)}+<n(i)>$ in the small $\omega$ limit. Therefore the interaction function is effectively increased for $C_1(i) \approx 0$, leading to the enhanced buffering. The reaction of the initially off gene is unaltered, as for $C_1(i) \approx 1$ $F(i)=<n(i)>^2+<n(i)> \sim const$, if $C_1(i)$ remains close to $1$. However if $\omega$ is very small (black dash-dot curve in Fig. \ref{omdepasym} (a)), the buffering proteomic cloud is not given a chance to form due to a very high degradation rate of proteins and gene 2 is simply repressed in a gradual transition. If $\omega_1$ is extremely small and $\omega_2$ large, the buffering proteomic cloud around gene 1 cannot form and the probability of it to be off in the forward transition decreases gradually. A buffering proteomic cloud exists around gene 2, hence the backward transition is reminiscent of the deterministic result (Fig. \ref{omdepasym} (b)).  The most interesting case is shown in Fig.  \ref{omdepasym} (c), where a large $\omega_1$ acts as a buffer against fluctuations in the number of proteins, which repress gene 1. For large production rates of repressors the probability of gene 2 to be on for the forward transition decreases faster than in the deterministic solution, however the buffering cloud repressing gene 1 allows gene 2 to remain in the on state. A buffering proteomic cloud does not form around gene 2 and it remains on until the number of proteins produced by gene 1 grows considerably, as the effective production rate, $X^{ad}_1$, is increased. The effective production rate of gene 1 must be very large to sustain a sufficient steady state number of proteins to repress gene 2 to the point that $C_1(1)<0.5$, which leads to switching. For the backward transition the lack of a buffering proteomic cloud around gene 2 results in destabilizing gene 1 for larger $X^{ad}_1$ effective production rates than for large $\omega_2$ values. These examples show how certain combinations of values of adiabaticity parameters can lead to a system with a larger switching region than the deterministic model predicts.
This property may be useful when engineering artificial switches. If one has a constraint on the production rates of the genes, one can use repressors with different binding affinities to achieve switching in the desired region of parameter space.\\
In this simple system slow unbinding from the DNA can compensate for the destabilizing of the DNA state by protein number fluctuations. As the probability of the initially active gene to be on gradually decreases, the initially repressed gene becomes active only once the probability of the other gene to be on has fallen bellow a certain values $\alpha$. The susceptibility of the system to protein number fluctuations may be estimated by the value of $\alpha$. For small $\omega$, which is still able to sustain a buffering proteomic cloud, this values tends to $0.5$. The incapability of the system to form a buffering proteomic cloud is much stronger if both adiabaticity parameters are small, since the reaction of both genes to the change in the number of proteins is hindered (Fig. \ref{omdepasym} (a)). DNA state fluctuations contribute to effectively faster protein number fluctuations, therefore the exact solution exhibits the very small $\omega$ characteristics, where a buffering proteomic cloud cannot form, for a slightly wider range of the adiabaticity parameter than one would expect with a Poissonian distribution (results not shown). Combining these observations a switch works most effectively if the change of the DNA state compared to the protein number fluctuations of one gene is sufficiently smaller than that of the other gene, to allow for effective buffering. \\ 
 \begin{figure}
\begin{minipage}[t]{.43\linewidth}
\includegraphics[height=3cm,width=4cm]{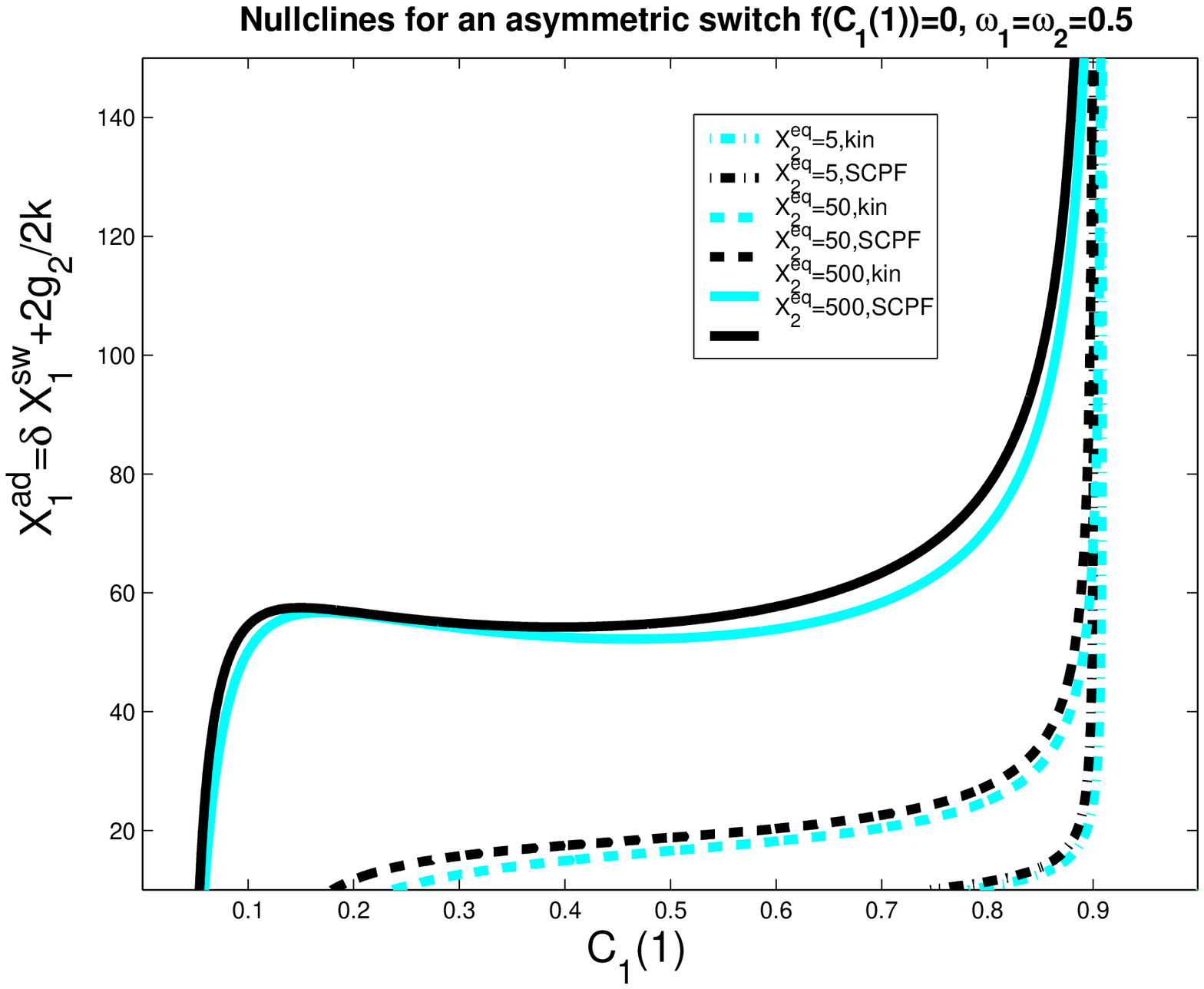}
\end{minipage}\hfill
\begin{minipage}[t]{.5\linewidth}
\includegraphics[height=3cm,width=4cm]{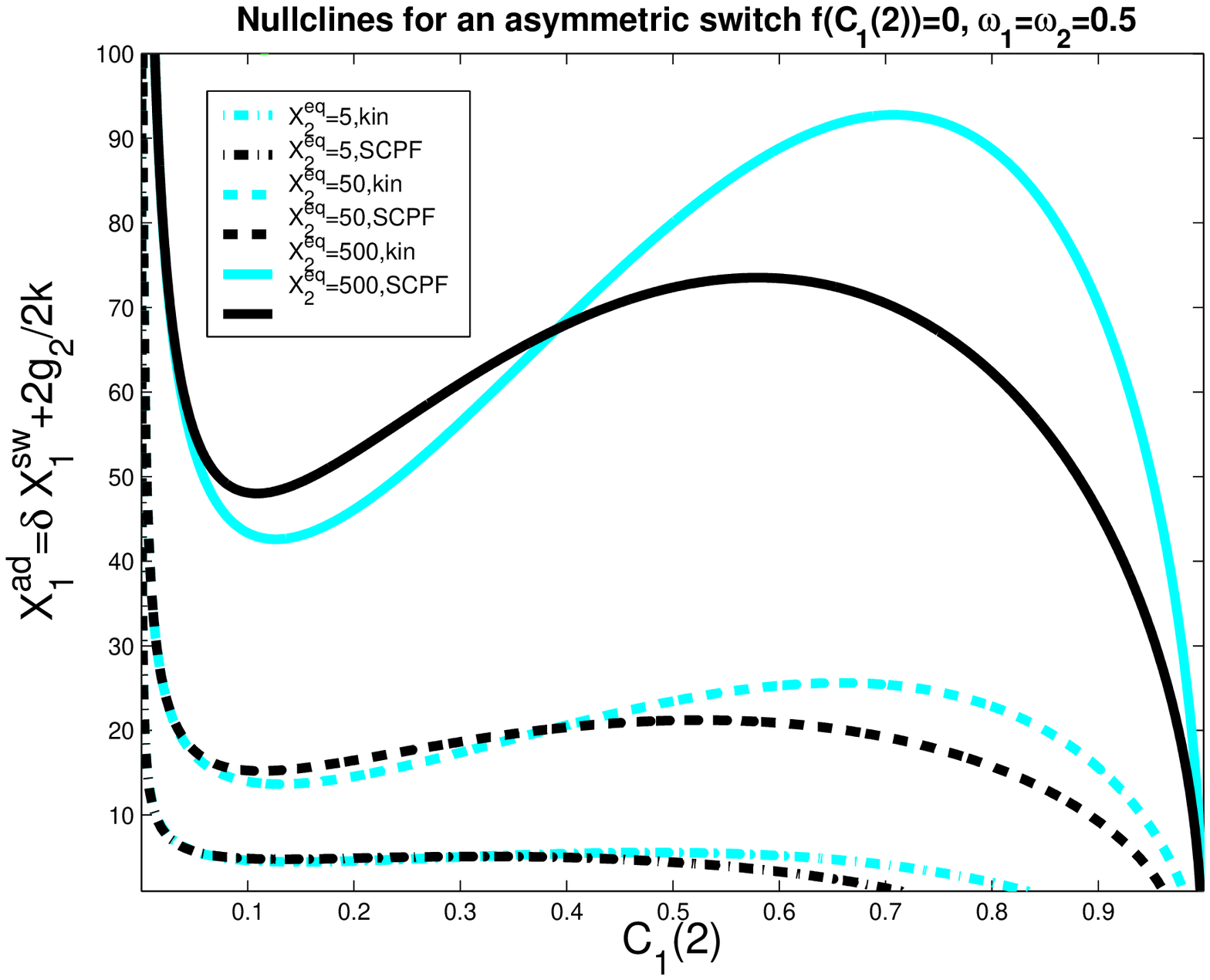}
\end{minipage}\hfill
\caption{Bifurcarion diagrams as a function of $X_1^{ad}=\delta X_1^{sw}+2 \frac{g_2}{2k}$, with $\frac{g_2(1)}{2k}=\frac{g_2(1)}{2k}=5$ (a) and $\frac{g_2(1)}{2k}=\frac{g_2(1)}{2k}=0.5$ (b) for $X_2^{eq}=5,50,500$ for exact solution of the SCPF and deterministic kinetic equations for an asymmetric switch. Nullclines for $C_1(1)$ (a) and $C_1(2)$ (b) with $\omega_1=\omega_2=0.5$ with $X^{eq}_1 = 1000$; $X_2^{ad}= \delta X_2^{sw}=80$.}
\label{asymbifback}
\end{figure}
 \textbf{The nonzero basal production rate}\\
The asymmetric switch in which both genes have a nonzero basal effective production rate proves to be susceptible to noise. In Fig. \ref{asymbifback}, we show the dependence of $C_1(1)$, with $\frac{g_2(1)}{2k}=\frac{g_2(2)}{2k}=5$ and $C_1(2)$, with $\frac{g_2(1)}{2k}=\frac{g_2(2)}{2k}=0.5$ in the small $\omega_i$ limit. The stochastic solutions converge to the deterministic solutions for large $\omega$. If gene 2 is initially in the on state, the majority of proteins are produced with the high fixed rate in the on state, as $g_1(2)>>g_2(2)$. The repression of gene 2 is in turn governed by the interaction function of gene 1. If $X^{ad}_1$ is small the number of proteins produced in the on and off states by gene 1 are comparable. As the number of proteins produced by gene 1 grows faster the larger $g_2$ is, gene 2 gets repressed more effectively for smaller $X^{ad}_1$ values. This results in a smaller number of repressors produced by gene 2 and the transition from gene 1 to be on to be off takes place for smaller $X^{ad}_1$ - effective growth rate values, than for small $g_2$.\\
The deterministic solution is much more influenced by the production of proteins in the off state than the stochastic solution. In the exact SCPF solution slow DNA unbinding rates compared to protein degradation rates are another means of control of the stability of the DNA state against random protein number fluctuations. The state of the system is far less influenced by the exact protein numbers than in the deterministic solution. So until the probability of a gene to be on is larger than that to be off, the fraction of proteins produced with a smaller effective production rate in the off state is treated as a random fluctuation by the system. Once again the SCPF system demonstrates its susceptibility to protein number fluctuations. \\
The influence of the off state protein production on the total repressor yield may also be seen in the fast decrease of $C_1(2)$ and increase of $C_1(1)$ in the forward transition. If $g_2$ is considerably large its effect can also be seen in the stochastic solution, hence even when gene 1 is in the on state, it never reaches $C_1(1)=1$, although gene 2 is totally repressed (Fig. \ref{asymbifback} a and results not shown for gene 2). The magnitude of the probability of gene 1 to be on for very large effective production parameters strongly depends on the the tendencies of the proteins to be unbound from gene 1. As $X^{eq}_1$ increases the asymptotic $X^{ad}_1$ limit of $C_1(1)$ becomes smaller, as it is effectively harder for repressors to stay bound to the DNA.  The gene is more likely to be in the off state, which however manages to sustain the necessary number of proteins produced by gene 1 to repress gene 2. As $g_2$ increases the region of bistability grows into areas of parameter space, in which the tendency of proteins to be unbound, $X^{eq}_2$, is larger than for small $g_2$. For small values of  $X^{eq}_2$ the number of repressors produced by gene 1 in the off state is sufficient to repress gene 2 and one observes a smooth and slow transition in terms of $X^{ad}_1$. If $g_2$ is considerably large the transition takes place for larger values of $X^{ad}_1$ in the stochastic solution than in the deterministic solution, hence showing the large buffering region the interplay of DNA and protein number fluctuations provides. This also results in an effective similarity of the deterministic and stochastic solution. In regions of parameter space, in which the change of DNA state is rapid, the deterministic and stochastic solutions differ, apart from the large $\omega$ limit. Most experimentally observed proteins have very small basal production rates, which seconds our analysis, that it is functionally unfavourable for large basal production to occur. The dependence on other parameters is analogous to the case without a basal production rate.\\
\begin{figure}
\begin{minipage}{.20\linewidth}
\includegraphics[height=3.0cm,width=3.5cm]{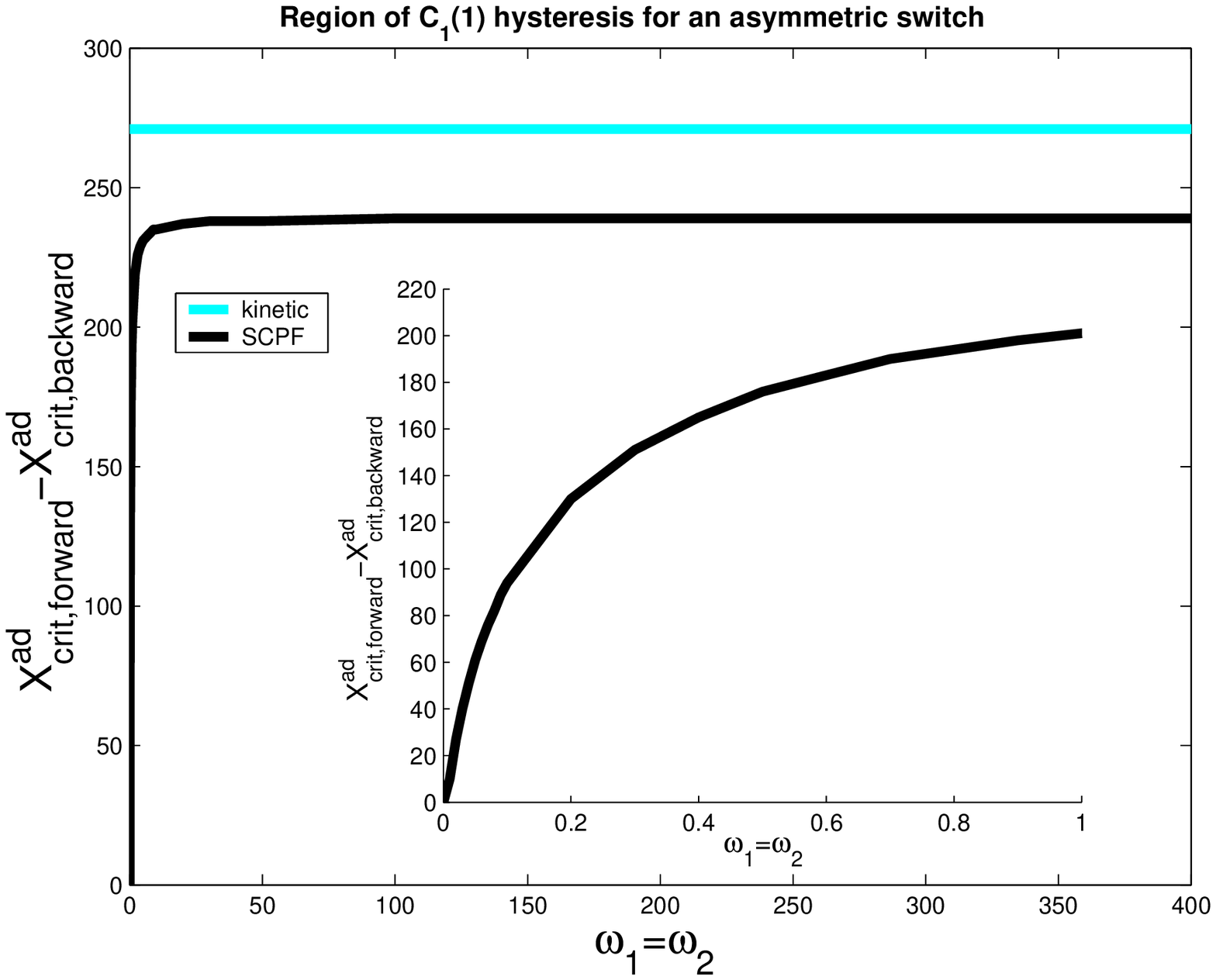}
\end{minipage}\hfill
\begin{minipage}{.55\linewidth}
\includegraphics[height=3.0cm,width=3.5cm]{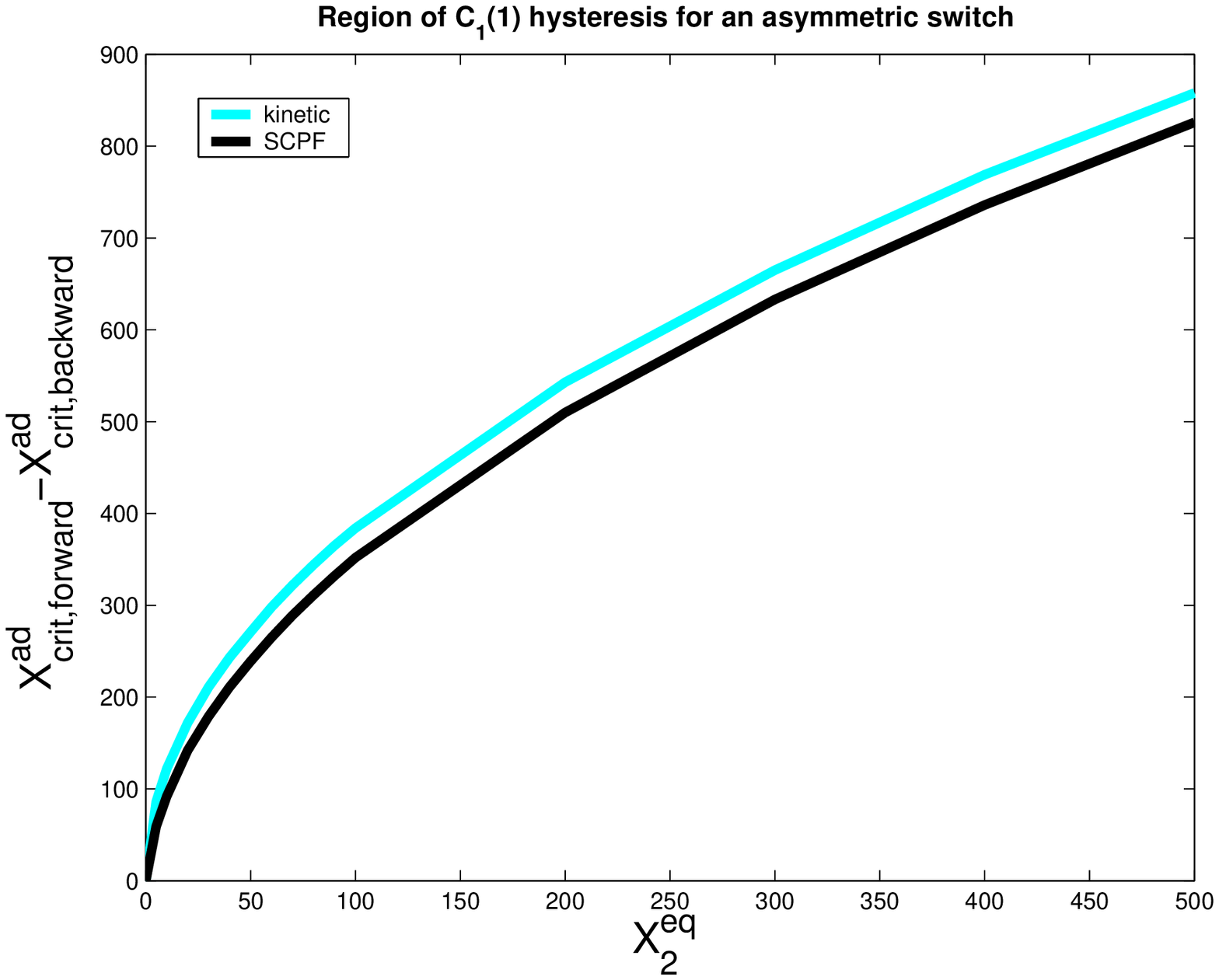}
\end{minipage}
\end{figure}
\begin{figure}
\begin{minipage}{.20\linewidth}
\includegraphics[height=3.0cm,width=3.5cm]{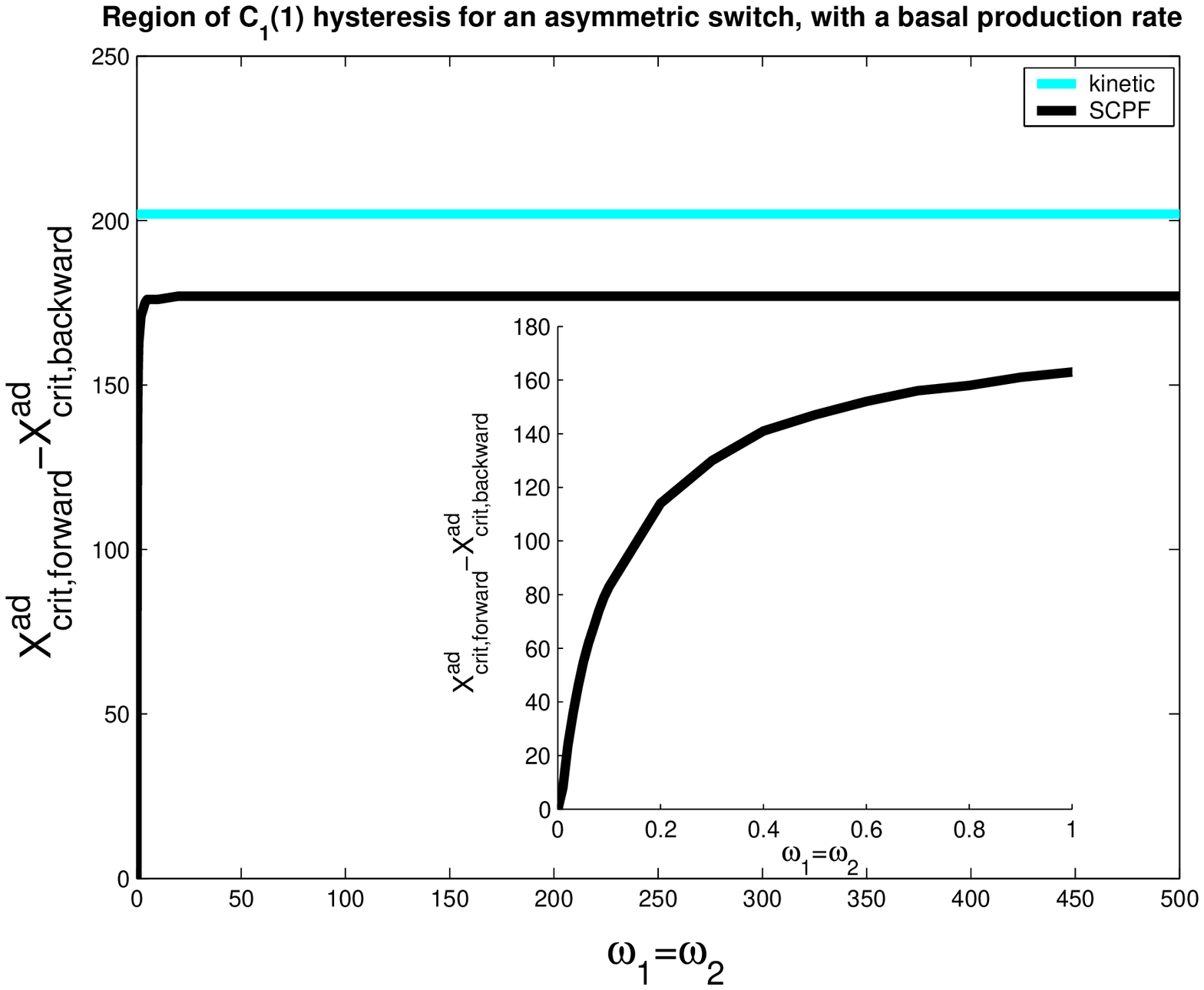}
\end{minipage}\hfill
\begin{minipage}{.55\linewidth}
\includegraphics[height=3.0cm,width=3.5cm]{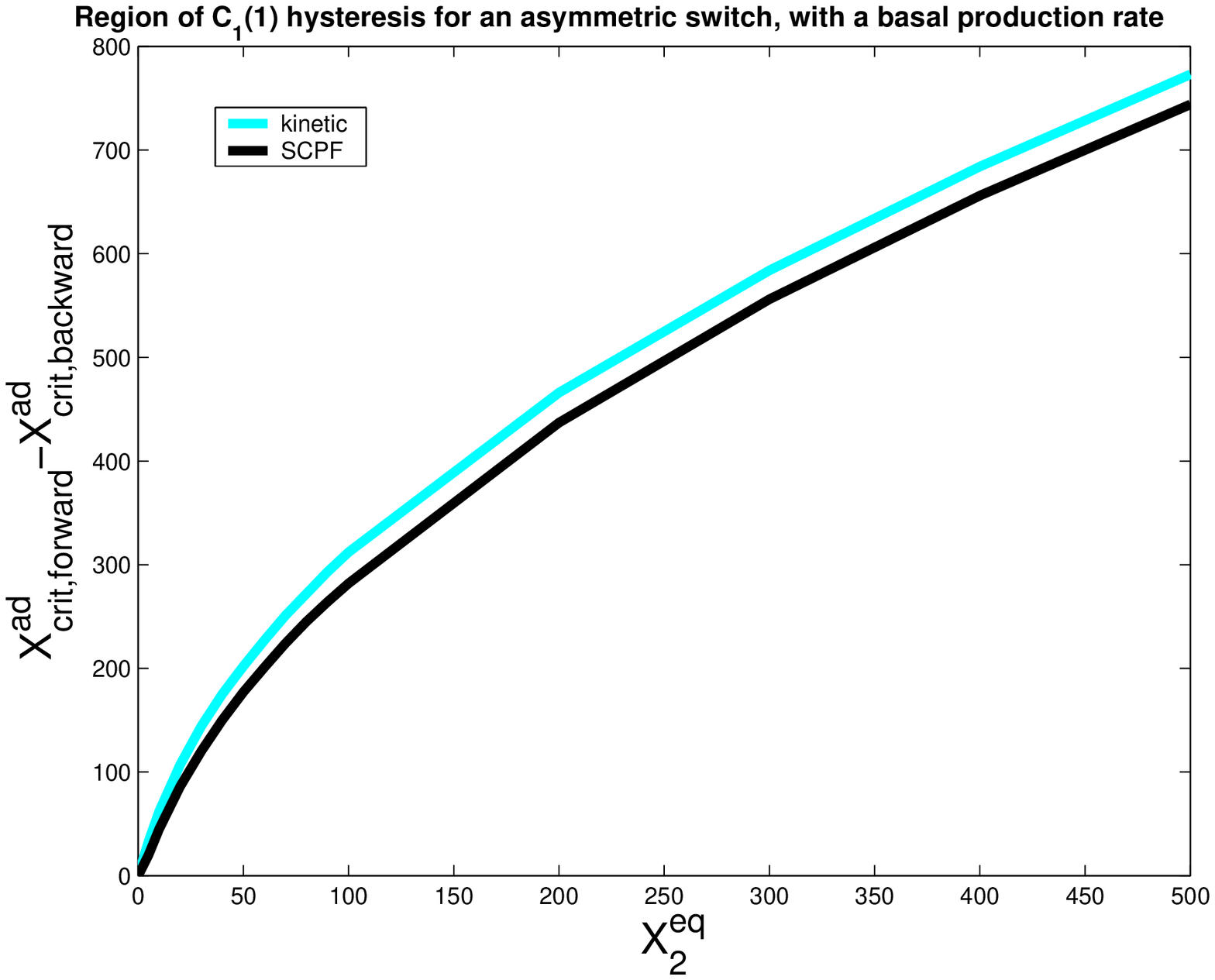}
\end{minipage}
\caption{Region of $C_1(1)$ hysteresis for an asymmetric switch for the SCPF and deterministic  approxiamtions as a function of $\omega_1=\omega_2$, with $X_2^{eq}=50$ (a) and $X_2^{eq}$ with $\omega_1=\omega_2=100$ (b). $X_1^{eq}=100$, $X_2^{ad}=\delta X_2^{sw}=80$, $X_1^{ad}=\delta X_1^{sw}$. The same comparison for a switch with a basal production level $\frac{g_2}{2k}=0.5$ as a function of $\omega_1=\omega_2$ (c) and $X_2^{eq}$ (d). $X_1^{eq}=100$, $X_2^{ad}=\delta X_2^{sw}=80$.}
\label{asymhisc11}
\end{figure}
\textbf{The region of bistability}\\
The backward transition, as already discussed, is analogous to the forward transition. In most cases, the regions of bistability (Fig. \ref{asymbif}) in parameter space are reduced in size by noise. When engineering artificial switches, one may be interested in making sure the forward and backward transition takes place for considerably different production rates. We therefore consider how the region of bistability, defined as the difference in the critical effective production rate for the forward and backward transition, depends on the parameters of the model. For the deterministic case the region of bistability depends on the tendencies that proteins are unbound from the DNA in a quadratic manner, as can easily be seen from the bifurcation equations \ref{bikin1}, \ref{bikin2} and is demonstrated in Fig. \ref{asymhisc11}. The SCPF solution shows the same behavior. For large values of the adiabaticity parameter the size of the region of bistability is independent of $\omega$, as is the form of the bifurcation curve (Fig. \ref{asymhisc11}). The approach to this plateau is very rapid and is given by the ratio of polynomials. However, the size of the region of bistability for the $\omega_1=\omega_2$ never reaches that of the deterministic solution, as even in the large $\omega$ limit the greater nonlinearity of the interaction function $F(i)$ results in a more complex SCPF curve which does not reduce to deterministic  solution, but $X^{ad}_1(C_1(2))\rightarrow \frac{1}{2}((((\frac{1}{C_1(1)}-1)X^{eq}_2)^{\frac{1}{2}}+1)^{\frac{1}{2}}-1)\frac{1}{2 C_1(1)} \neq \frac{{X^{eq}_2}^{\frac{1}{2}}}{2} (1+\frac{( 2 X^{ad}_2 C_1(2))^2}{X^{eq}_1})(\frac{1}{C_1(2)}-1)^{\frac{1}{2}}$. This effect is true for both curves, as the presented graphs show $C_1(1)$ hysteresis and the chosen equations $C_1(2)$. The same behavior is observed for the case with a nonzero basal production rate. The increase with $X^{eq}_2$ is slightly slower in the $g_2 \neq 0$ case as the bifurcation curve is smaller by $\abs{\frac{2 g_2}{2 k} (C_{1 f}(i)-C_{1 in}(i))-\frac{1}{2} ln \frac{C_{2 f}(i)}{C_{1 in}(i)}}$.\\
\textbf{Summary}\\
After the transition, the number of proteins produced by the now on gene, follows a linear dependence on $X^{ad}$, similarly to the symmetric switch. The number of proteins in the cell is independent of the DNA dynamical characteristics, as those remain constant in that region of parameter space. The number of proteins of the off gene, rapidly falls before the transition takes place. Based on the bifurcation diagram of Fig. \ref{asymbif} the phase transition is discontinuous, for a certain region of the parameter space, where switching may occur. That region may be roughly estimated by the parameters of the genes which must be competitive, $(\frac{X_1^{ad}}{X_2^{ad}})^2 \approx \frac{X_2^{eq}}{X_1^{eq}}$. This has a major implication for biological systems, such as the $\lambda$ phage, where many mechanisms are used to achieve balance between two genes. The first order phase transition, as opposed to the second order present in the symmetric system, is a results of the breaking of symmetry and is clearly seen in the evolution of probability distributions in phase space (Fig. \ref{asympr}). The gene that is on after the transition rapidly increases its probability of being on, whereas the off gene decreases with a rapid drop in the number of proteins it produces.\\ 
\subsection*{The Case when Proteins bind as Monomers}
\begin{figure}
\begin{minipage}[t]{.43\linewidth}
\includegraphics[height=3cm,width=4cm]{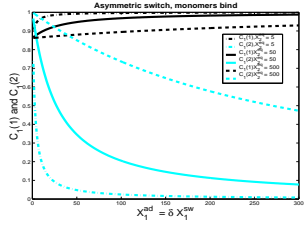}
\end{minipage}\hfill
\caption{Probability of genes in an asymmetric switch to be active when proteins
 bind as monomers, for different values of $X_2^{eq}$. $X_1^{eq}=1000$,$X_2^{ad}
=\delta X_2^{sw}=80$.}
\label{p1asym}
\end{figure}

The equations presented above can easily be augmented to describe the binding of monomers or higher order oligomers by changing the form of the binding term to $h_i n^p_{3-i} $, where $p=1$ for monomers. The equations remain solvable for any value of $p$.\\
\textbf{Monomers do not make good repressors/activators}\\
The behavior of the system is quite different if we consider the case when proteins bind as monomers. For a symmetric switch there is no region of the parameter space, in which one observes switching. The SCPF equations may be reduced to a single quadratic equation:\\
\begin{equation}
2 \delta X^{sw}{C_1(i)}^2+(X^{eq}+X^{ad}-\delta X^{sw})C_1(i)-X^{eq}=0
\label{p1}
\end{equation}
which has at most only one positive solution. Therefore the probability of one gene to be in the active state is always equal to that of the other to be in the active state and no switching is observed. The equation (\ref{p1}) is independent of $\omega$, the adiabaticity parameter, therefore it is solely a consequence of the lack of nonlinearity in the binding of proteins and cannot be influenced by very slow DNA unbinding rates. By writing down deterministic  equations we can also show that when proteins bind as monomers switching does not occur. A similar equation to (\ref{p1}), also independent of $\omega$, holds for asymmetric switches. It also has one positive solution, therefore the parameters of the model predetermine the solution and each gene has a probability to be on determined by its kinetic rates. Since the rates are different for the two genes, the gene with the larger production rate will be in the active state, repressing the weaker gene (Fig. \ref{p1asym}).
In naturally occurring biological switches and those developed experimentally proteins bind as dimers, or higher order multimers (Ptashne, 1992).  We see cooperativity contributes to improving the efficiency of a switch. A switch controlled by monomers is shown to react ineffectively to changes in the repressor concentration, just as in the case of the asymmetric switch in our model discussed above. Monomers do not have the ability to stabilize a broken symmetry state, therefore the solution is fragile to kinetic rates and inefficient. Effectively monomers do not make good repressors/activators. Ptashne and Gann (Ptashne and Gann, 2002) explain the cooperativity process between two monomers by claiming that one monomer bound to the DNA increases the ``local concentration'' of proteins around the binding site through weak protein-protein interaction, thus causing the second to bind cooperatively. Our model lacks spatial dependence, therefore shows this effect need not be thought of as due to changes in local concentration, but actually is required by the insufficient nonlinearity for monomers, which cannot produce bistability.\\
\begin{figure}
\begin{minipage}[t]{.43\linewidth}
\includegraphics[height=3cm,width=4cm]{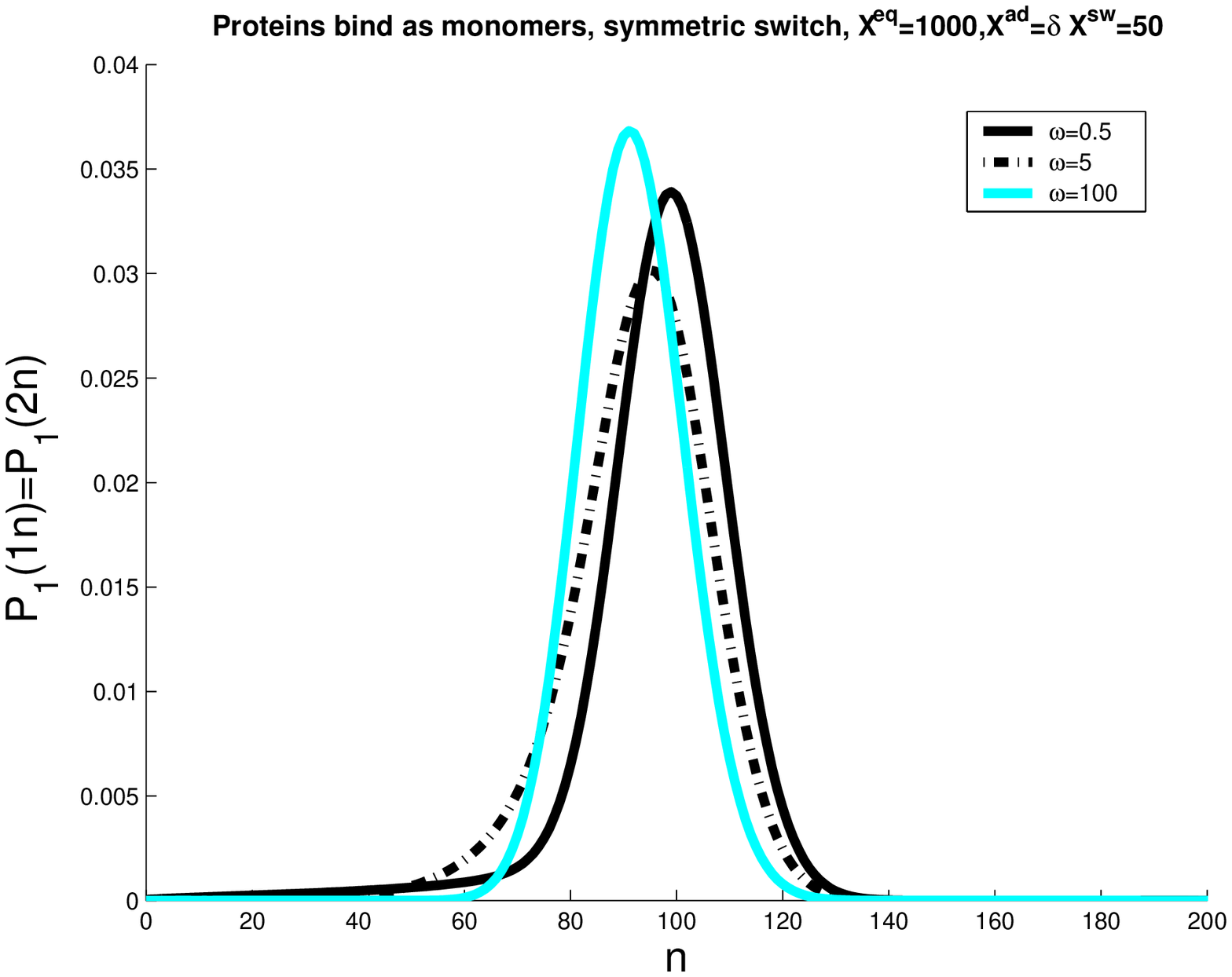}
\end{minipage}\hfill
\begin{minipage}[t]{.50\linewidth}
\includegraphics[height=3cm,width=4cm]{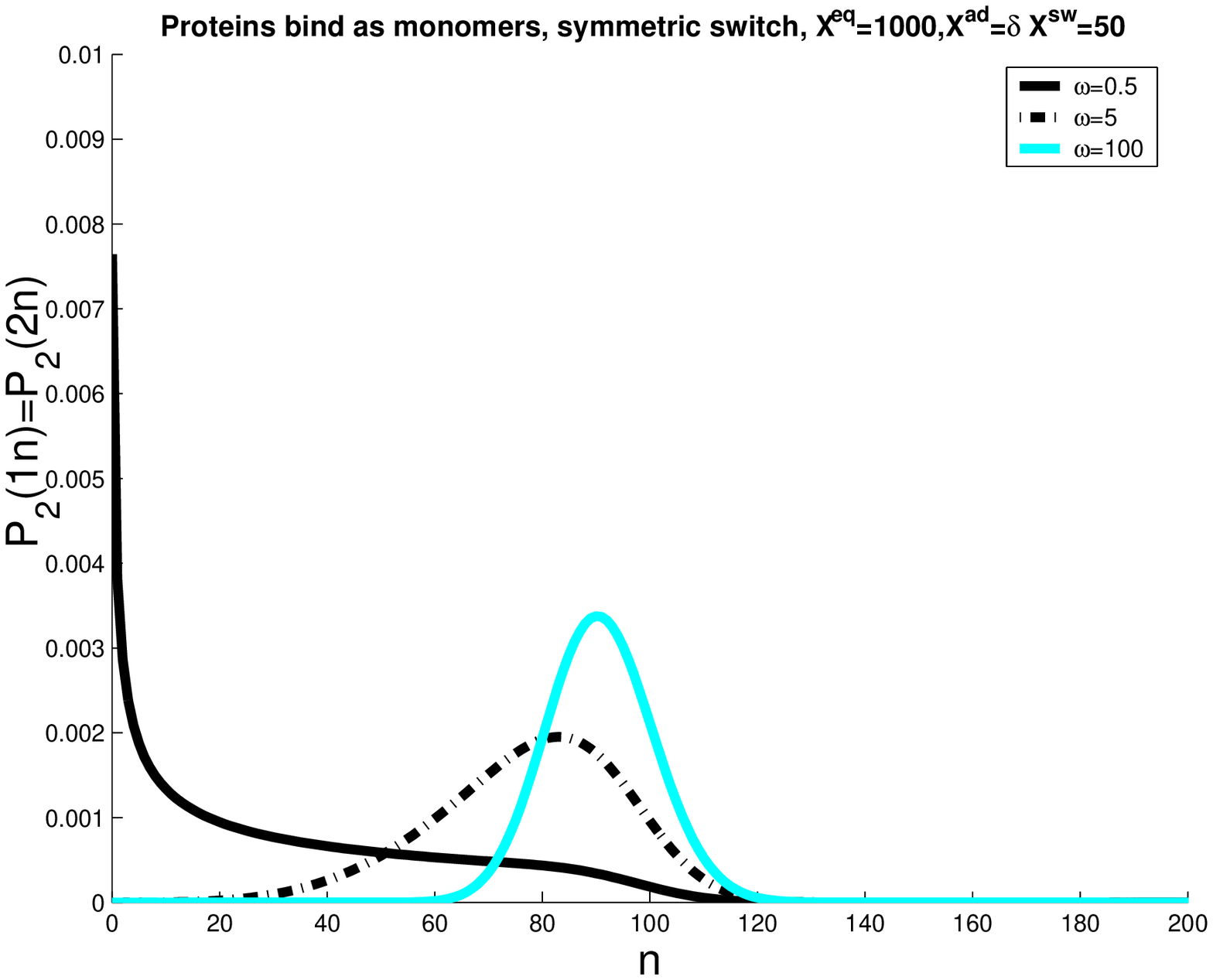}
\end{minipage}
\caption{Probability distributions for the gene to be in the on state (a) and off state (b) for a gene in the active state for different values of the adiabaticity parameter $\omega=0.5,5,100$. $X^{eq}=1000$, $X^{ad}=\delta X^{sw}=50$, when proteins bind as monomers to a symmetric switch.}
\label{evmon}
\end{figure}
\textbf{Bimodal probability distribution}\\
Although the probabilities of the two genes to be on are equal for the whole region of parameter space and the mean number of both types of proteins in the cell is the same as in the deterministic  case, the probability distributions are bimodal when the DNA unbinding rates are slower than the protein number fluctuations. The mechanism of this small $\omega$ behavior has already been discussed on the example of the symmetric switch when proteins bind as dimers. This is analogous to the case when DNA fluctuations induce a probability distribution with two peaks for the single gene with an external inducer (Cook et al., 1998). In fact the SCPF approximation has reduced this two gene system to an effective one gene system with an external inducer. A bimodal distribution in the small $\omega$ case is also observed for the asymmetric switch, when proteins bind as monomers.\\
\subsection*{The Case when Proteins bind as Higher Order Oligomers}
\begin{figure}
\begin{minipage}[t]{.43\linewidth}
\includegraphics[height=3cm,width=4cm]{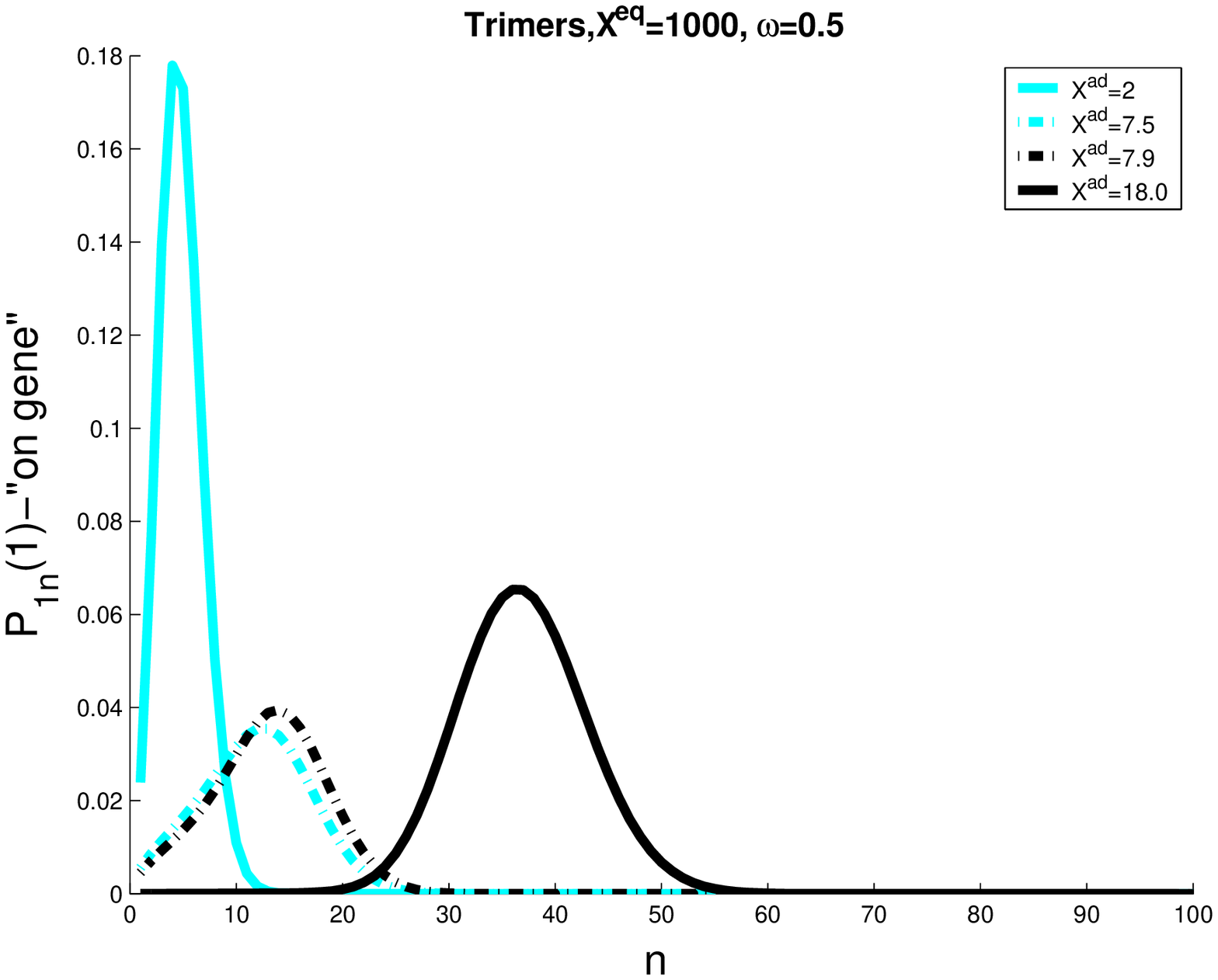}
\end{minipage}\hfill
\begin{minipage}[t]{.5\linewidth}
\includegraphics[height=3cm,width=4cm]{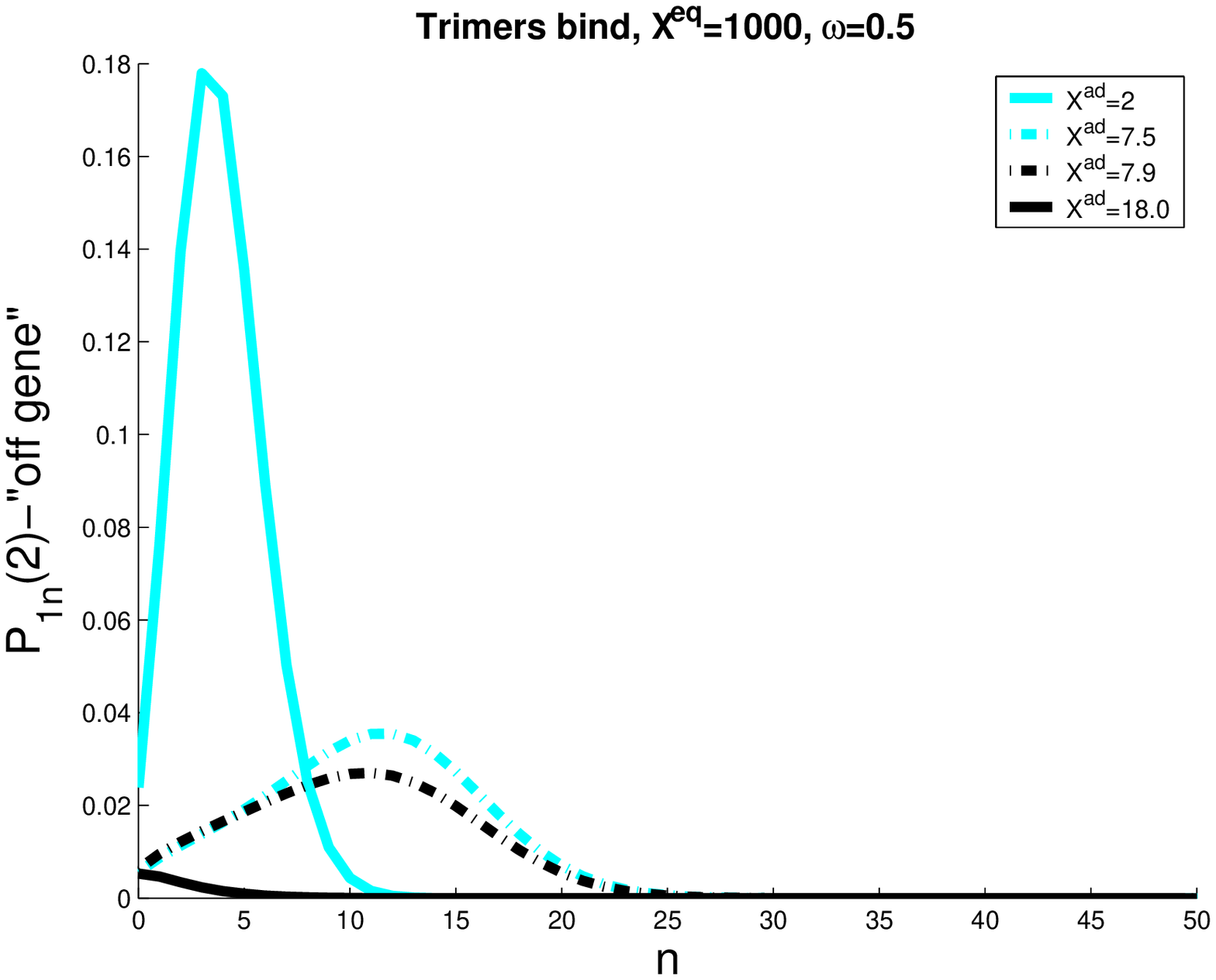}
\end{minipage}
\caption{The evolution of the probability distribution for the probability of the gene that will be active (a) and inactive (b) after the bifurcation to be on as function of $X^{ad}$ for a switch when proteins bind as trimers, $X^{eq}=1000$,$\omega=0.5$.}
\label{hoprotri}
\end{figure}

\begin{figure}
\begin{minipage}[t]{.43\linewidth}
\includegraphics[height=3cm,width=4cm]{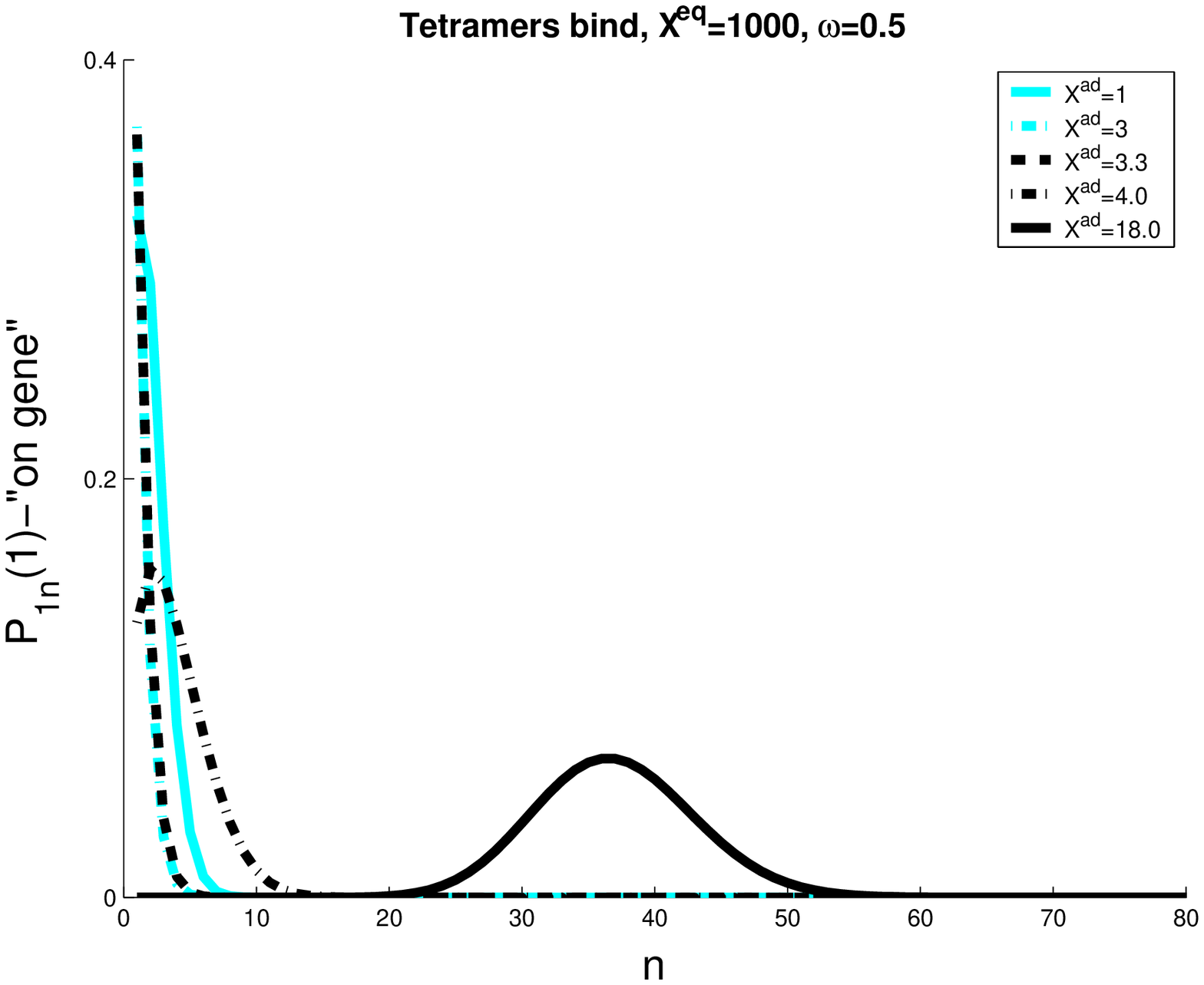}
\end{minipage}\hfill
\begin{minipage}[t]{.5\linewidth}
\includegraphics[height=3cm,width=4cm]{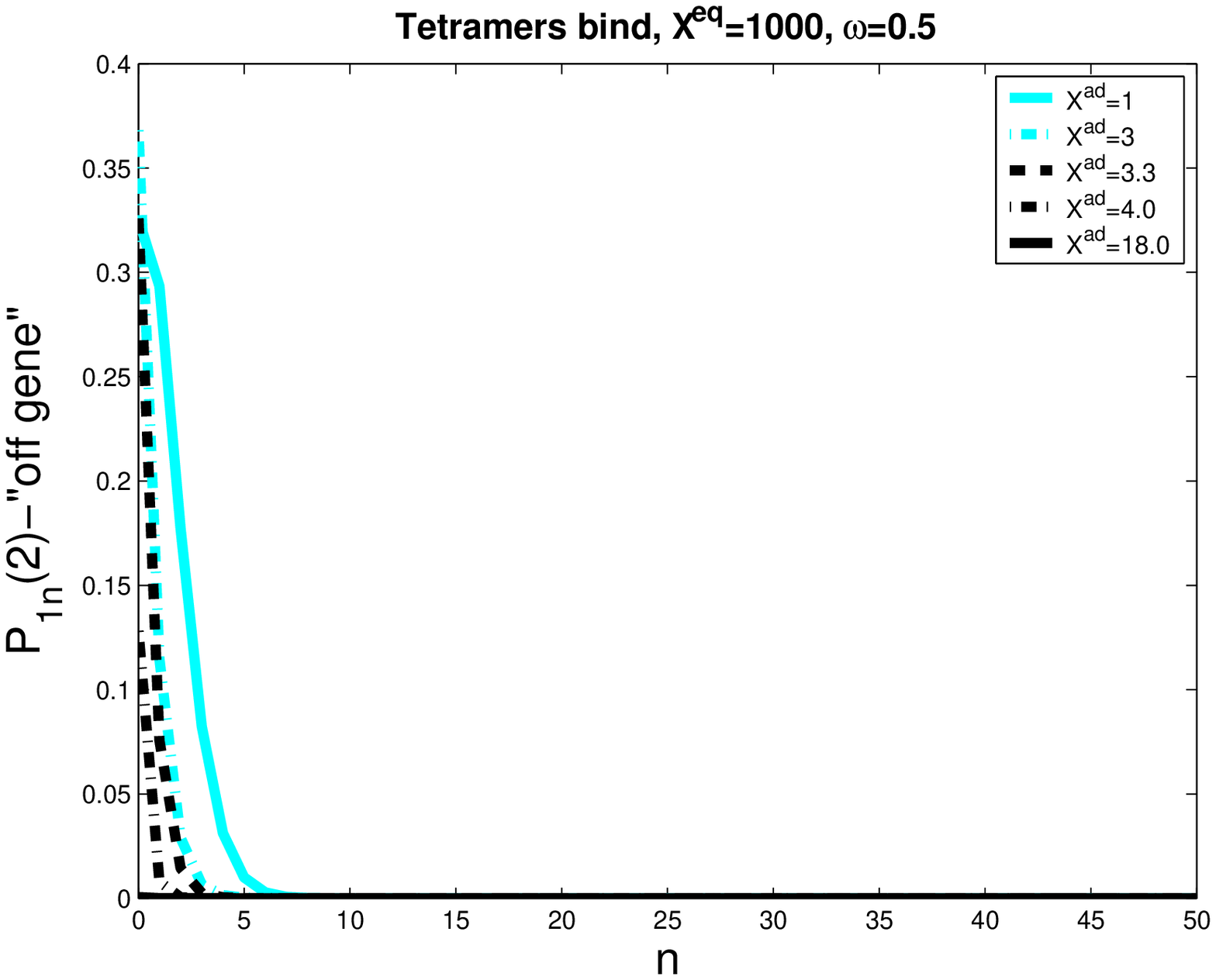}
\end{minipage}
\caption{The evolution of the probability distribution of the gene that will be active (a) and inactive (b) after the bifurcation to be on as a function of $X^{ad}$ for a switch when proteins bind as tetramers, $X^{eq}=1000$,$\omega=0.5$.}
\label{hoprotet}
\end{figure}

Switches in which effector proteins bind as higher order oligomers are omnipresent in nature and have been realized experimentally in artificial switches (McLure and Lee, 1998). We considered the binding of trimers ($h_i(n_{3-i})=h_in^3_{3-i}$) and tetramers ($h_i(n_{3-i})=h_in^4_{3-i}$) in symmetric switches. The equations of motion have the same form as before, but the interaction function $F(i)$ accounts for the higher moments. For proteins binding as $k^{th}$ order oligomers it has the form $F(i)=C_1(i) <n^{k}_1(i)>+C_2(i) <n^{k}_2(i)>$. As shown when discussing the dimer binding switch, the $k^{th}$ order moments have a simple form in the creation operator representation.\\ 
\begin{figure}
\begin{minipage}[t]{.43\linewidth}
\includegraphics[height=3cm,width=4cm]{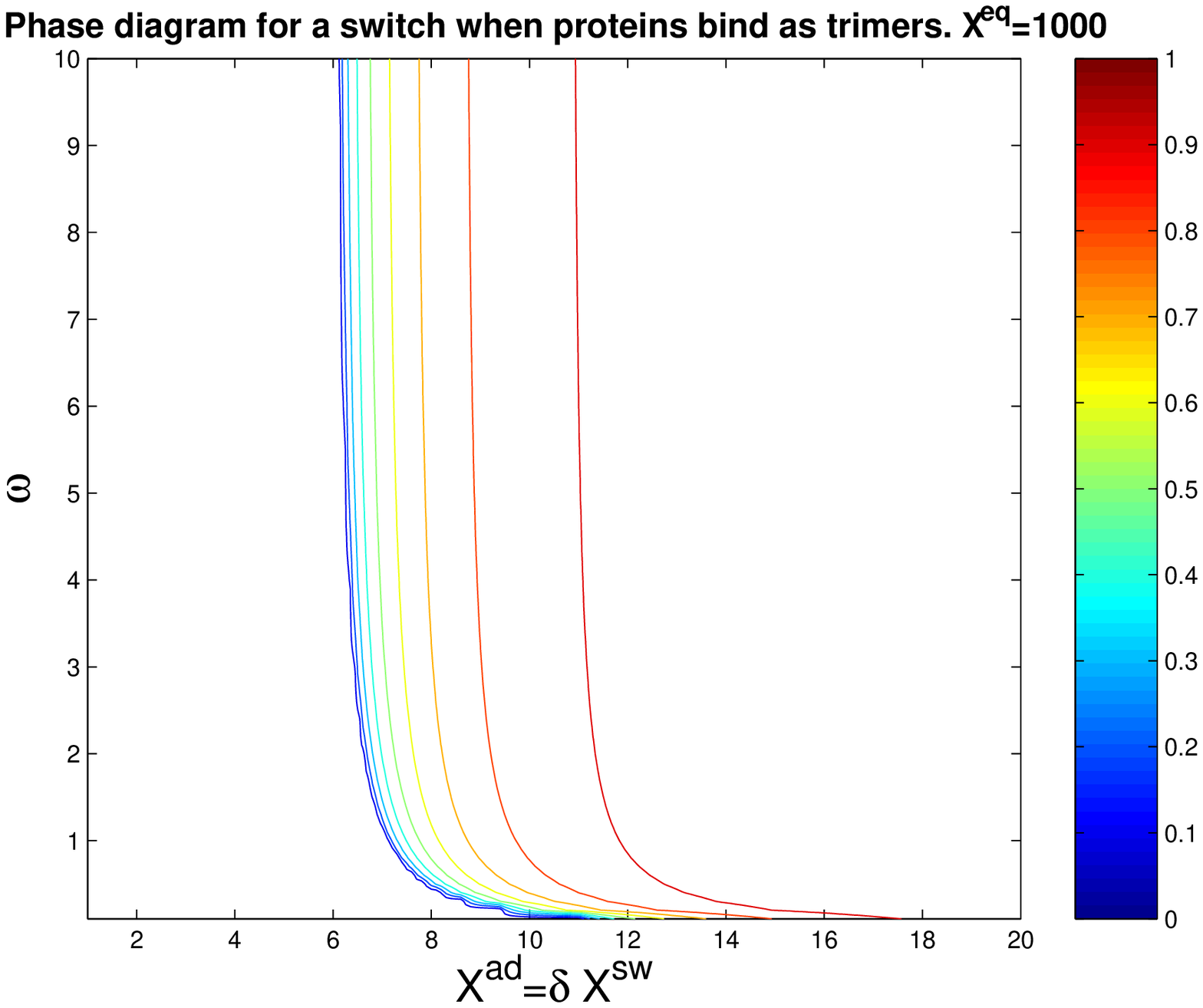}
\end{minipage}\hfill
\begin{minipage}[t]{.5\linewidth}
\includegraphics[height=3cm,width=4cm]{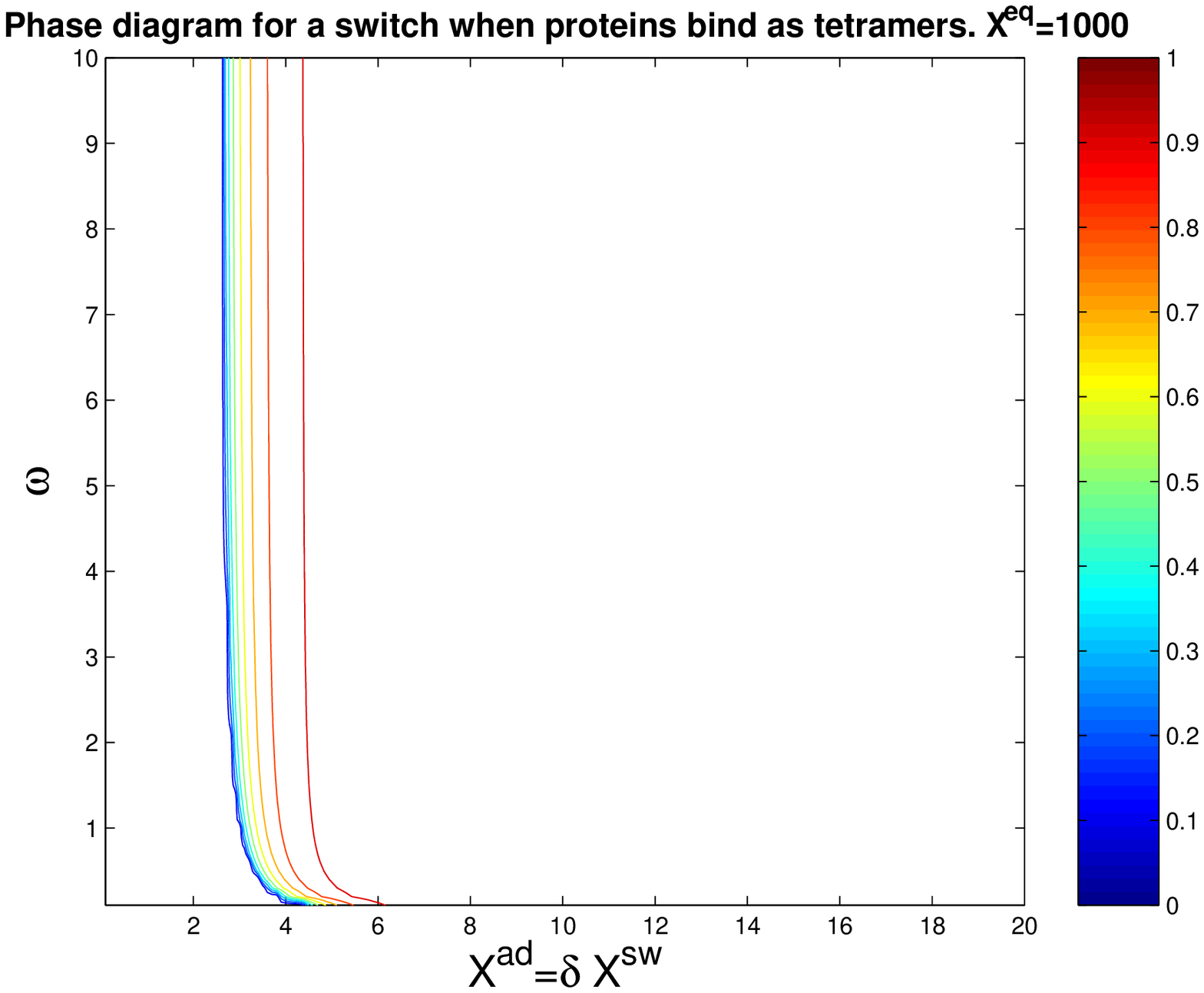}
\end{minipage}
\caption{Phase diagram for the SCPF approximation for a single symmetric switch to which proteins bind as trimers (a) and tetramers (b), with $X^{eq}=1000$. Contour lines mark values of $\Delta C$.}
\label{hoph}
\end{figure}
\begin{figure}
\begin{minipage}[t]{.43\linewidth}
\includegraphics[height=3cm,width=4cm]{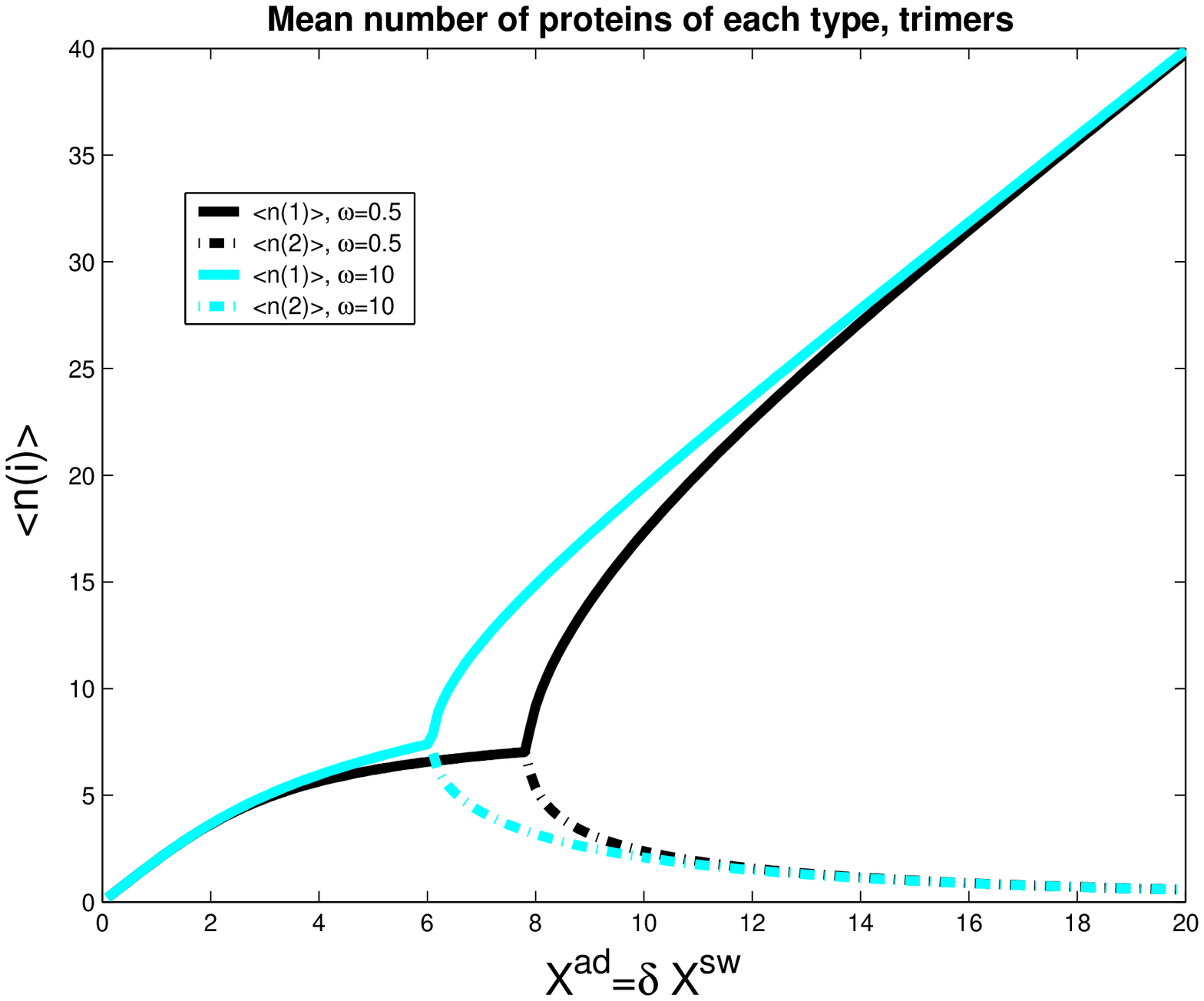}
\end{minipage}\hfill
\begin{minipage}[t]{.5\linewidth}
\includegraphics[height=3cm,width=4cm]{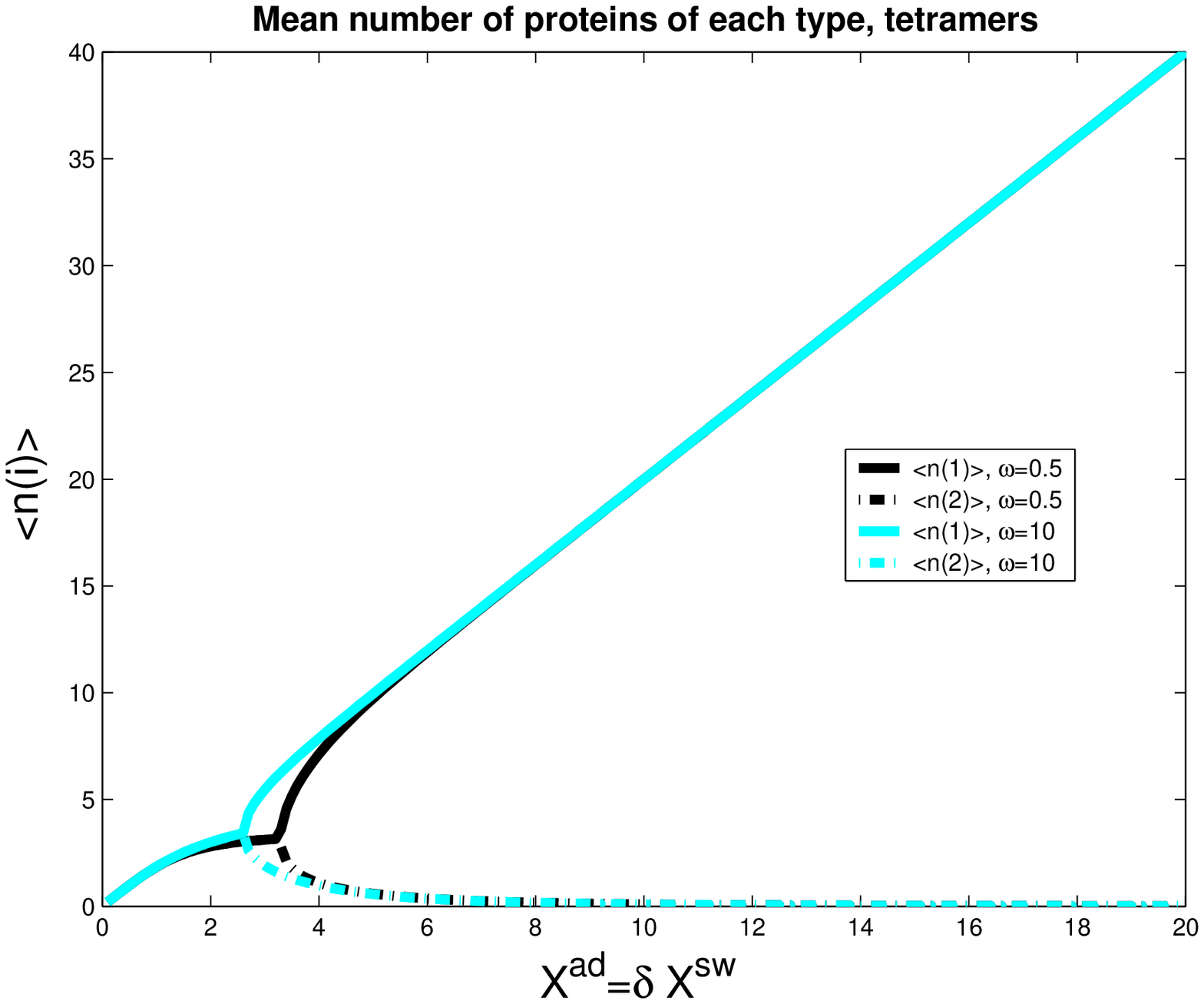}
\end{minipage}
\caption{Mean number of proteins in the cell, for each type when proteins bind as trimers (a) and tetramers (b), $\omega=0.5,10$, $X^{eq}=1000$, symmetric switch. }
\label{hopro}
\end{figure}
\textbf{The general mechanism}\\
From Fig. \ref{hoph} one notes that in order for the system to act as a bistable switch a considerably smaller number of reservoir proteins is needed than in the case of the dimer binding switch. As the multimericity number grows the area of bistability of the switch in parameter space grows. Since we assumed only one type of protein repressed a given gene, binding of higher order multimers is an effective model of cooperativity. Therefore we expect the system to have a larger region of bistability the higher the order of the binding multimer. The evolution of the system in parameter space when trimers bind is qualitatively similar to the dimer binding scenario (Fig. \ref{hoprotri}). Fast DNA unbinding rates stabilize the system and the bifurcation takes place for smaller effective production rates, for large $\omega$ than for small $\omega$ (Fig. \ref{hopro}). The critical number of proteins necessary for the bifurcation to take place is independent of the adiabaticity parameter and decreases with multimericity: $<n>_c=32$ for dimers binding, $<n>_c=8$ for trimers binding and $<n>_c=4$ for tetramers binding. This along with the narrow probability distributions (Fig. \ref{hoprotet}), small $\omega$ dependence when tetramers bind (Fig. \ref{hoph}), shows that one binding event determines the result, hence DNA binding rates do not play a role. Once there are $<n>_c$ proteins of a given type in the cell, a tetramer repressor will bind and stay bound. In the deterministic case the probability of the genes needs to fall to $(p-1)/p$, where $p$ is the order of multimerization of the repressor, for the bifurcation to take place. That along with the need for the number of repressors to be comparable with the tendency for proteins to be unbound from the DNA sets the critical number of proteins necessary for the bifurcation. Hence the bifurcation occurs when both genes are more probable to be on than off, for both tetramers and trimers. Therefore for the tetramer system a large buffering proteomic cloud is not needed to stabilize the DNA binding state of the switch and the characteristics of the system are practically independent of the adiabaticity parameter.\\
\textbf{Tetramer binding results in nearly deterministic characteristics}\\
In naturally occurring systems the production of the critical number of proteins is slowed down by relatively high multimerization rates and spatial dependence arising from the need of a large number of particles to diffuse together. These elements, which we neglect in our simple model constitute what might be called the cost of multimerization. This analysis also explains why most repressors and activators bind as dimers and tetramers, not trimers or pentamers. The effect of trimers binding is not different from that of dimers: a buffering proteomic cloud needs to be formed, the state of the system is quite influenced by noise, the switching region (region in $X^{ad}$ parameter space from the bifurcation point to $\Delta C>0.9$) is quite large. Yet in a real system there is an effective cost of trimerization: the energy of trimer formation and a need for the diffusion of particles. For tetramers the effect of stochasticity becomes negligible. Effectively one tetramer is sufficient for the bifurcation to take place. The binding of tetramer repressors may be thought of as a mechanism for increasing the deterministic nature of the switch.\\
\textbf{Binding of higher order oligomers as a competitve mechanism}\\
This analysis, although it neglects some important features, allows for a more quantitative formulation of cooperativity. Since most biological switches are asymmetric, cooperativity is also used as a means of making genes with smaller chemical rates more competitive. Tetramer binding seems to have a different role than that of lower order multimers. It may be used by genes which need to react to very small concentrations of proteins, for example they turn on degradation mechanisms when even a small number of toxic molecules is present. Or they may act as an extra mechanism stabilizing the existent state of a gene, as seems to be the case for the $cI$ gene of the $\lambda$ phage. It seems tetramers are used as having either a stabilizing role or that of a drastic, all or none response to the protein distributions in the system. This formulation of the problem is naturally oversimplified, but it allows for general observations.\\ 
\subsection*{The Case when Proteins are Produced in Bursts}
\begin{figure}
\begin{minipage}[t]{.25\linewidth}
\includegraphics[height=3cm,width=2.9cm]{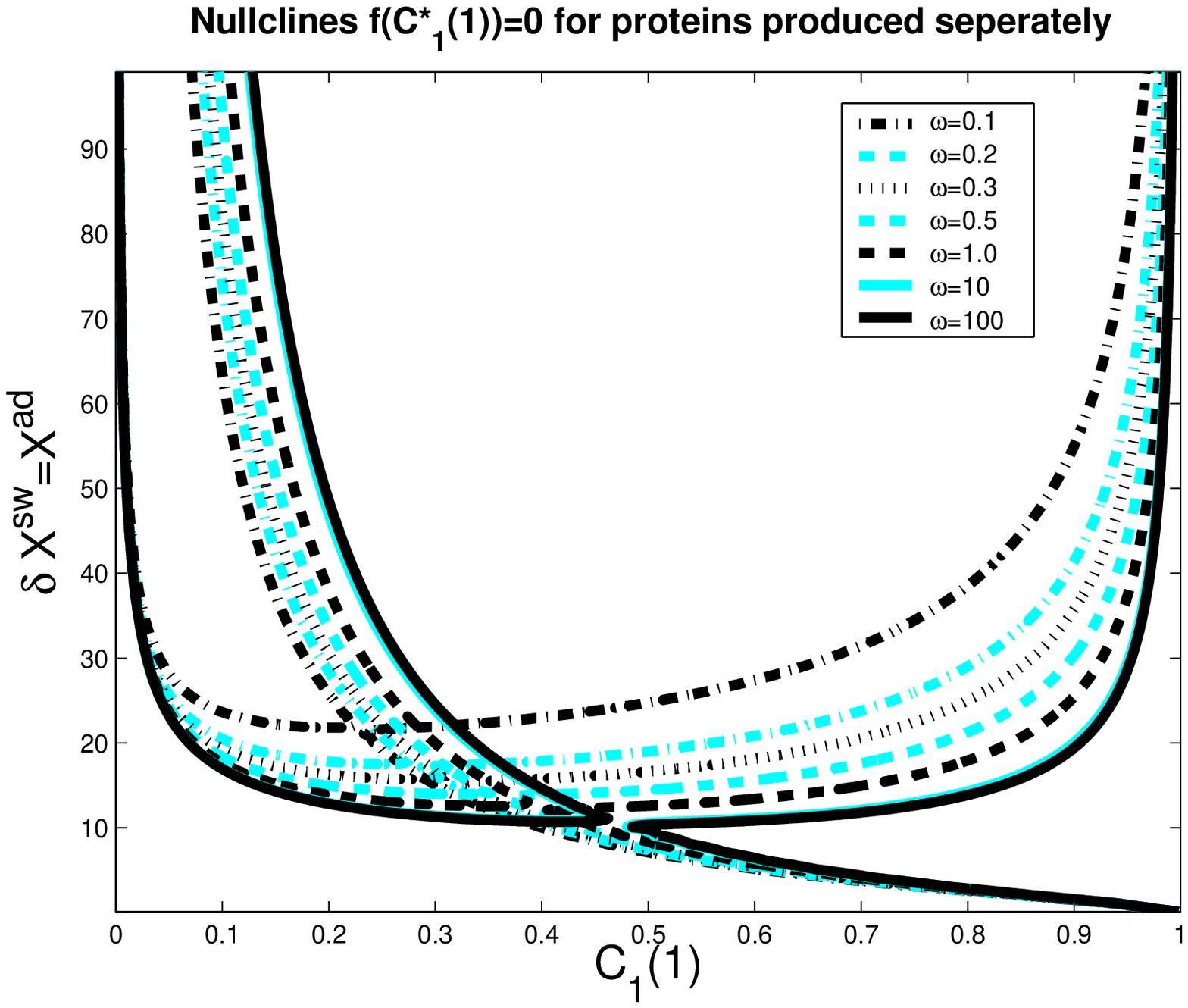}
\end{minipage}\hfill
\begin{minipage}[t]{.25\linewidth}
\includegraphics[height=3cm,width=2.9cm]{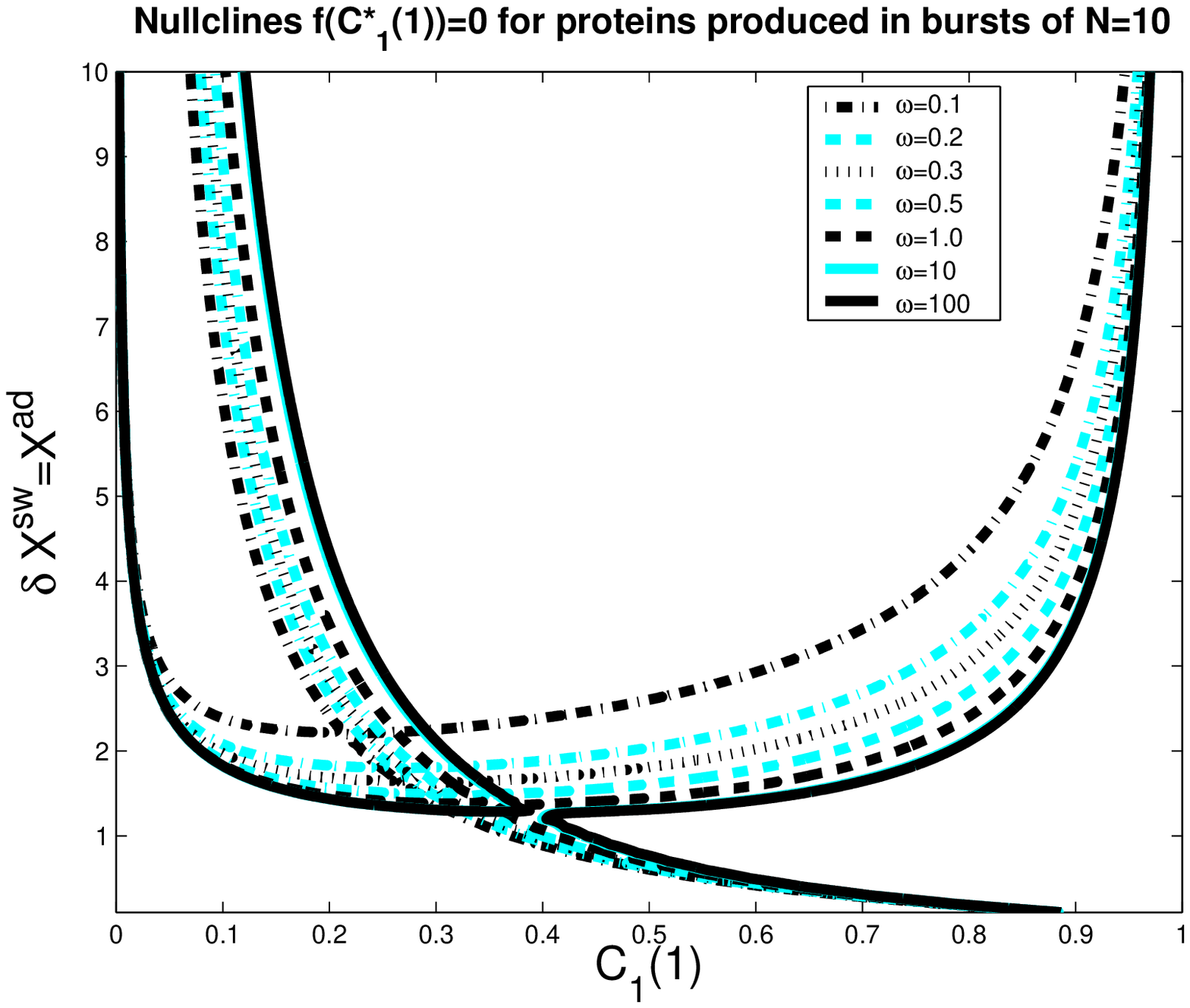}
\end{minipage}\hfill
\begin{minipage}[t]{.35\linewidth}
\includegraphics[height=3cm,width=2.9cm]{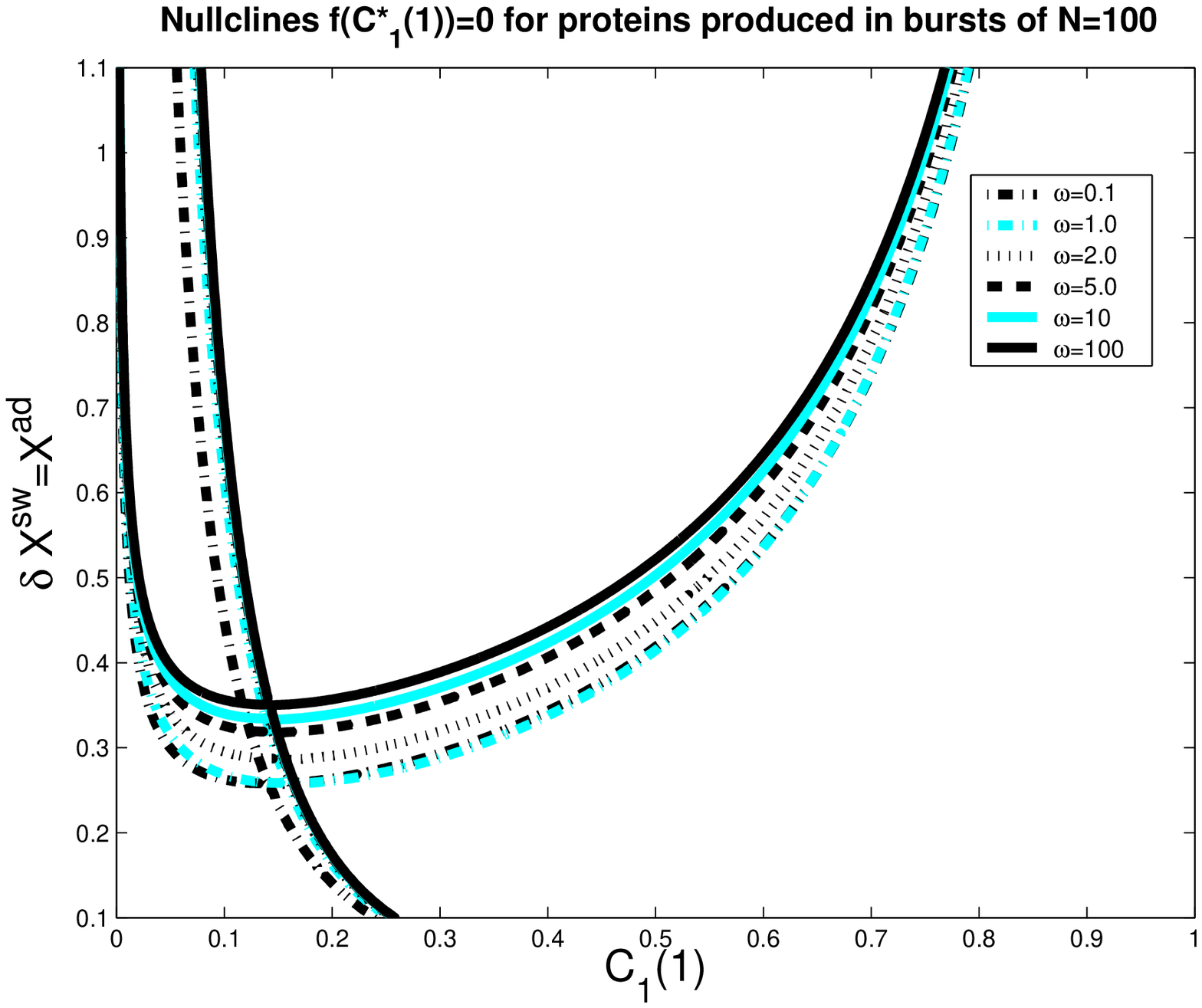}
\end{minipage}
\caption{Bifurcation curves for proteins produced separately $N=1$ (a), in bursts of $N=10$ (b) and $N=100$ (c) as a function of $X^{ad}=\delta X^{sw}$ for different values of the adiabaticity parameter $\omega=0.1,1,10,100$.}
\label{burstpromnull}
\end{figure}

\begin{figure}
\begin{minipage}[t]{.43\linewidth}
\includegraphics[height=3cm,width=4cm]{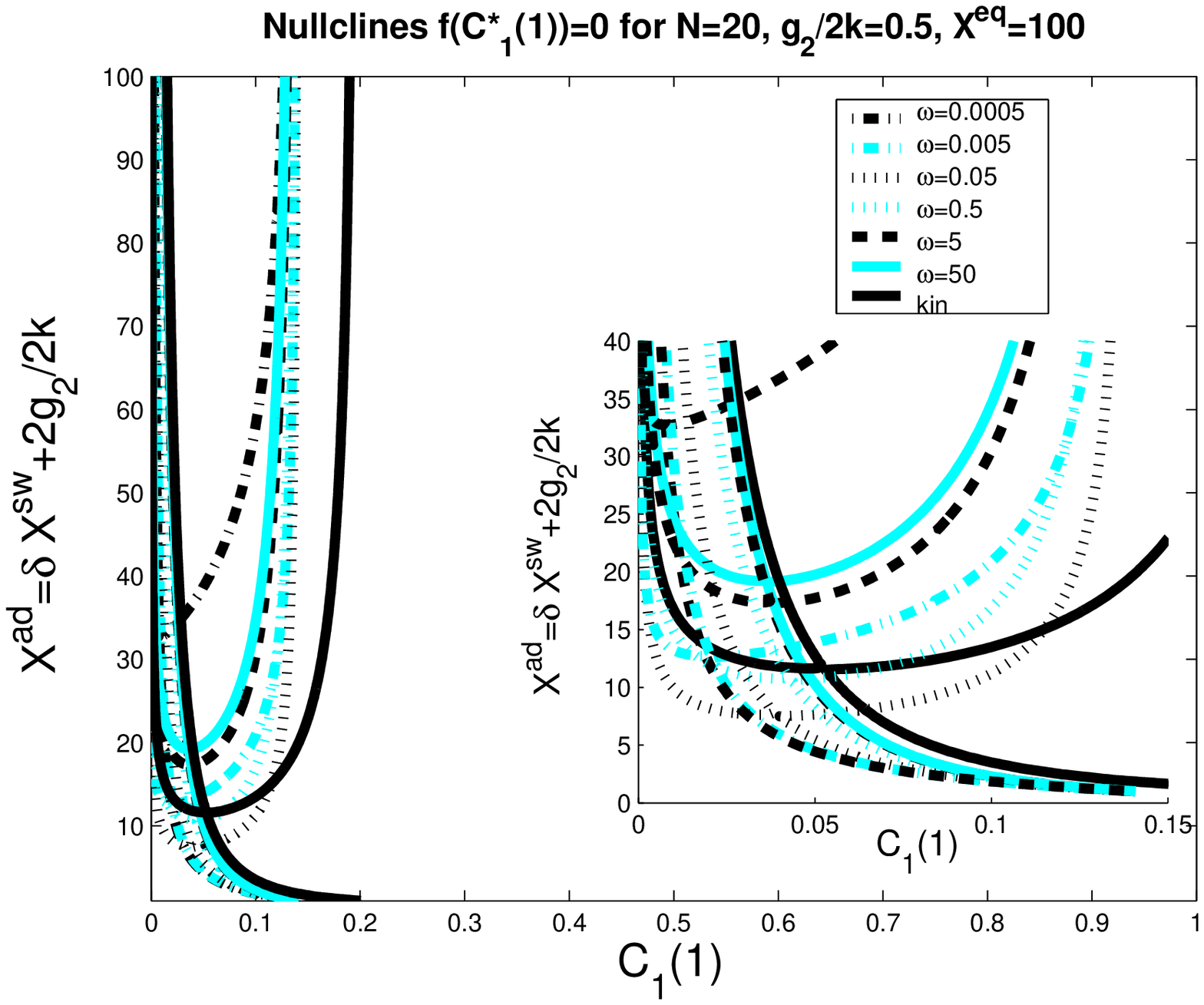}
\end{minipage}\hfill
\begin{minipage}[t]{.5\linewidth}
\includegraphics[height=3cm,width=4cm]{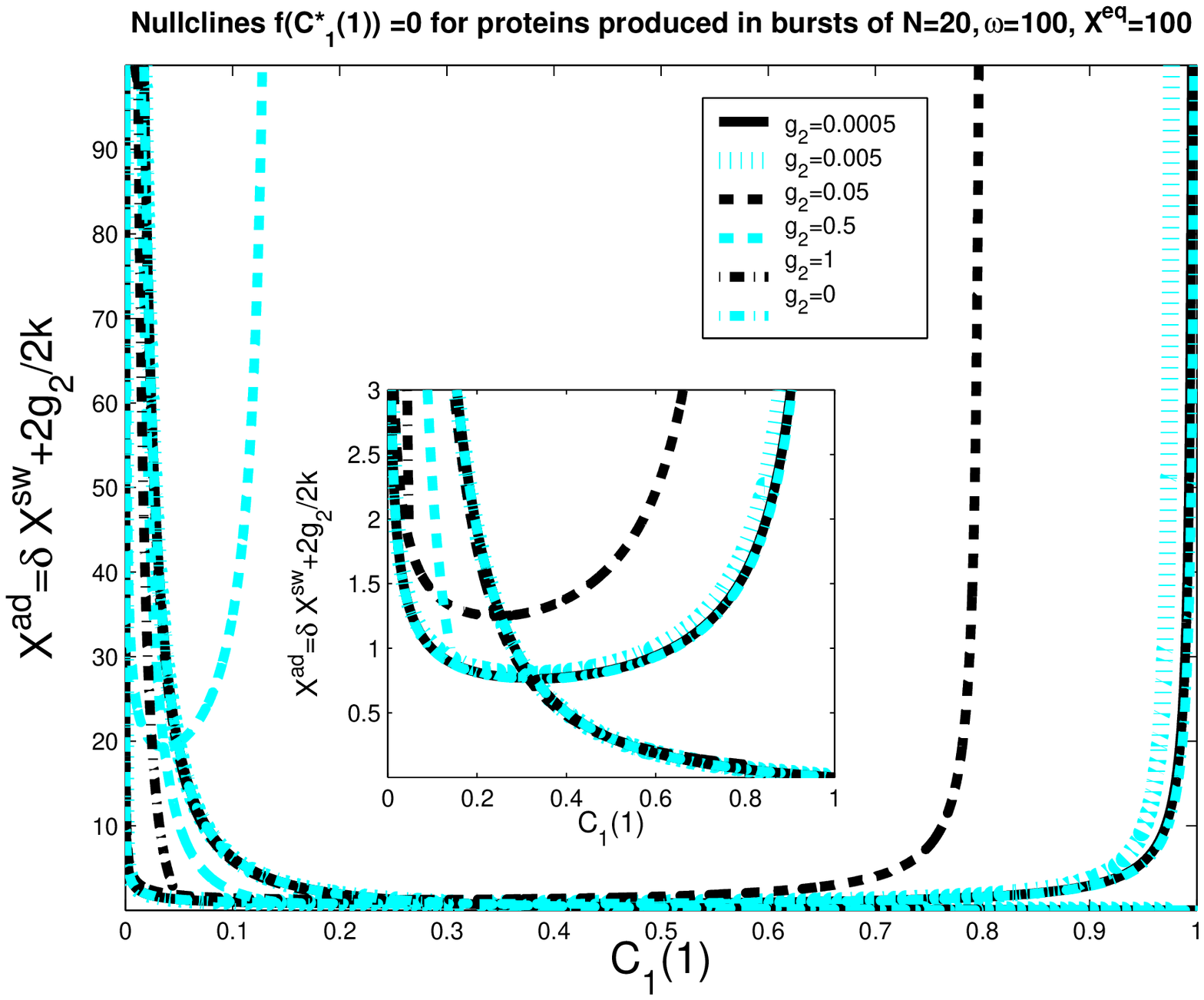}
\end{minipage}
\caption{Bifurcation curves as a function of $X^{ad}=\delta X^{sw}$, $X^{eq}=100$, $\omega=100$, $N=100$, for different basal effective production rate $g_2/2k=0.0005,0.005,0.05,0.5,1,0$  (a). Comparison of $\omega$ dependence with deterministic solution. $N=20$, $g_2/2k=0.5$, $X^{eq}=100$, $\omega=0.0005, 0.005, 0.05,0.5,5,50$ (b).}
\label{burstprndep1}
\end{figure}

Many proteins in biological systems, for example the Cro protein in $\lambda$ phage are produced in bursts of $N$ of the order of tens. We consider a symmetric switch, where proteins bind as dimers and are produced in bursts of $N$. The master equation in this case has the form:
\begin{eqnarray*}
\frac{\partial P_1(n_i)}{\partial t}&=&g_1(i)[P_1(n_i-N)-P_1(n_i)]+\\
&&+k_i [(n_i+1)P_1(n_i+1)-n_iP_1(n_i)]+\\
&&-h_i n^2_{3-i} P_1(n_i)+f_iP_2(n_i)\\
\frac{\partial P_2(n_i)}{\partial t}&=&g_2(i)[P_2(n_i-N)-P_2(n_i)]+\\
&&+k_i [(n_i+1)P_2(n_i+1)-n_iP_2(n_i)]+\\ 
&&+h_i n^2_{3-i} P_1(n_i)-f_iP_2(n_i)
\end{eqnarray*}
for $n \geq N$. For $n < N$ the equations have the form.
\begin{eqnarray*}
\frac{\partial P_1(n_i)}{\partial t}&=&-g_1(i)P_1(n_i)+\\
&&+k_i [(n_i+1)P_1(n_i+1)-n_iP_1(n_i)]+\\
&&-h_i n^2_{3-i} P_1(n_i)+f_iP_2(n_i)\\ 
\frac{\partial P_2(n_i)}{\partial t}&=&-g_2(i)P_2(n_i)+\\
&&+k_i [(n_i+1)P_2(n_i+1)-n_iP_2(n_i)]+\\ 
&&+h_i n^2_{3-i} P_1(n_i)-f_iP_2(n_i)
\end{eqnarray*}
Following the same procedure as for the the single protein production case, we get the following equations of motion for the first three moments:
\begin{eqnarray*}
\frac{\partial C_1(i)}{\partial t}=-h_i F(3-i) C_1(i)+f_i C_2(i) &\\
\frac{\partial C_2(i)}{\partial t}=h_i F(3-i) C_1(i)-f_i C_1(i) &\\
\frac{\partial C_1(_i) <n_1(i)>}{\partial t}=[N g_1(i)-k_i <n_1(i)>] C_1(i)&+&\\
-h_i F(3-i) <n_1(i)> C_1(i)+f_i <n_2(i)> C_2(i)&&\\
\frac{\partial C_2(_i) <n_2(i)>}{\partial t}=[N g_2(i)-k_i <n_2(i)>] C_2(i)&+&\\
+h_i F(3-i) <n_1(i)> C_1(i)-f_i <n_2(i)> C_2(i)&&\\
\frac{\partial C_1(_i) <n^2_1(i)>}{\partial t}=g_1(i)[2N <n_1(i)>+N^2] C_1(i)&+&\\
+k_i [-2 <n^2_1(i)>+<n_1(i)>] C_1(i)&+&\\ 
-h_i F(3-i) <n^2_1(i)> C_1(i)+f_i <n^2_2(i)> C_2(i)&& \\
\frac{\partial C_2(_i) <n^2_2(i)>}{\partial t}=g_2(i)[2N <n_2(i)>+N^2] C_2(i)&+&\\
+k_i [-2 <n^2_2(i)>+<n_2(i)>] C_1(i)&+&\\ 
+h_i F(3-i) <n^2_1(i)> C_1(i)-f_i <n^2_2(i)> C_2(i)&&\\
\end{eqnarray*}
where $F(i)=C_1(i) <n^2_1(i)>+C_2(i) <n^2_2(i)>$ as before. Writing out $N^2=N(N-1)+N$ and subtracting the $<n_j(i)>$ equations from $<n^2_j(i)>$ we get the equations of motion for the previously defined creation operators $a$. Due to the form of $F(i)$ for the dimer binding case only the first three moments are relevent. However generally this procedure can be carried out for higher moments, yielding an expression for the $m^{th}$ creation operator moment in the steady state of the form:
\begin{eqnarray*}
<a^m_{1i}>&=&(N X_i^{ad}+N \delta X_i^{sw})(1-\frac{\omega_i C_2(i)}{\omega_i+mC_1(i)})<a^{m-1}_1>+\\ 
&+&(N X_i^{ad}-N \delta X_i^{sw})\frac{\omega_i C_2(i)}{\omega_i+mC_1(i)}<a^{m-1}_2>+
\\ 
&+&\frac{N^{m-1}-1}{2}(N X_i^{ad}-N \delta X_i^{sw}(1-\frac{\omega_i C_2(i)}{\omega_i+mC_1(i)}))
\\ 
<a^m_{2i}>&=&(X_i^{ad}-\delta X_i^{sw})(1-\frac{\omega_i C_1(i)}{\omega_i+mC_1(i)})<a^{m-1}_2>+ \\  &+&(X_i^{ad}+\delta X_i^{sw})\frac{\omega_i C_1(i)}{\omega_i+mC_1(i)}<a^{m-1}_1>+
\\ 
&+&\frac{N^{m-1}-1}{2}(N X_i^{ad}-N \delta X_i^{sw}(1-\frac{\omega_i C_1(i)}{\omega_i+mC_1(i)}))
\\ 
\end{eqnarray*}
To consider the binding of higher order oligomers when proteins are produced in bursts one simply accounts for the changed form of $F(i)$ as discussed in the previous section.\\
\begin{figure}
\begin{minipage}[t]{.43\linewidth}
\includegraphics[height=3cm,width=4cm]{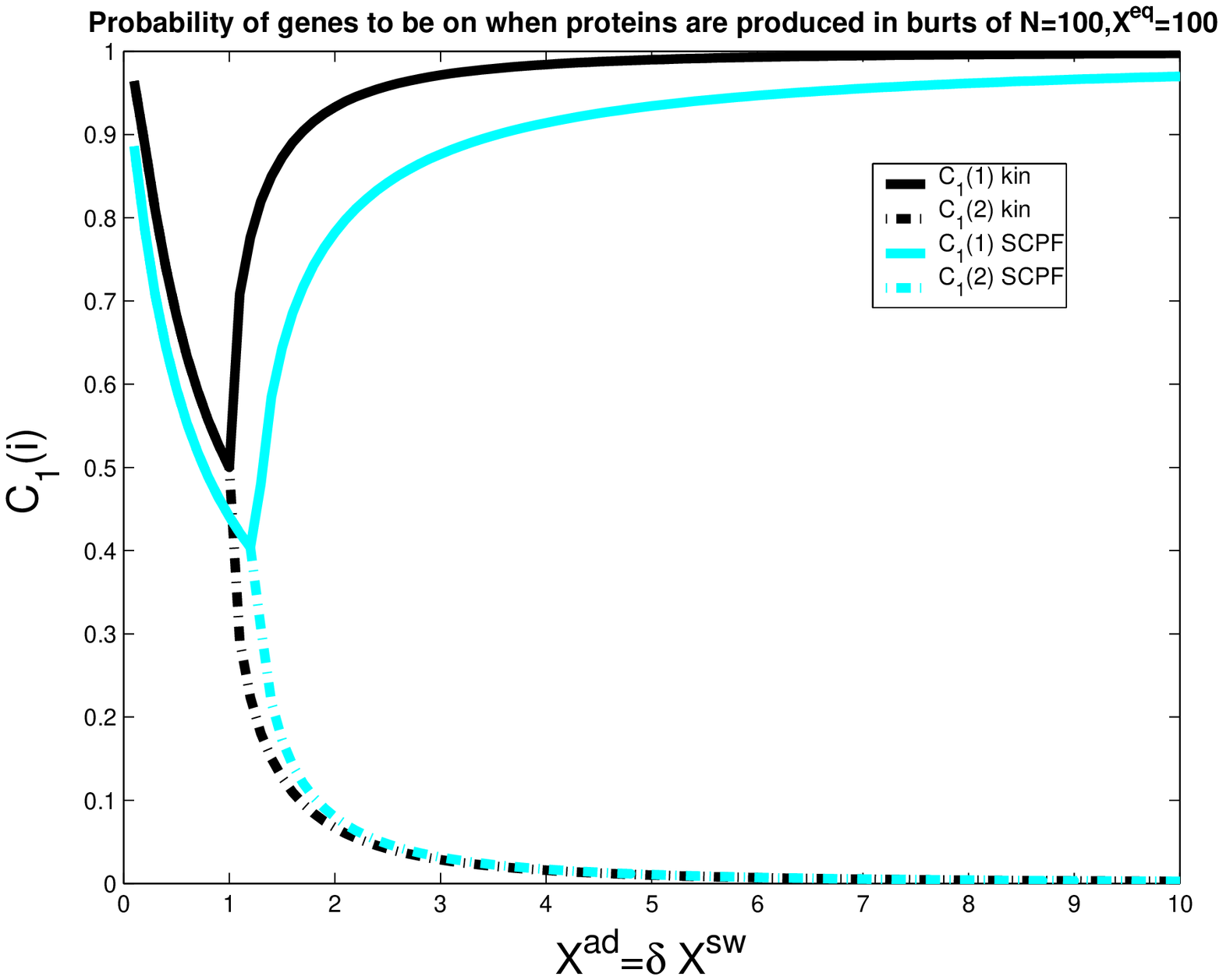}
\end{minipage}\hfill
\begin{minipage}[t]{.5\linewidth}
\includegraphics[height=3cm,width=4cm]{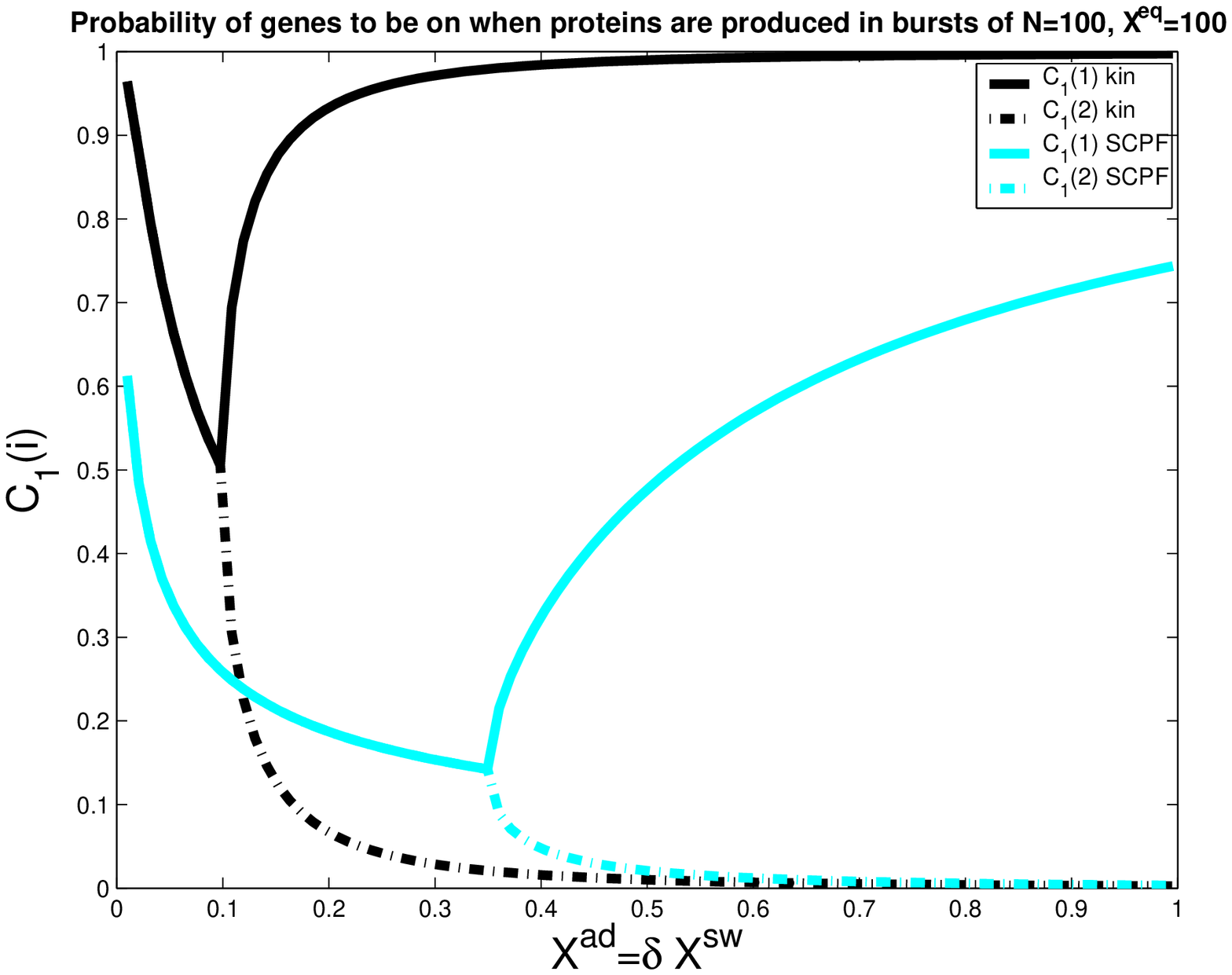}
\end{minipage}
\caption{Probability that gene $i$ is on when proteins are produced in bursts of $N=10$ (a) and $N=100$ (b), symmetric switch proteins bind as dimers, $X^{eq}=100$, $\omega=100$. Comparison of deterministic and stochastic solutions. }
\label{burstnum}
\end{figure}
\begin{figure}
\begin{minipage}[t]{.43\linewidth}
\includegraphics[height=3cm,width=4cm]{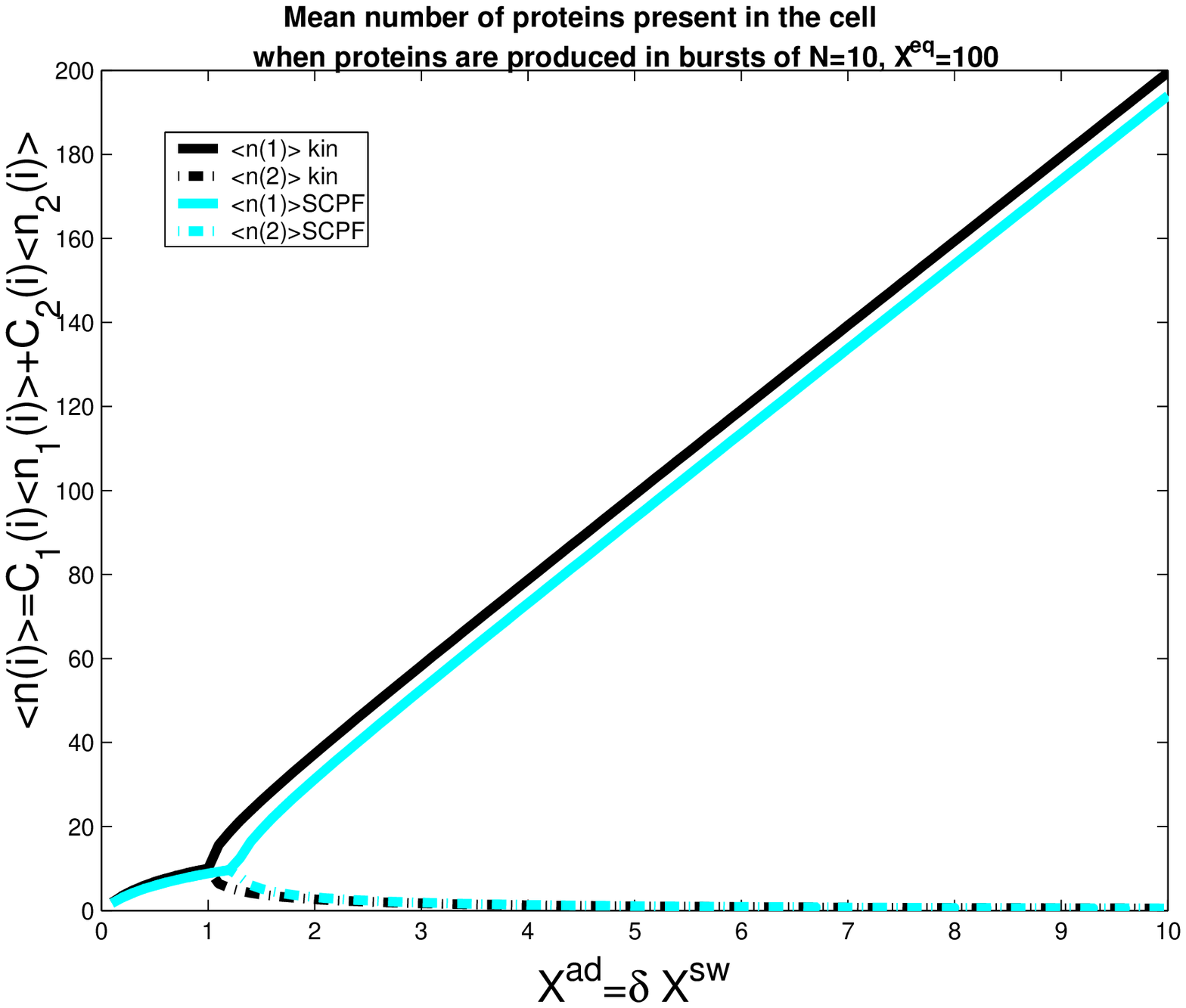}
\end{minipage}\hfill
\begin{minipage}[t]{.5\linewidth}
\includegraphics[height=3cm,width=4cm]{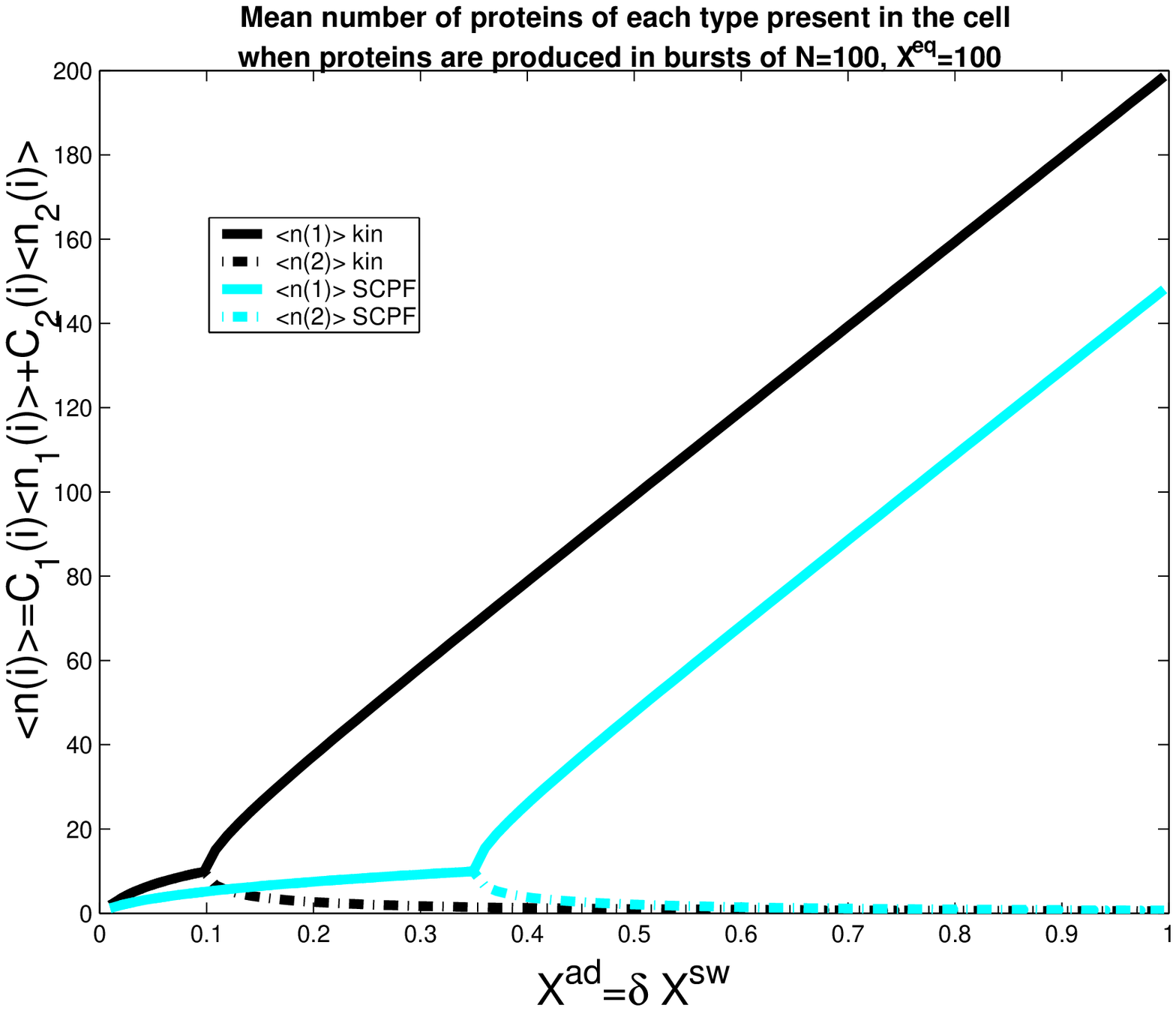}
\end{minipage}
\caption{Mean number of proteins of each type present in the cell when proteins are produced in bursts of $N=10$ (a) and $N=100$, symmetric switch proteins bind as dimers, $X^{eq}=100$, $\omega=100$. Comparison of deterministic and stochastic solutions. }
\label{burstpro}
\end{figure}
\begin{figure}
\begin{minipage}[t]{.43\linewidth}
\includegraphics[height=3cm,width=4cm]{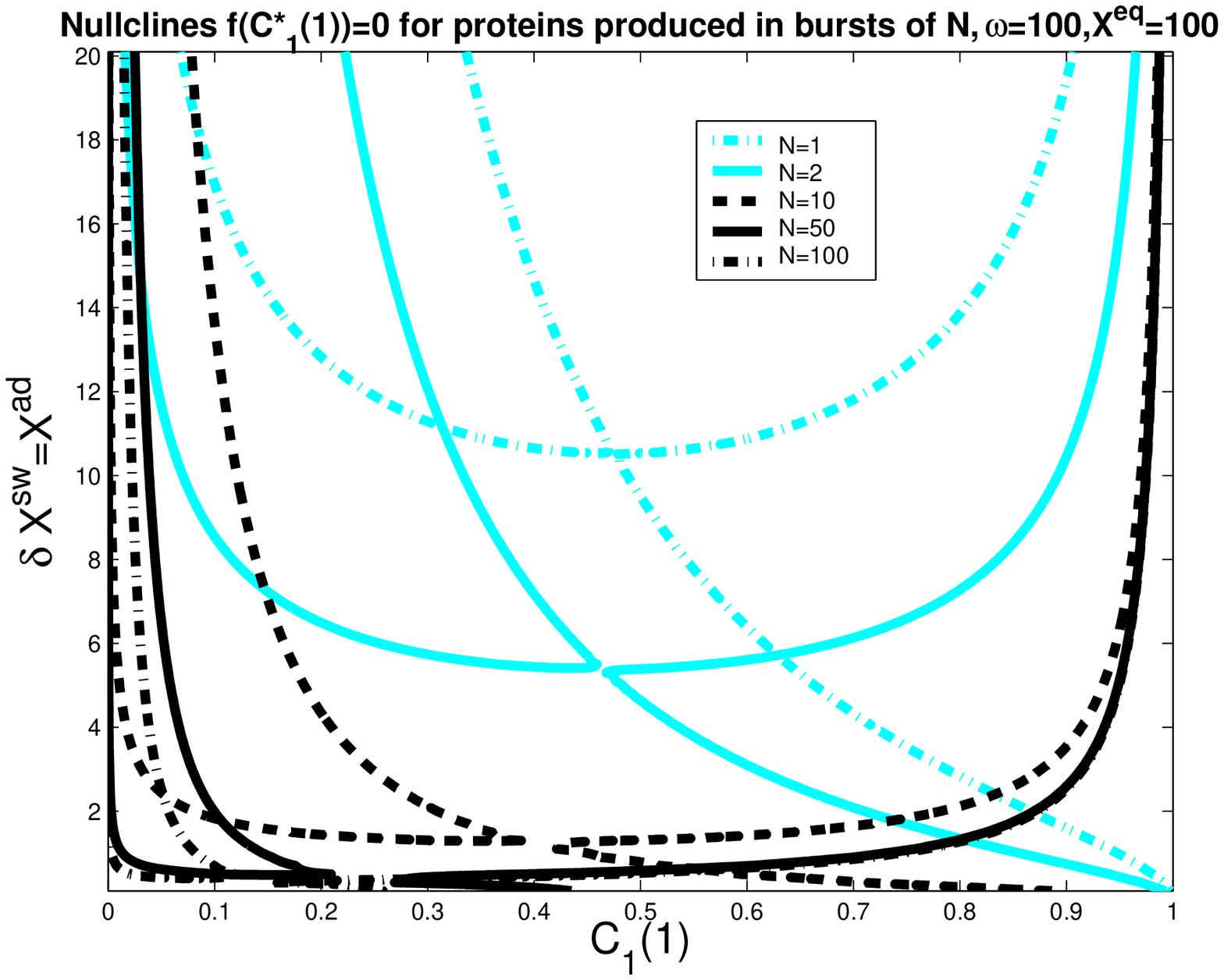}
\end{minipage}\hfill
\begin{minipage}[t]{.43\linewidth}
\includegraphics[height=3cm,width=4cm]{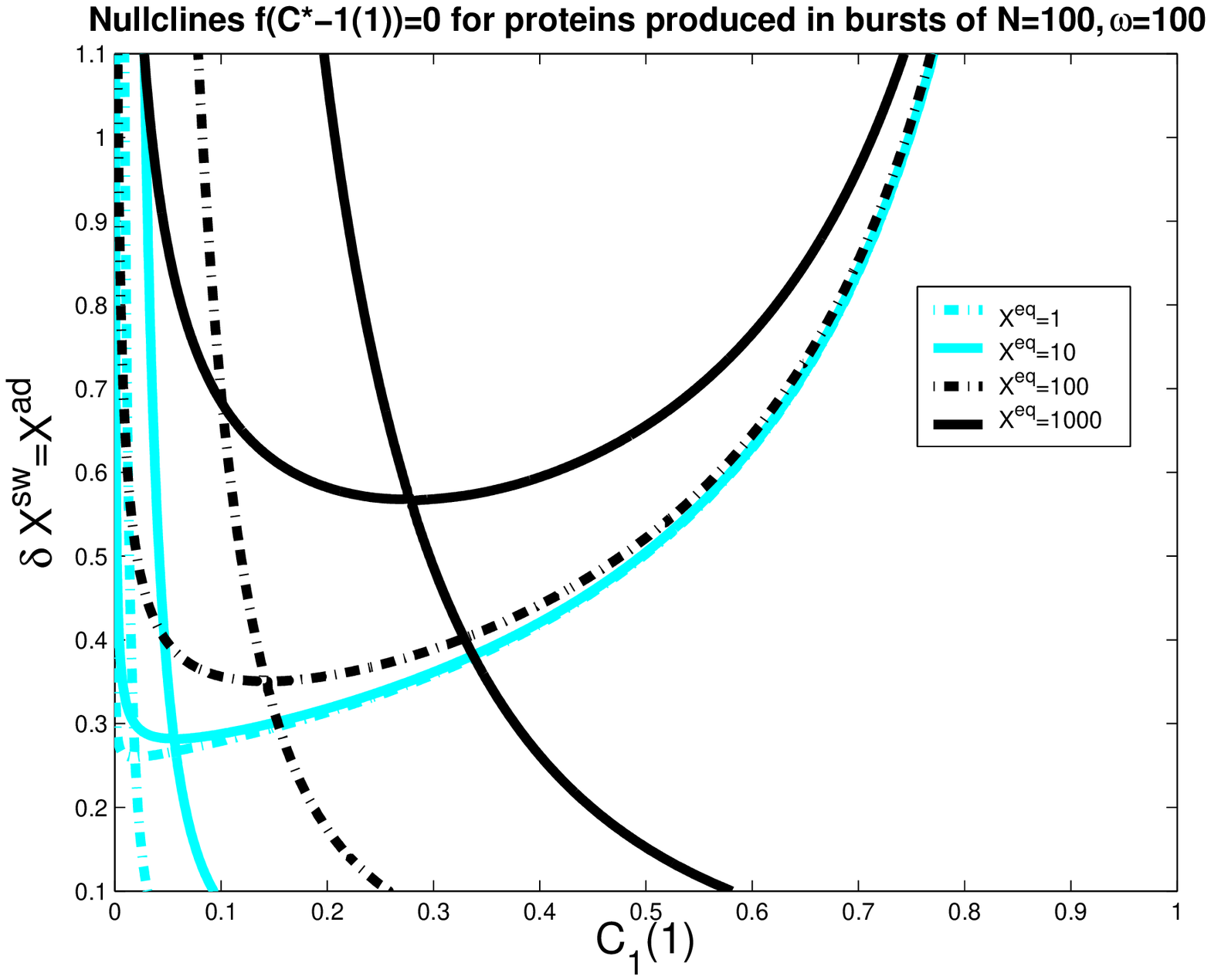}
\end{minipage}
\caption{Bifurcation curves as a function of $X^{ad}=\delta X^{sw}$, $\omega=100$ for different burst size values $N=1,2,5,10,50,100$, with $X^{eq}=100$ (a) and for proteins produced in bursts of $N=100$ (b) for different values of $X^{eq}=1,10,100,1000$.}
\label{burstprndep}
\end{figure}
\textbf{The general mechanism}\\
We discuss the effect of bursting phenomena on the example of a symmetric toggle switch when proteins bind as dimers, as that can offer the most insight, when compared to previous results. In this case switching takes place for much smaller values of the effective production rate parameter $X^{ad}$ compared to when proteins are produced separately. Therefore even in the large $\omega$ limit, noise resulting from large protein number fluctuations plays a role in defining the region of stability of the switch, as the criterion of large $X^{ad}$ is not reached. The number of proteins in the cell when the bifurcation occurs is determined by the tendency that proteins are unbound from the DNA and does not change when proteins are produced in bursts. For the rates discussed in Fig. \ref{burstnum} and Fig. \ref{burstpro} the critical mean number of proteins present in the cell at which the bifurcation occurs is $n_c=10={X^{eq}=100}^{\frac{1}{2}}$. If proteins are produced in bursts of $N=10$, as in the left hand figures, this value of $n_c$ is achieved when $X^{ad}>1$, that is proteins must get produced at a higher rate than they are destroyed to be able to sustain the steady state number of $10$ proteins in the cell. In the figures on the right hand side of Fig. \ref{burstnum} and Fig.\ref{burstpro} proteins are produced in bursts of $N=100$. In this case even when the degradation rate is larger that the production rate, the critical steady state number of proteins necessary for the bifurcation to take place, can be reached and a bistable switch is possible. A bistable switch can exist even if the degradation rate exceeds the production rate for burst sizes present in biology. For $X^{eq}=100$, the order of the tendencies for proteins to be unbound from the DNA in the $\lambda$ phage, the value of N for which $X_c^{ad}<1$ is smaller than $N=20$, the burst size for Cro proteins in the $\lambda$ phage. $X^{ad}$ at the critical point decreases as function of N  (Fig. \ref{burstprndep}) and depends on the tendency that proteins are unbound from the DNA $X^{eq}$ (Fig. \ref{burstprndep} (b)) and the adiabaticity parameter, $\omega$ (Fig. \ref{burstpromnull}). \\
If proteins are produced individually the span of the non-adiabatic regime is clear from Fig. \ref{burstpromnull}. It corresponds to $\omega<1$. The bifurcation curves show small discrepancies for larger values of the adiabaticity parameter. However for larger burst sizes there is a continuous change in the form of the bifurcation curves with $\omega$. All of the solutions differ substantially from the deterministic treatment, as shown in Fig. \ref{burstnum}.\\
\textbf{The influence of the adiabaticity parameter on the bifurcation mechanism}\\
Contrary to the $N=1$ case, the effective production rate at the bifurcation point $X_c^{ad}$, grows smaller with the increase of the adiabaticity parameter, for considerably large burst sizes, as in the $N=100$ example in Fig. \ref{burstpromnull}. In this case each gene produces a large number of repressors at a time.  The bifurcation takes place in a region with $X^{ad}<1$, which corresponds to very small effective production rates, which denote very large death rates. Therefore in the region of parameter space before the bifurcation takes place both genes remain repressed ($C_1(i)<0.5$) in the steady state, as opposed to the provisionally discussed situations, in which both genes had equal probabilities to be active ($C_1(i)>0.5$). For large $N$ bursts, the bifurcation takes place when one of the genes becomes unrepressed in the steady state. That is when the repressor cloud buffering the DNA becomes destabilized, not when the cloud forms as in the smaller $N$ examples. For large $N$ bursts, if the rate of unbinding from the DNA is fast compared to the protein degradation rate, larger effective production rates are needed for the buffering proteomic cloud to stabilize the DNA state, than for small $\omega$ (Fig. \ref{burstpromnull} (c)). The larger $X^{ad}$ is, the more repressor molecules are present in the system, which corresponds to larger protein number fluctuations, which are necessary for one of the genes to become unrepressed. For slower DNA unbinding rates, the buffering proteomic cloud is smaller, since the protein number reaches a steady state before the DNA state does. Therefore the buffering proteomic cloud is  destabilized at smaller values of $X^{ad}$. Hence, in the case of small $\omega$ the unrepressing bifurcation takes place for smaller effective production rates than for large $\omega$. However if the unbinding rate from the DNA is very small, $\omega<0.01$, $X^{ad}_c$ as a function of the adiabaticity parameter grows again, as this corresponds to effectively large death rates, which need very high production rates to sustain a proteomic cloud buffering the DNA.  If the effective production rate is too small in this case, the steady state number of proteins is too small to form the buffering proteomic cloud, although the burst size is enormous. In the very small $\omega$ limit the bifurcation cloud needs to be formed for the bifurcation to be possible, as in the mechanism present in the small N case. The value of $X^{ad}$ at the bifurcation point in both the large and small $\omega$ limit is strongly governed by protein and DNA binding state fluctuations in the system. For this reason the deterministic  solution fails. It assumes the incorrect mechanism, in which the bifurcation is a result  of repressing one of the genes. This can happen if the death rate of proteins is slow enough to allow for the existence of $<n(i)_c>$ repressor molecules in the system at very small production rates ($C_1(1)^{biff, kin}=0.5$) (Fig. \ref{burstnum}). One can see that the order of taking the adiabatic limits in the steady state for proteins produced in large bursts is subtle and depends strongly on the parameters of the system, as the bifurcation is governed mainly by relative protein and DNA fluctuations, both of which are very large. Furthermore, the deterministic solution is closer to  the small $\omega$ limit, which corresponds to slow DNA unbinding rates compared to protein number fluctuations. Deterministic results may therefore be misleading in the bursting situation, even for large $\omega$.\\
\begin{figure}
\begin{minipage}[t]{.23\linewidth}
\includegraphics[height=3cm,width=4cm]{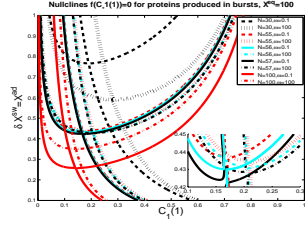}
\end{minipage}
\caption{Bifurcation curves for proteins produced in bursts of $N=30,55,56,57,100$ as a function of $X^{ad}=\delta X^{sw}$ for different values of the adiabaticity parameter $\omega=0.1,100$.}
\label{burstnombiff}
\end{figure}
Figure \ref{burstnombiff} shows explicitly how the steady state comes about as a result of different mechanisms depending on the burst number $N$ and how the order of reaching the steady state by the protein and DNA binding site dynamics changes depending on $\omega $. The other parameters were chosen so the bifurcation would take place for $X^{ad}<1$ for all examples. For small burst sizes, slower DNA unbinding rates require larger effective production rates to reach the steady state number of proteins necessary to form the buffering proteomic cloud than for large $N$. For larger burst sizes, faster DNA unbinding rates destabilize the buffering cloud of proteins for smaller effective production rates than in the small $N$ case. The inset shows the transition region between the two possible mechanisms.\\
\begin{figure}
\begin{minipage}[t]{.43\linewidth}
\includegraphics[height=3cm,width=4cm]{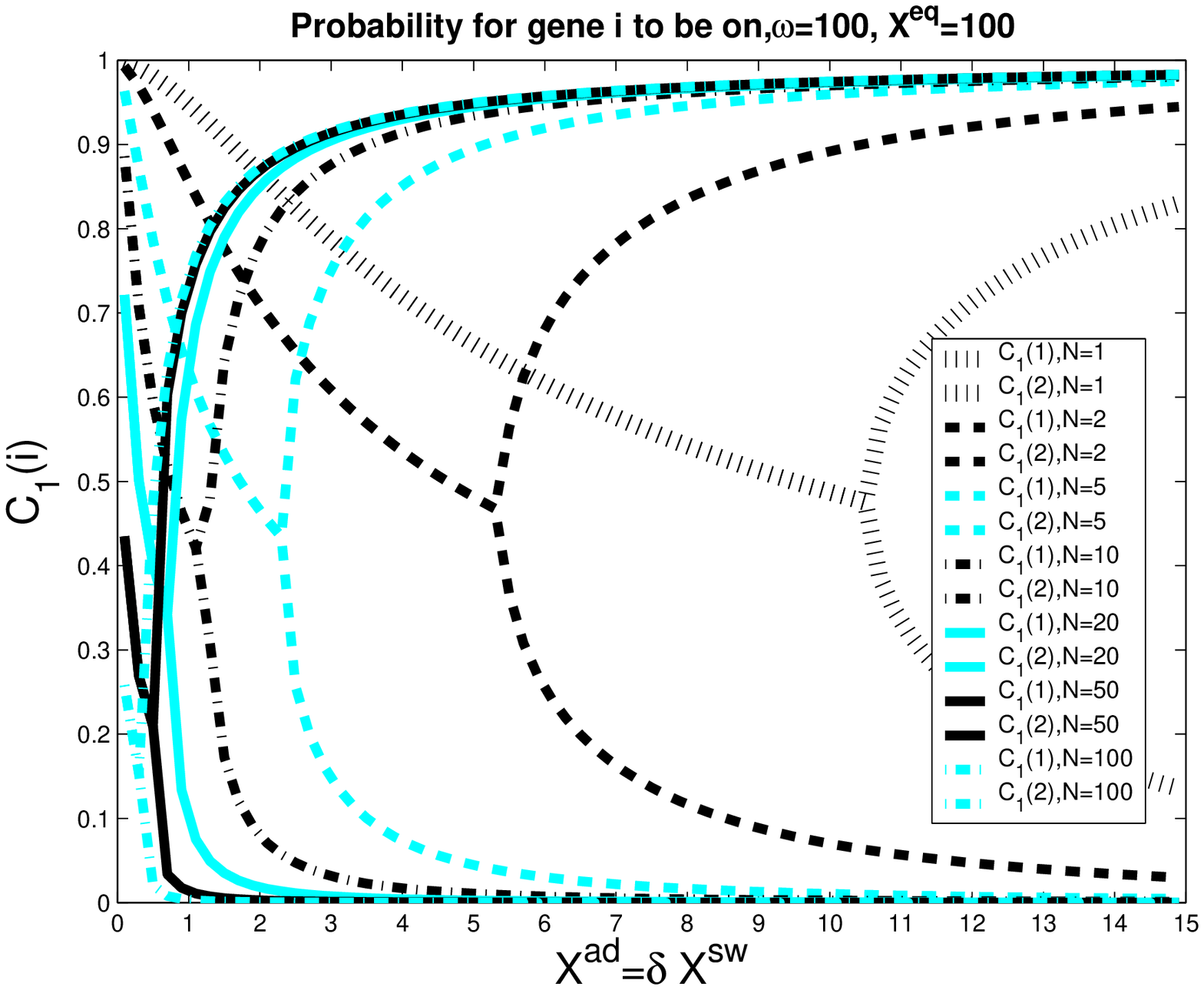}
\end{minipage}\hfill
\begin{minipage}[t]{.5\linewidth}
\includegraphics[height=3cm,width=4cm]{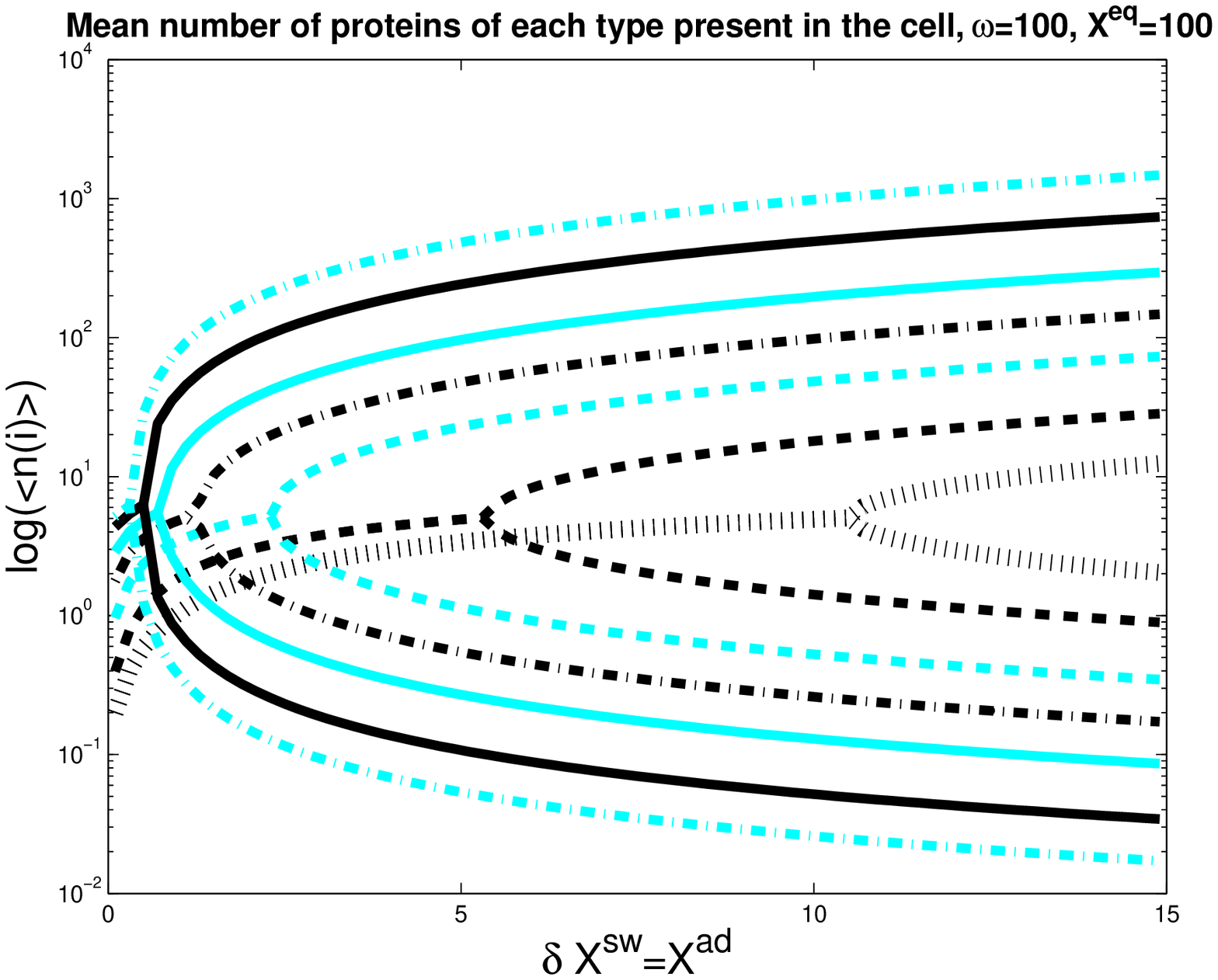}
\end{minipage}
\caption{Probability for genes to be on in the SCPF approximation as a function of $X^{ad}=\delta X^{sw}$, with $X^{eq}=100$, $\omega=100$ for $N=1,2,5,10,20,50,100$ (a) and the mean number of proteins present in the cell (b).)}
\label{burstnfun}
\end{figure}
\textbf{Consequences of bifurcation at smaller $X^{ad}$ values}\\
The divergence from the deterministic solution at the bifurcation point increases with the burst size, as is expected due to the enormous noise effect due to large $N$, on a system with a constant and independent of the burst size number of proteins at the bifurcation point (Fig. \ref{burstnfun}). As already noted the number of proteins in a cell, is in the range of tens to hundreds, even if they are produced in bursts. Figure \ref{burstnfun} shows that this number is reached for smaller effective production rates for larger burst sizes than for small $N$ values. Therefore systems where proteins are produced in bursts display smaller values of $X^{ad}$ and are more susceptible to noise if the number of proteins in the cell is to be of the order which is observed experimentally. Furthermore the noisy burst systems even for very large values of $X^{ad}$ do not converge as closely to the deterministic solution as they do for the single protein production example. This can be seen from the form of the steady state moment equations. The interaction function F(i) for the $N=1$ case in the limit of large $\omega$ and $X^{ad}$ converges to $F(i) \rightarrow <n(i)>+<n(i)>^2$ whereas the deterministic solution corresponds to $F(i)= <n(i)>^2$. Therefore for large mean values of proteins the two are equal. However in the case when $N>1$, $F(i) \rightarrow <n(i)>(1+\frac{N-1}{2})+<n(i)>^2$, which requires $N<<2<n(i)>$ for the effect of bursting to be negligible at very large N. The values of the effective production rate that correspond to values of the proteins seen experimentally seem to be small. Therefore we can say that effectively the role of bursting is to enable for the existence of a  bistable solution at lower effective production rates, which determines a region of parameter space which has been previously unstudied. In this region one cannot make the adiabatic assumption that the change in the DNA state can be integrated out due to a separation of timescales. That assumption leads to erroneous results, predicting a region of bistability where explicit treatment of both timescales suggests monostability. Furthermore, for very large $N$, the region of bistability decreases with the adiabaticity parameter, making the disagreement of the stochastic solutions with those of the deterministic rate equations larger. The adiabatic approximation and the full solutions converge only in the regime of large $\omega$ and $X^{ad}$, the second of which is never fulfilled at the bifurcation point or for biological concentration for systems in which proteins are produced in large bursts.\\ 
\textbf{Dependence on the DNA Binding Coefficient}\\
Just as increasing the burst size, decreasing the tendency for proteins to not be bound to the DNA results in a different switching mechanism. The probability of the genes to be on falls to far smaller values than the $0.5$ of the $N=1$ case. If the burst size is large both genes have a very low probability of being on before the critical number of proteins necessary for bifurcation is achieved. The same effect is observed if proteins are more likely to bind to the DNA (small $X^{eq}$) (Fig. \ref{burstprndep} (b)). When the genes are more probable to bind a repressor and successful unbinding events are rare, earlier bifurcations in terms of $X^{ad}$ result. As $X^{eq}$ increases, the probability of the genes to be on at the bifurcation point decreases as repressors have a higher tendency of unbinding.\\
For very high values of the adiabaticity parameter, corresponding to high unbinding rates form the DNA binding site, the stable solution which corresponds to the off state and the unstable state merge and the system is monostable again, with only the on state present. This limit is also reached by keeping $X^{ad}$ fixed but taking the burst size $N \rightarrow \infty$.\\
\begin{figure}
\begin{minipage}[t]{.43\linewidth}
\includegraphics[height=3cm,width=4cm]{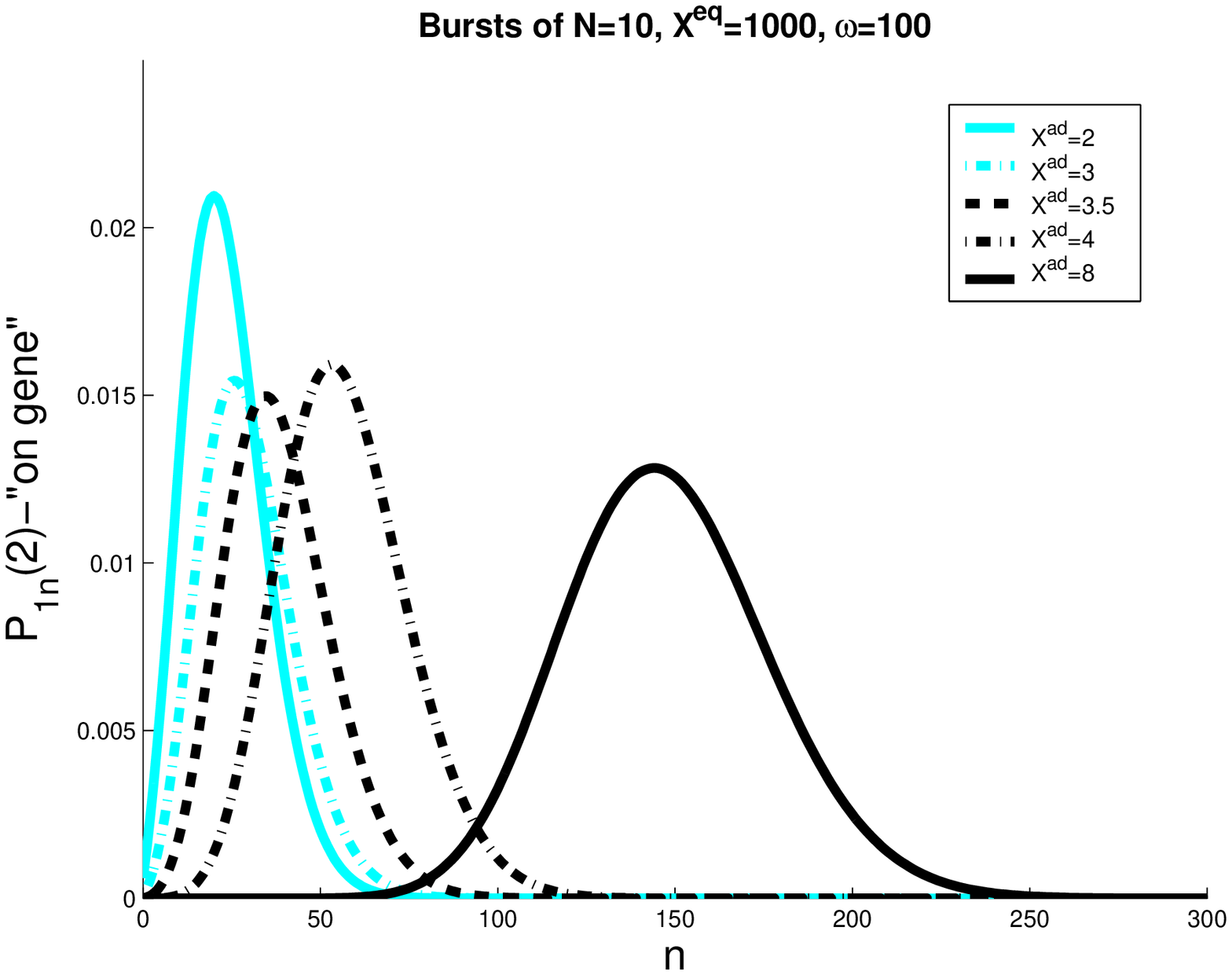}
\end{minipage}\hfill
\begin{minipage}[t]{.5\linewidth}
\includegraphics[height=3cm,width=4cm]{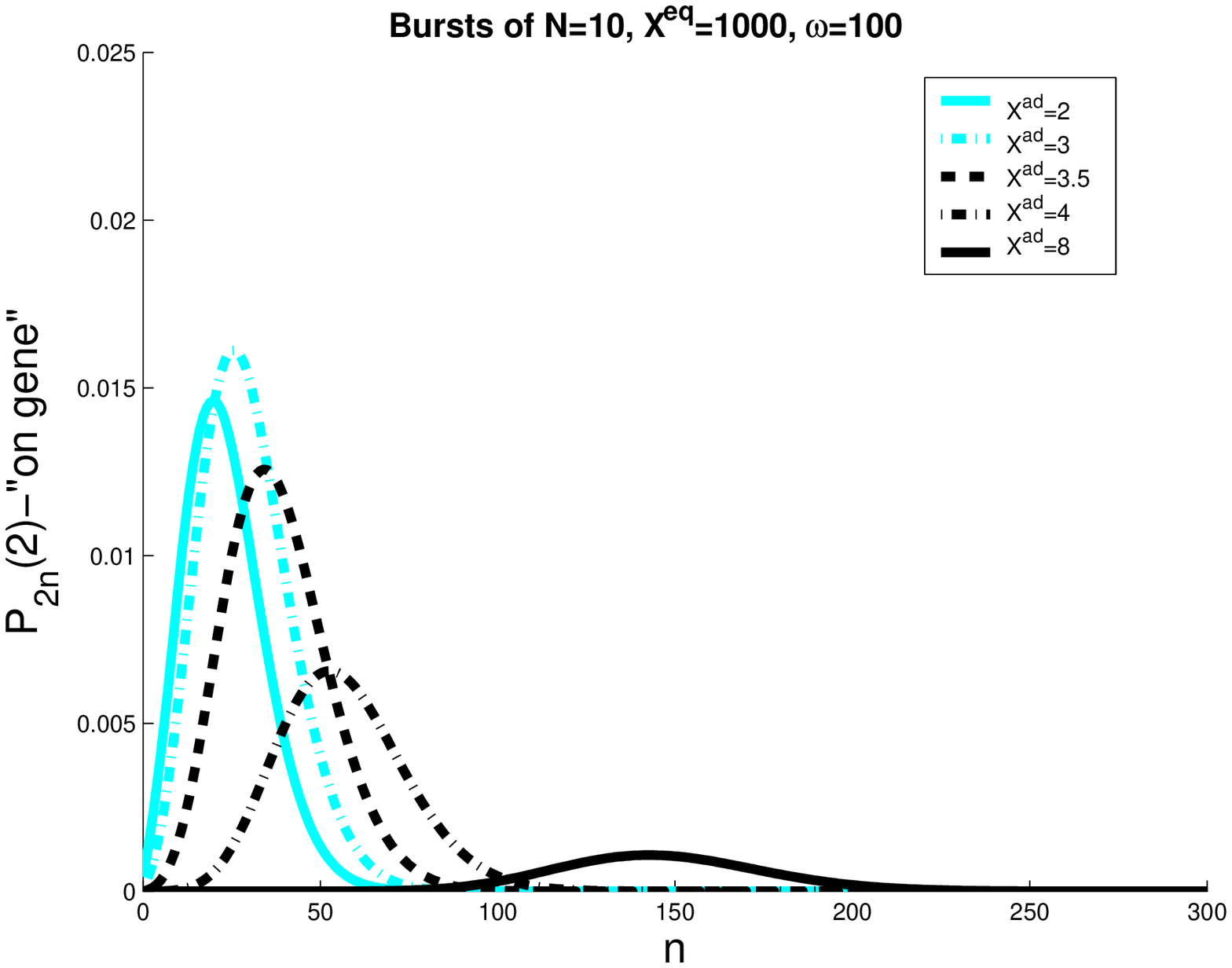}
\end{minipage}
\caption{The evolution of the probability distribution of the gene that is on after the bifurcation, to be on (a) and off (b) as a function of $X^{ad}$ for a switch when proteins are produced in bursts of $N=10$, $X^{eq}=1000$, $\omega=100$. Bifurcation point at $X^{ad}=\delta X^{sw}=35$.}
\label{burstprdist}
\end{figure}
\begin{figure}
\begin{minipage}[t]{.43\linewidth}
\includegraphics[height=3cm,width=4cm]{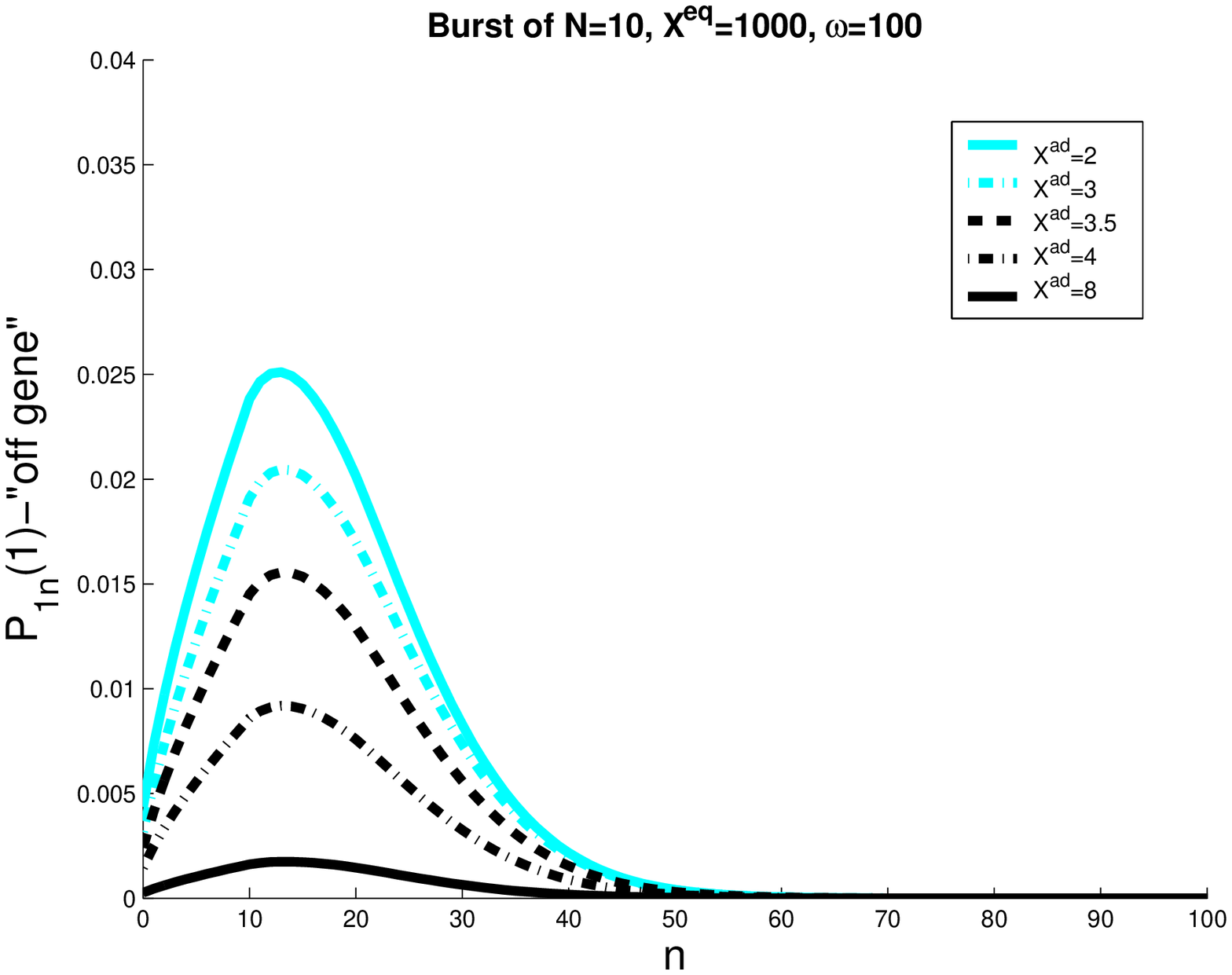}
\end{minipage}\hfill
\begin{minipage}[t]{.5\linewidth}
\includegraphics[height=3cm,width=4cm]{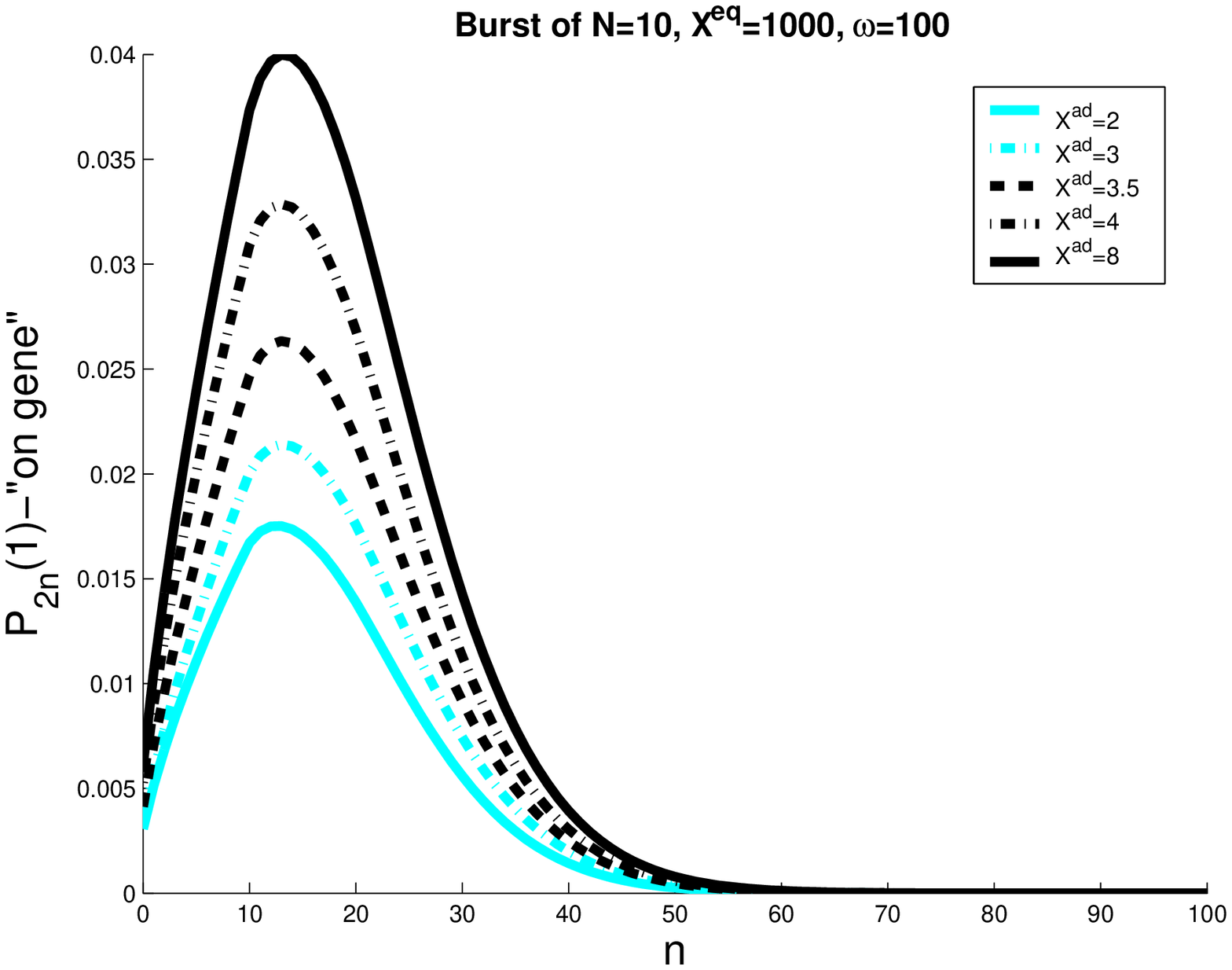}
\end{minipage}
\caption{The evolution of the probability distribution of the gene that is off after the bifurcation, to be on (a) and off (b) as a function of $X^{ad}$ for a switch when proteins are produced in bursts of $N=10$, $X^{eq}=1000$,$\omega=100$. Bifurcation point at $X^{ad}=\delta X^{sw}=35$.}
\label{burstprdist2}
\end{figure}
\textbf{Probability distributions}\\
In the case of the rates used in Fig. \ref{burstprdist} and Fig. \ref{burstprdist2}, $n_c=32$ is the same as for $N=1$, but we note a tenfold decrease in $X_c^{ad}$ compared to when proteins are produced separately. When proteins are produced in bursts, the probability distributions have tails towards larger $n$, as opposed to the distributions for individual protein production. The mean number of proteins in the system for given states of the switch is similar to that of the $N=1$ case, however the distributions with bursts are much broader, as could be expected. In this case even very fast unbinding rates from the DNA cannot correct for the enormous protein number fluctuations and one must explicitly keep track of the change of the DNA binding state. A system in which proteins are produced in bursts is very noisy, especially compared to the nearly deterministic case of proteins binding as tetramers.\\ 
\begin{figure}
\begin{minipage}[t]{.43\linewidth}
\includegraphics[height=3cm,width=4cm]{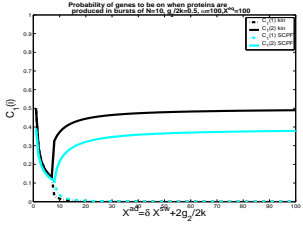}
\end{minipage}\hfill
\begin{minipage}[t]{.5\linewidth}
\includegraphics[height=3cm,width=4cm]{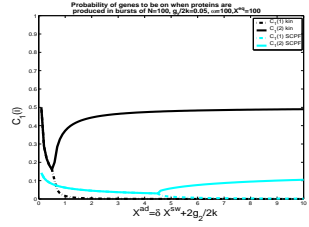}
\end{minipage}
\caption{Probability that gene $i$ is on when proteins are produced in bursts of $N=10$ with a basal effective production rate $g_2/2k=0.5$ (a) and $N=100$, with a basal effective production rate $g_2/2k=0.05$ (b), symmetric switch proteins bind as dimers, $X^{eq}=100$, $\omega=100$. Comparison of deterministic and stochastic solutions.}
\label{burstg2higho}
\end{figure}
\begin{figure}
\begin{minipage}[t]{.43\linewidth}
\includegraphics[height=3cm,width=4cm]{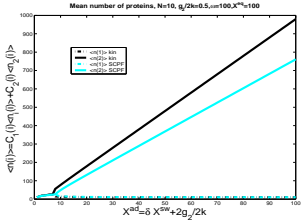}
\end{minipage}\hfill
\begin{minipage}[t]{.5\linewidth}
\includegraphics[height=3cm,width=4cm]{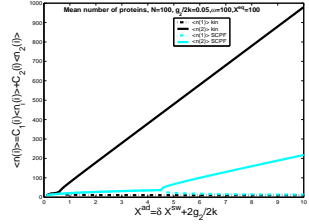}
\end{minipage}
\caption{Mean number of proteins produced by each gene when proteins are produced in bursts of $N=10$ (a) with a basal effective production rate $g_2/2k=0.5$ and $N=100$, with a basal effective production rate $g_2/2k=0.05$, symmetric switch proteins bind as dimers, $X^{eq}=100$, $\omega=100$. Comparison of deterministic and stochastic solutions.}
\label{burstpro1}
\end{figure}
\begin{figure}
\begin{minipage}[t]{.43\linewidth}
\includegraphics[height=3cm,width=4cm]{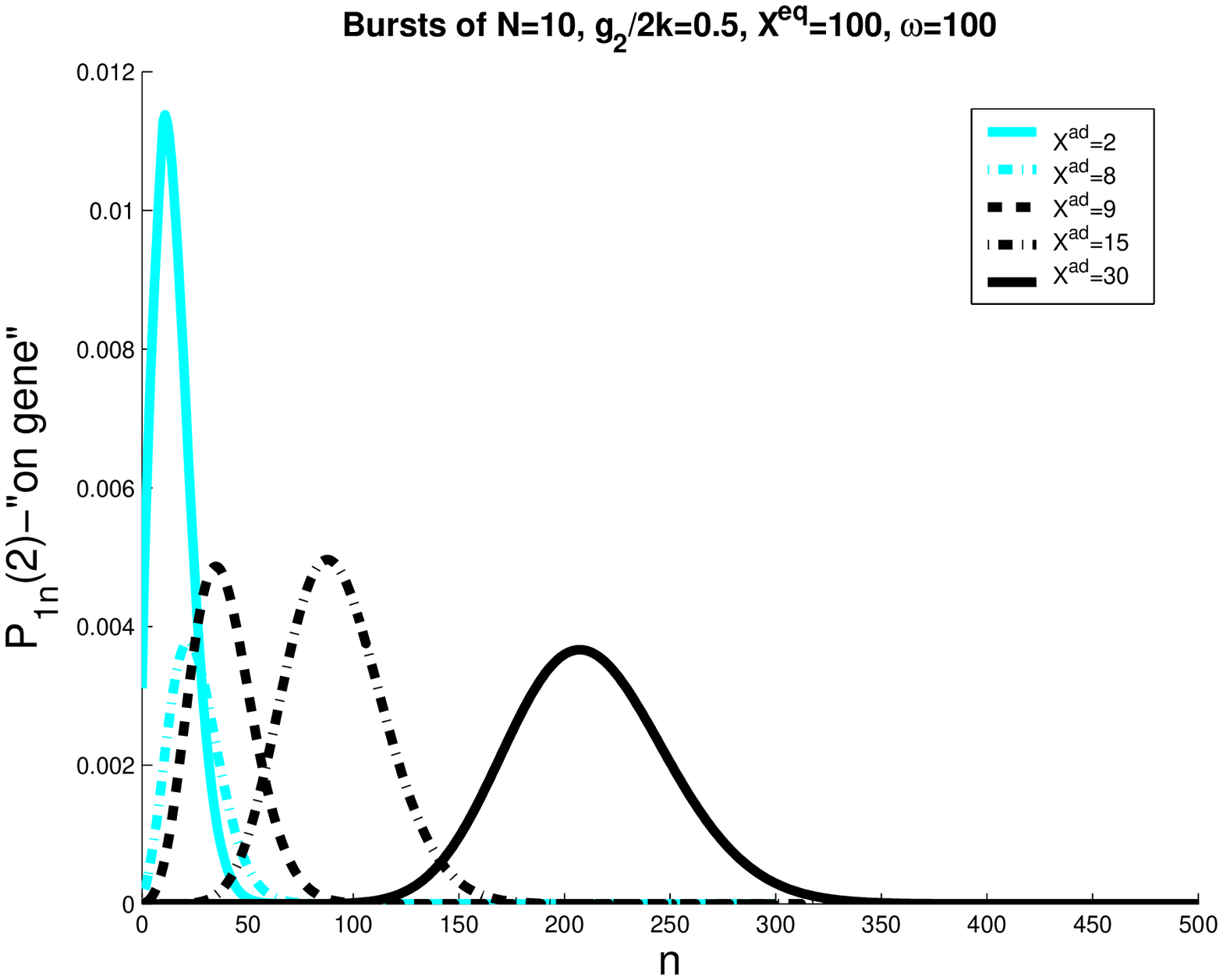}
\end{minipage}\hfill
\begin{minipage}[t]{.5\linewidth}
\includegraphics[height=3cm,width=4cm]{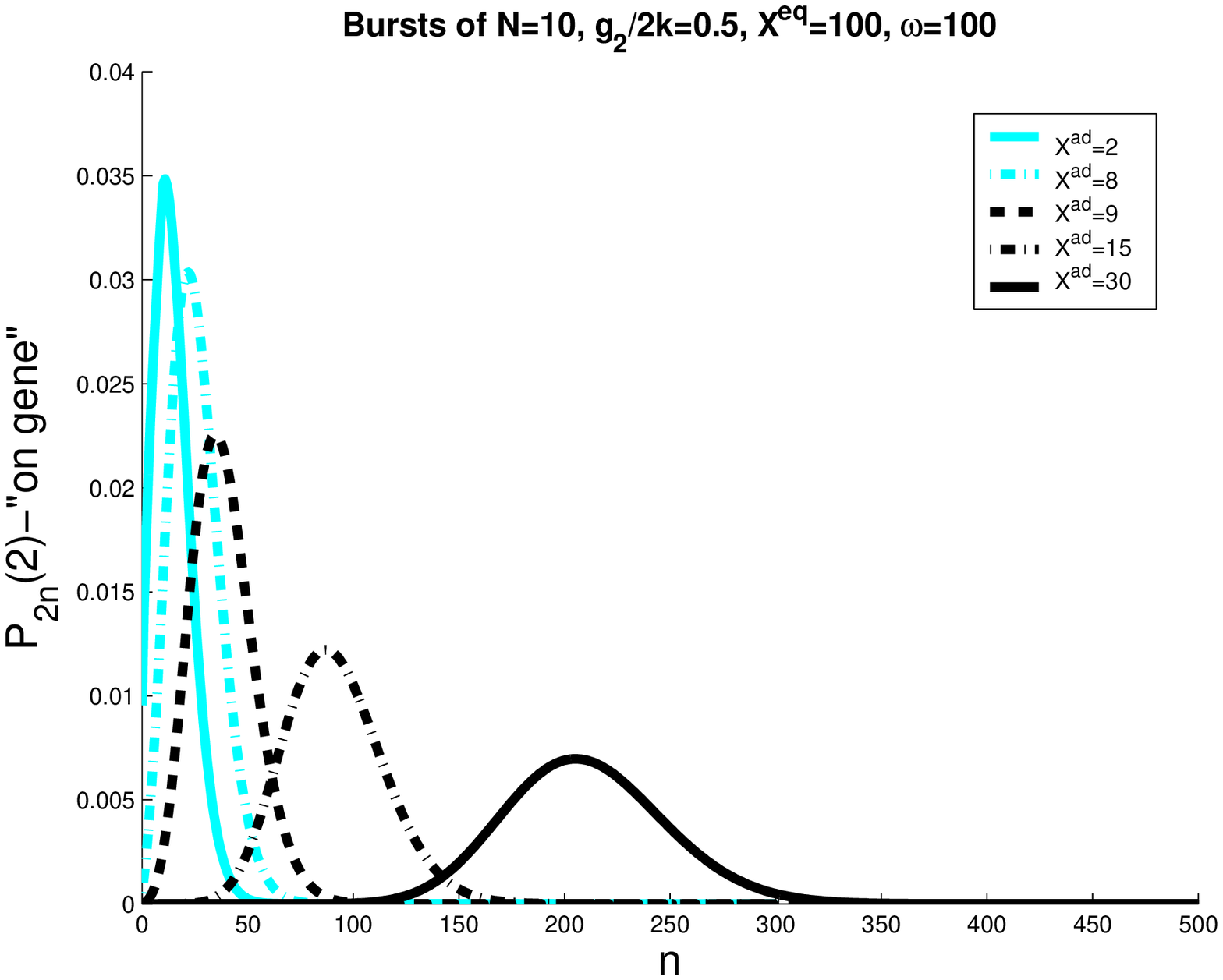}
\end{minipage}
\caption{The evolution of the probability distribution of the gene that is on after the bifurcation, to be on (a) and off (b) as a function of $X^{ad}$ for a switch when proteins are produced in bursts of $N=10$ with a basal effective production rate $g_2/2k=0.5$, $X^{eq}=100$, $\omega=100$. bifurcation point at $X^{ad}=\delta X^{sw}+2g_2/2k=8$.}
\label{burstprdist01}
\end{figure}
\begin{figure}
\begin{minipage}[t]{.43\linewidth}
\includegraphics[height=3cm,width=4cm]{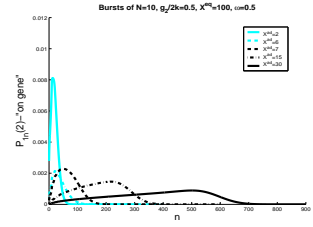}
\end{minipage}\hfill
\begin{minipage}[t]{.5\linewidth}
\includegraphics[height=3cm,width=4cm]{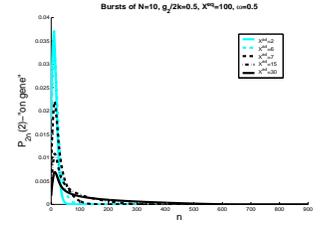}
\end{minipage}
\caption{The evolution of the probability distribution of the gene that is on after the bifurcation, to be on (a) and off (b) as a function of $X^{ad}$ for a switch when proteins are produced in bursts of $N=10$ with a basal effective production rate $g_2/2k=0.5$, $X^{eq}=100$, $\omega=0.5$. Bifurcation point at $X^{ad}=\delta X^{sw}+2g_2/2k=6$.}
\label{burstprdist4}
\end{figure}
\textbf{Nonzero basal effective production rate}\\
If there is a nonzero basal production rate the difference between the deterministic and stochastic solutions is also qualitative even for relatively small burst sizes. In this case proteins are also produced in the off state, so there the number of repressors produced by the off gene after the bifurcation is nonzero, but equal to the burst size $N$, since $<n(i)>=N(X^{ad}+\delta X^{sw}(2 C_1(i)-1)) \rightarrow^{C_1(1)\rightarrow 0} N 2 \frac{g_2}{2 k}$. This number is equal for both the stochastic and deterministic solutions and is equal to $10$ in the examples presented in Fig. \ref{burstpro1}. So production in bursts maintains a high level of repressor proteins, even for very small $\frac{g_2}{k}$ values if the burst size is large. When using experimental data one must be very careful to consider the burst size when assuming the basal production level is zero. Furthermore, the value of the interaction function of the gene in the off state ($C_1(i)\sim 0$) for the stochastic case is much larger than for the deterministic case, due to the multiplication of $<n(i)>^2$ which gives $F(i) \rightarrow <n(i)>^2(1+\frac{k}{2 g_2})+N \frac{g_2}{2 k}$, for large $\omega$, the effect of which is shown in Fig. \ref{burstg2higho}. The number of repressor proteins produced by the off gene decreases as $g_2 \rightarrow 0$, as expected and the probability of the on gene to be active tends to one, as is shown in Fig. \ref{burstprndep1} (a). The dependence of the effective production rate at which the bifurcation occurs on the adiabaticity parameter is analogous to that of $g_2=0$ case. The very small $\omega$ cases are shown explicitly in Fig. \ref{burstprndep1}. The probability distributions for the gene which is active after the bifurcation in the on and off state are presented in Fig. \ref{burstprdist01}, for large unbinding rates from the DNA, and Fig. \ref{burstprdist4}, for small unbinding rates from the DNA. They exhibit maxima around $2 X^{ad}$ for the on state and $2 \frac{g_2}{2 k}$ for the off state and display behavior analogous to that of proteins produced separataly, apart from the different curvature of the slopes for $n<N$ and $n>N$. For small $\omega$ values the protein numbers reach a steady state before the DNA states, hence we observe bimodal probability distributions. The mechanism of competition in this noisy burst system is different than in the single protein production case. If the gene is in the on state, probability states with higher $n$ values are strongly occupied and there is hardly any probability flux into the lower $n$ states. In the off state however, a flux pushes the system into the lower $n$ states, essentially trapping it there, hence the difference in the slopes, as can be seen in Fig. \ref{burstprdist4}. This is also true for the $g_2=0$ system when proteins are produced in bursts.\\
 \subsection*{Limitations of the SCPF Treatment}
The examples presented above cover a large class of two gene switches, all of which are exactly solvable within the SCPF approximation. An exact solution may be obtained within this approximation for systems of genetic networks and switching cascades. However the SCPF approximation does not allow for an exact analytical solution of all systems. If we try to model one of the simplest natural systems where regulation is achieved by means of a switch, that is the $\lambda$ switch, we encounter a problem. The genes in the $\lambda$ switch, apart from having a toggle like regulation, also exhibit auto-regulation, that is cI proteins can bind to OR3, repressing the cI gene, and the Cro proteins can bind to OR1 or OR2, enabling the RNA polymarase from transcribing the Cro gene (Ptashne, 1992), (Ptashne and Gann, 2002). If we expand the master equation to account for self-regulation we add a $h_i n_i^p$ binding term to the $P_i(n_i)$ equations. Therefore the $k^{th}$ moment equation will display a dependence on the $k+p^{th}$ moment and the set of equation will not exhibit closure. One can find the probability distribution for a single self-regulating single gene. However if we consider as system like the $\lambda$ phage, where self regulation is also combined with regulation by another gene, the problem is no longer solvable exactly and demands a cutoff of the hierarchy or other approximations. We can nevertheless treat these systems using the variational method, as proposed by Sasai and Wolynes (Sasai and Wolynes, 2003). The fact that self-regulation renders the system incompletely solvable within the SCPF approximation, is not surprising, since it corresponds to the exact solution for such a system. Gene $i$ is influenced only by the number of proteins it produces. It is independent of the state of the other gene. Therefore, as one would expect the full solution should depend on all moments of the distribution of gene $i$. However for systems such as the $\lambda$ phage, we can treat all inter gene regulation effects exactly and truncate the self-regulation equation at the highest order of the inter gene interaction, which would be six, corresponding to, for example, 3 cI proteins binding to the 3 operator sites. 
\subsection*{Conclusions}
The self-consistent proteomic field approximation for stochastic switches reproduces many intuitive notions about their behavior. It proves to be a a very powerful tool that allows for the consideration, of all but one, of the basic building blocks of more general switches and networks. A switch with a self-repressing/activating gene cannot be solved exactly within the SCPF approximation, as in this case the approximation is equivalent to the full solution. Therefore the probability distribution is determined by an infinite number of moments. The probability distributions obtained for the systems considered in this paper are not symmetric and exhibit long tails. This anticipates problems for using the variational principle for finding probability distributions when one accounts for correlations between the two states. The possibility to expand this method to consider networks and cascades will allow for are more realistic treatment of complex systems with emergent behavior at low computational costs.\\
 One can account for the mRNA step in the system by a adding a deterministic step which using a deterministic kinetic rate equation translates the number of mRNA molecules into proteins produced in bursts. This is a valid procedure, as as separately shown by (Thattai and van Oudenaarden, 2001) and (Swain et al., 2002), transcription noise is just amplified in the translation process. Therefore treating the mRNA step deterministically simply introduces another constant into the discussed case of proteins produced in bursts. Therefore the presented treatment of proteins produced in bursts with a modified effective production rate is a simple model of including mRNA in the system. Of course, the effect of mRNA is much more complicated, as it also introduces, for example time delay, between binding and production. This model in the present state neglects these effects.\\
Our analysis of a large class of switches, shows how particular elements contribute to the emergent behavior of functioning switches. Comparison of the stochastic and deterministic treatments of a single gene switch shows convergence in the region of fast rates of unbinding from the DNA compared to protein number fluctuations and large effective production rates. For symmetric switches when proteins are produced separately the two solutions converge after the bifurcation, but often differ when defining the region of parameter space, where the bifurcation occurs. The agreement between the deterministic and stochastic solutions, is especially good for symmetric switches, with $N=1$ and a non-zero basal production rate. However even though the mean repressor protein levels in the cell are similar in both approximations, the probability distributions are broad and far from Poissonian, i.e. they are not completely characterized by these means. If the adiabaticity parameter is small ($\omega<1$) the protein number state reach a steady state before the DNA binding state and we observe a bimodal probability distribution. For the symmetric switch noise has a destructive effect on the region of bistability.  Increasing the adiabaticity parameter facilitates the formation of a buffering proteomic cloud around a gene, which leads to repression at lower effective production rates than for small $\omega$.\\
As was already mentioned, the symmetric switch is hard to design and build experimentally. The asymmetric switch, which is the experimental toy system, is much more susceptible to noise than the symmetric switch and stochasticity has not only the destructive effect on the region of stability one might expect, but also introduces new phenomena and can be utilized to increase the bistable region. This is of fundamental importance, since experimentally one deals with asymmetric switches and these offer greater possibilities in artificially engineering new systems. As can also be learned from the asymmetric switch as well as from the analysis of binding of different oligomers, the region of bistability of a switch grows with increasing the interaction function. When creating artificial switches, one may argue a large region of bistability may be desired, so the switch reacts by the forward or backward transition to very specific concentrations or production levels of a protein. If the experimental setup constrains the protein production rates, this can also be achieved by modifying the adiabaticity parameters of the system, which ensures the transition remains rapid and effective. Asymmetric switches, exhibit first order phase transitions. This size of the region of phase space, in which the forward and backward transitions occur grows with the tendency that proteins are unbound from the DNA of both genes. Large adiabaticity parameters stabilize the buffering proteomic cloud around the repressed gene and lead to the formation of an effectively repressing cloud for smaller numbers of repressors, in the forward transition, than for small $\omega$, for the active gene.\\
Experimental data available at this point  (Darling et al., 2000), suggest biological switches function in regions of high adiabaticity parameters from the deterministic point of view. Nevertheless, even for large values of adiabaticity parameters one must account for the DNA binding site fluctuations explicitly when proteins are produced in bursts. The deterministic solutions give qualitatively wrong results in biologically relevant areas of parameter space. The stochastic solutions for large burst sizes suggest that the bifurcation of the solution is a result of destabilizing of the repressor cloud buffering the DNA, not  formation of the cloud as for smaller burst systems. The probability distribution therefore exhibit tails towards large $n$ values, not as in the small $N$ case towards small $n$ values. The deterministic kinetics remains unchanged for large burst sized, unlike the stochastic kinetics, hence presenting results derived from a wrong mechanism. The definition of the adiabatic limit, when proteins are produced in bursts is not clear as in the $N=1$ case, when it corresponds simply to $\omega<1$. This ambiguity does not allow one to integrate out the degrees of freedom corresponding to the change in DNA binding site occupation. Such an approximation leads one to erroneously identify the regions of bistability. The switch with a nonzero basal production rate when proteins are produced in bursts results in probabilities to be on and mean numbers of proteins in the cell very different from those of the deterministic solution, even for small effective basal production rates. If proteins are produced in bursts assuming that a small effective basal production rate may be approximated by a zero rate may be misleading. Binding of proteins produced in bursts results in a bifurcation transition for smaller values of the effective production rate. It is also a mechanism for making two genes in an asymmetric switch more competitive.\\
Binding of higher order oligomers leads to results closer to those of deterministic treatments, with narrower probability distributions. This can be experimentally used to stabilize DNA binding states. In this simple model tetramers seem to be the most optimum binders,. The close to deterministic all or nothing switching they offer may be worth the effective cost of the energy of multimerization and diffusion of particles. Binding of higher order oligomers may be viewed as a simple model of cooperativity, which increases the competitiveness of genes in an asymmetric switch. Within the SCPF approximation monomers do not make good switches due to lack of nonlinearity in protein concentration. They do not exhibit a region of bistability. This model neglects any structural DNA-protein interactions and spatial dependence. Hence this conslusion is simply a result of the lack of cooperativity in the system. For small adiabaticity parameters, they do however exhibit bimodal probability distributions, unlike in the large $\omega$ limit. \\ 
The thorough investigation of different components of gene regulatory networks using the self-consistent proteomic field approximation provides a tool kit for engineering new switches and networks. Based on our analysis, if one would want to build a strong component of a switch out of a gene with relatively small chemical parameters, one could use components that utilize binding of tetramers and that produce proteins in bursts. This is what the $Cro$ gene in the $\lambda$ switch uses.\\
\noindent{\bf Acknowledgements:}
AMW and PGW were supported by the Center for Theoretical Biological
Physics through National Science Foundation Grants PHY0216576 and
PHY0225630. MS was supported by ACT-JST project of Japan Science and Technology 
Corporation and by grants from the Ministry of Education, Culture, Sports, 
Science, and Technology, Japan.
\subsection*{References}
\noindent Ackers, G. K., Johnson, A. D., and M. A. Shea. 1982. Quantitative model for gene regulation by lambda-phage repressor. {\em PNAS}, 79:1129--1133.\\
\noindent  Arkin, A., Ross, J., and H.H. McAdams. 1998. Stochastic kinetic analysis of developmental pathway bifurcation in phage lambda-infected escherichia coli cells. {\em Genetics}, 149:1633--1648.\\
\noindent  Aurell E., Brown S, Johanson J, and K. Sneppen. 2002. Stability puzzles in phage lambda. {\em PRE}, 65:051914--1--051914--9.\\
\noindent  Buchler N. E., Gerland U., and T. Hwa. 2003. On schemes of combinatorial transcription logic. {\em PNAS}, 100:5136--5141.\\
\noindent  Bialek, W. 2001. Stability and noise in biochemical switches. {\em Advances in Neural Information Processing}, 13:103--109.\\
\noindent  Becskei, A., Seraphin, B., and L. Serrano. 2001. Positive feedback in eukaryotic gene networks: cell differentiation by graded to binary response conversion. {\em EMBO J}, 20:2528--2535.\\
\noindent  Cook D. L., Gerber, A. N., and S. J. Tapscott. 1998. Modeling stochastic gene expression: Implications for haploinsufficiency. {\em PNAS}, 95:15641--15646.\\
\noindent  Darling, P. J., Holt, J. M., and G. K. Ackers. 2000. Coupled energetics of lambda cro repressor self-assembly and site-specific dna operator binding ii: Cooperative interactions of cro dimers. {\em JMB}, 302:625--638.\\
\noindent  Doi., M. 1976. Stochastic theory of diffusion-controlled reaction. {\em J Phys A}, 9:1479--1495. \\
\noindent  Gillespie, D. T. 1977. Exact stochastic simulation of coupled chemical-reactions. {\em J. Phys. Chem}, 81:2340--2361.\\
\noindent  Hasty, J., Issacs, F., Dolnik, M., McMillen, D., and J. J. Collins. 2001. Designer gene networks: Towards fundamental cellular control. 11:207--220.\\
\noindent  Hasty, J., McMillen, D., Issacs, F., and J. J. Collins. 2001. Computational studies of gene regulatory networks: In numero molecular biology. {\em Nature Reviews Genetics}, 2:268--279.\\
\noindent  Hasty, J., Pradines, J., Dolnik, M., and J. J. Collins. 2000. Noise-based switches and amplifiers for gene expression. {\em PNAS}, 97:2075--2080.\\
\noindent  Kepler, T. B., and T. C. Elston. 2001. Stochasticity in transcriptional regulation: Origins, consequences, and mathematical representations. {\em Biophys J}, 81:3116--3136.\\
\noindent  McAdams, H. H, and A. Arkin. 1997. Stochastic mechanisms in gene expression. {\em PNAS}, 94:814--819.\\
\noindent McLure, K. G., and P. W. Lee. 1998. How p53 binds dna as a tetramer. {\em EMBO J.}, 17:3342--3350.\\
\noindent  Paulsson, J., Berg, O. G., and M.Ehrenberg. 2000. Stochastic focusing: Fluctuation-enhanced sensitivity of intracellular regulation. {\em PNAS}, 97:7148--7153.\\
\noindent  Ptashne, M., and A. Gann. 2002. {\em Genes and Signals}. Cold Spring Harbor Laboratory Press, New York.\\
\noindent  Ptashne, M. 1992. {\em A genetic switch}. Cell Press and Blackwell Science, 2nd edition.\\
\noindent  Sneppen, K., and E. Aurell. 2002. Epigenetics as a first exit problem. {\em PRL}, 88:048101--1--048101--4.\\
\noindent  Swain, P. S., Elowitz, M. B., and E.D. Siggia. 2002. Intrinsic and extrinsic contributions to stochasticity in gene expression. {\em PNAS}, 99:12795--12800.\\
\noindent Sasai, M. and P. G. Wolynes. 2003. Stochastic gene expression as a many-body problem. {\em PNAS}, 100:2374--2379.\\
\noindent Thattai, M., and A. van Oudenaarden. 2001. Intrinsic noise in gene regulatory networks. {\em PNAS}, 98:8614--8619.\\
\noindent  Vilar, J. M. G, Guet, C. C., and S. Leibler. 2003. Modeling network dynamics: the lac operon, a case study. {\em JCB}, 161:471--476.\\
\noindent  Zeldovich, Y. B., and A. A. Ovchinikov. 1978. Mass-action law and kinetics of chemical-reactions with allowance for thermodynamic density fluctuations. {\em Sov. Phys. JETP}, 74:1588--1598.\\
\end{document}